\pdfoutput=1
\documentclass[12pt]{article}

\usepackage[margin=1in]{geometry}
\usepackage{amsmath,amssymb,amsfonts,amsthm,mathtools}
\usepackage{bm}
\usepackage{mathrsfs}

\usepackage{graphicx}
\graphicspath{{figs/}}
\usepackage{booktabs}
\usepackage{longtable}
\usepackage{multirow}
\usepackage{caption}
\usepackage{subcaption}
\usepackage{float}
\usepackage{algorithm}
\usepackage[noend]{algpseudocode}

\usepackage{enumitem}
\usepackage{xcolor}
\usepackage{url}
\usepackage{etoolbox}
\usepackage{pdfpages}

\usepackage[authoryear,round]{natbib}
\usepackage[colorlinks=true,allcolors=blue]{hyperref}

\newcommand{\e}{\mathbb{E}}
\numberwithin{equation}{section}
\allowdisplaybreaks[4]

\theoremstyle{plain}
\newtheorem{theorem}{Theorem}
\newtheorem{proposition}{Proposition}
\newtheorem{corollary}{Corollary}
\newtheorem{assumption}{Assumption}
\theoremstyle{definition}

\newcommand{\SuppNum}[1]{\textcolor{blue}{#1}}
\newcommand{\SuppKernelSupportAssumption}{\SuppNum{S5}}

\newcommand{\SuppMeanSurfaceAssumption}{\SuppNum{S7}}
\newcommand{\SuppDependenceAssumption}{\SuppNum{S8}}
\newcommand{\SuppCovarianceAssumption}{\SuppNum{S9}}

\newcommand{\SuppLongitudinalConsistencyCorollary}{\SuppNum{S1}}
\newcommand{\SuppOracleToFeasibleProposition}{\SuppNum{S3}}
\newcommand{\SuppSensitivityLPProposition}{\SuppNum{S4}}

\title{When Treatments Spill Over: A Neural Framework for Spatiotemporal Causal Effect Estimation}
\author{%
  Shaohui Zhang\textsuperscript{1},
  Jiasheng Shi\textsuperscript{2}\thanks{Corresponding author:
    \href{mailto:shijiasheng@cuhk.edu.cn}{shijiasheng@cuhk.edu.cn}.},
  and Jing Huang\textsuperscript{1}\\[1ex]
  \small\textsuperscript{1}Department of Biostatistics, Epidemiology and Informatics,\\
  \small University of Pennsylvania, Philadelphia, PA, USA\\
  \small\textsuperscript{2}School of Data Science,
  The Chinese University of Hong Kong, Shenzhen, Guangdong, China}
\date{}

\begin{document}

\maketitle

\begin{abstract}
Real-world public health policies are often enacted by local jurisdictions, but their effects can ripple through connected populations. For causal evaluation, this creates a difficult spatiotemporal problem: intermittent policy updates create path-level treatments, and because spillover depends on neighboring treatment paths along with the focal-unit's own history, controlling time-varying confounding requires high-dimensional, multimodal information from both sources. To address this challenge, we develop a neural spatiotemporal spillover framework that extends potential-outcome causal analysis to networked policy-systems with unit-level effects and structured interference. The framework indexes each unit's potential outcomes through own-treatment and neighborhood spillover-exposure paths, yielding direct, spillover, and total effects. We introduce an LSTM-based representation-learning encoder that integrates each unit's own and neighborhood histories into compact sequential summaries for downstream modeling and causal estimation. These summaries feed a self-normalized kernel inverse-probability-weighting estimator for continuous spillover exposure. Beyond theoretical guarantees, simulations show improved performance over outcome-regression baselines; in a practically motivated county-level COVID-19 policy study, the framework finds both business-and-economic restrictions and education-and-childcare policies associated with lower subsequent case burden, with larger estimates for business-and-economic restrictions. Overall, it provides a scalable, policy-interpretable framework for evaluating unit-level effects in high-dimensional, multimodal spatiotemporal systems when interventions spill over across space and time.
\end{abstract}

\noindent\textbf{Keywords:} High-dimensional and Multi-Modal Data, Network Interference, Policy Evaluation, Neural Representation Learning, Spatiotemporal Causal Inference, Spillover Effects

\section{Introduction}

Public health policies are often designed and implemented by local jurisdictions, yet they evolve over time and their effects spill over across connected populations. The COVID-19 pandemic made this tension especially visible: U.S. counties and states repeatedly updated nonpharmaceutical interventions (NPIs), including business restrictions, mask mandates, school policies, and gathering limits, in response to local disease burden and administrative constraints. Because jurisdictions were linked through commuting and travel, economic activity, regional health-care systems, and shared transmission dynamics, a policy adopted in one place could shape outcomes in neighboring or otherwise connected regions \citep{holtz2020interdependence,chang2021mobility,giffin2022estimating,chang2023estimating}. Evaluating such policies therefore requires causal methods that account simultaneously for temporal carryover and spatial spillover.

This spatiotemporal interference structure changes the causal object. Rather than treating policy as a one-time binary exposure, both own treatment and neighborhood spillover exposure should be represented as paths: policy updates and the resulting changes in neighborhood exposure occur at discrete administrative decision times, while outcomes and covariates evolve between decisions and may respond with delay \citep{flaxman2020estimating,hale2021global}. The estimand is therefore a contrast among potential outcomes indexed jointly by own-treatment and neighborhood spillover-exposure paths. This formulation allows direct, spillover, and total effects to be defined as policy-relevant causal contrasts rather than as generic predictive associations.

Several causal-inference literatures address parts of this challenge. For dynamic treatments with time-varying confounding, g-methods, including marginal structural models estimated by inverse-probability-weighting (IPW), address feedback in which covariates may be affected by prior treatment while also influencing future treatment assignment \citep{robins1997causal,robins2008estimation,cole2008constructing,imai2015robust}. Interference methods formalize direct and spillover effects through partial interference, causal diagrams, exposure mappings, and weighting-based estimators \citep{halloran1995causal,hudgens2008toward,hong2006evaluating,ogburn2014causal,aronow2017estimating,forastiere2021identification}. Recent work further combines temporal dependence and interference through treatment-history estimands, stochastic interventions, and spatiotemporal causal estimands \citep{wang2021causal,papadogeorgou2022causal,zhou2024estimating}. Together, these developments provide rigorous foundations, but they often require one or more analyst-specified components: exposure mappings, assignment models, spatial smoothing structures, or low-dimensional summaries of the relevant history. 

The burden of specifying these components becomes especially acute in modern public-health studies, where treatment paths are long, neighborhood exposure is continuous, and valid adjustment must track focal-unit and neighborhood histories over time. These histories may include trajectories of prior covariates, outcomes, and treatments, as well as structured network-connectivity information and, increasingly, unstructured modalities such as text, images, or electronic health records (EHRs) \citep{veitch2020adapting,zhang2023causal,egami2022make}. The resulting adjustment problem is therefore sequential, cross-unit, high-dimensional, and multimodal. To alleviate this burden, causal representation learning offers a natural alternative to fully handcrafted summaries by using generative and encoding-based methods that learn compact representations of complex covariate information for adjustment and effect estimation \citep{louizos2017causal,liu2024encoding,liu2026ai}. Flexible machine-learning methods have also been developed for sequential causal estimation, including recurrent marginal structural weighting, adversarial representation learning, and neural g-computation \citep{lim2018forecasting,bica2020estimating,li2021g}, with emerging neural or network-aware approaches addressing causal estimation under interference \citep{guo2024estimating,ma2020causal}. However, existing work mainly targets own-treatment effect estimation without structured interference or focuses on prediction-oriented representations, leaving a gap in statistical inference for path-indexed direct and spillover effects in spatiotemporal interference settings with continuous neighborhood exposure and distinct time scales for treatment decisions and outcome evolution.

Bridging this gap, we develop a neural spatiotemporal spillover framework that integrates an LSTM-based history encoder with a self-normalized kernel IPW estimator. Building on the path-indexed potential-outcome formulation above, the introduced encoder serves as a flexible upstream dimension-reduction component for causal adjustment, compressing multi-source histories into compact sequential summaries for direct and spillover treatment effect modeling rather than outcome prediction alone. In turn, the self-normalized kernel IPW estimator uses these summaries to weight neighborhood spillover exposure, with kernel smoothing accommodating exposure continuity and self-normalization yielding a stabilized H\'ajek-type construction. This weighting-based construction keeps estimation tied to causal adjustment rather than pure outcome regression, and supports uncertainty quantification under explicit assumptions on the spatiotemporal interference structure, overlap, and the adequacy of the learned summaries. The resulting framework bridges the representational flexibility of neural encoders with the statistical structure of causal weighting, producing policy-interpretable and uncertainty-aware direct, spillover, and total effect estimates that can inform the evaluation and adaptation of public-health interventions in connected populations.

Our contributions are fivefold. First, we formulate a path-indexed potential-outcome framework for spatiotemporal policy evaluation under structured interference, characterizing direct, spillover, and total effects through contrasts of own-treatment and neighborhood spillover-exposure paths. Second, we develop an LSTM-based representation-learning strategy that summarizes high-dimensional, multimodal focal-unit and neighborhood histories for causal adjustment. Third, we propose a self-normalized kernel IPW estimator for continuous neighborhood spillover exposure and establish its theoretical and inferential properties, providing a basis for uncertainty quantification. Fourth, simulation studies show improved finite-sample performance over outcome-regression baselines. Fifth, in a county-level COVID-19 policy application, the framework yields policy-interpretable evidence that both business and economic restrictions and education and childcare policies are associated with lower subsequent case burden, with larger estimates for business and economic restrictions, suggesting that intervention timing and the transmission channels targeted by a policy are central to evaluating public-health responses in connected populations. Together, these contributions provide a scalable, policy-interpretable, and uncertainty-aware causal framework for evaluating unit-level effects in high-dimensional, multimodal spatiotemporal systems when interventions spill over across space and time.

\section{Causal Formulation and Estimands for Spatiotemporal Spillover}

\subsection{Temporal network data and exposure paths}

\begin{figure}[ht]
    \centering
    \includegraphics[width=0.96\linewidth]{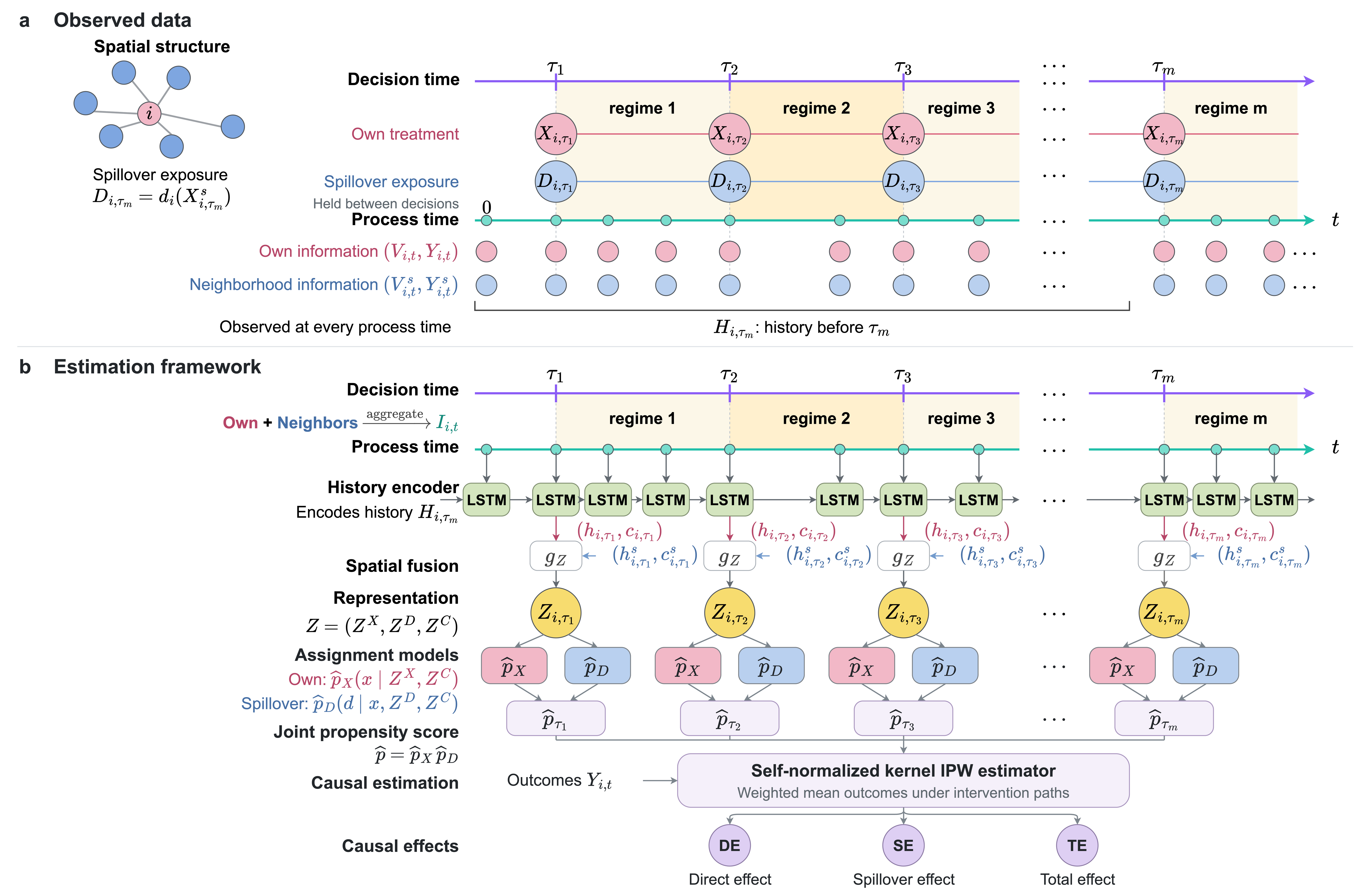}
    \captionsetup{font=small}
    \caption{Temporal network data and causal estimation framework.
    \textbf{(a)} Own treatment and neighborhood spillover exposure are updated at decision times and held fixed between decisions, while covariates, outcomes, and neighborhood information evolve over process time.
    \textbf{(b)} Estimation framework. An LSTM encodes pre-decision histories, and $g_Z$ combines focal-unit and neighborhood states for joint propensity-score estimation. A self-normalized kernel IPW estimator uses these scores and observed outcomes to estimate direct (DE), spillover (SE), and total (TE) effects.}
    \label{fig:background}
\end{figure}


Consider a longitudinal network setting with units \(i\in\{1,\ldots,n\}\)
observed over process times \(t\in\{0,1,\ldots,T\}\). Treatment decisions are
made only at a subset of these times, $1 \leq \tau_1 < \tau_2 < \cdots < \tau_L \leq T$, which we refer to as decision times. For convenience, set \(\tau_{L+1}=T\) as the terminal process-time boundary. This two-scale structure, illustrated in Figure~\ref{fig:background}(a), reflects a common feature of policy applications in which policies are updated only at discrete decision times and remain in force between decisions, whereas outcomes, covariates, and neighborhood summaries continue to evolve over process time.

Across these two time scales, unit \(i\) receives an own treatment
\(X_{i,\tau_m}\in\{0,1\}\) at decision time \(\tau_m\), while its covariates
\(V_{i,t}\) and outcome \(Y_{i,t}\) are observed at process time \(t\). Beyond
this temporal structure, units are connected through a prespecified network. Let $\mathcal G_n$ denote the $\sigma$-field generated by the observed network and
the baseline information treated as fixed throughout the analysis. Unless
otherwise stated, all expectations, probabilities, and asymptotic statements
are understood conditionally on $\mathcal G_n$. Let
\(\mathcal N^1(i)\) denote the set of first-order neighbors of unit \(i\), and let \(N_i=|\mathcal N^1(i)|\). We denote the corresponding neighborhood treatments, covariates, and outcomes by \(X^s_{i,\tau_m}\), \(V^s_{i,t}\), and \(Y^s_{i,t}\), respectively, with \(X^s_{i,\tau_m}=(X_{j,\tau_m}:j\in\mathcal N^1(i))\) and analogous definitions for \(V^s_{i,t}\) and \(Y^s_{i,t}\).

To represent neighboring treatment exposure on a common scale, we introduce an
exposure mapping with fixed output dimension \(q\), defined by
\begin{equation*}
    d_i:\{0,1\}^{N_i}\rightarrow\mathbb R^q, \qquad D_{i,\tau_m}=d_i(X^s_{i,\tau_m}).
\end{equation*}
This yields a common \(q\)-dimensional spillover-exposure summary even when the number of first-order neighbors varies across units. Scalar examples include the proportion of treated neighbors, \(N_i^{-1}\sum_{j\in\mathcal N^1(i)}X_{j,\tau_m}\), and the weighted average \(\sum_{j\in\mathcal N^1(i)}w_{ij}X_{j,\tau_m}\), where \(\sum_{j\in\mathcal N^1(i)}w_{ij}=1\) and the weights may encode spatial proximity, mobility flow, tie strength, or other network connectivity.

Write \(X_{i,\tau_{1:m}}=(X_{i,\tau_1},\ldots,X_{i,\tau_m})\) for the
own-treatment history up to \(\tau_m\), and define \(D_{i,\tau_{1:m}}\)
analogously for the spillover-exposure history. We use
\(V_{i,0:t}\), \(Y_{i,0:t}\), \(V^s_{i,0:t}\), and \(Y^s_{i,0:t}\) for
process-time histories. The information available immediately before treatment
assignment at decision time \(\tau_m\) is collected in
\[
H_{i,\tau_m}
=
\left(
X_{i,\tau_{1:m-1}},
D_{i,\tau_{1:m-1}},
Y_{i,0:\tau_m-1},
Y^s_{i,0:\tau_m-1},
V_{i,0:\tau_m-1},
V^s_{i,0:\tau_m-1}
\right),
\]
where, by convention, \(X_{i,\tau_{1:0}}\) and \(D_{i,\tau_{1:0}}\) denote empty histories. Within a period the treatment decision precedes the outcome, so \(Y_{i,t}\) responds to every decision made at \(\tau_m\leq t\). This pre-decision history serves as the adjustment history in the subsequent causal formulation and as the high-dimensional longitudinal object summarized by the representation encoder in Section~3.

\subsection{Potential outcomes and causal effects}

Building on the exposure paths above, potential outcomes are indexed by the two
intervention histories relevant for unit \(i\). For an outcome time $t$, let
$M_t=\max\{m:\tau_m \leq t\}$ index the most recent decision time at or before $t$, and
write $Y_{i,t}(x_{\tau_{1:M_t}},d_{\tau_{1:M_t}})$ for the potential outcome of
focal unit $i$ under own-treatment path $x_{\tau_{1:M_t}}$ and
spillover-exposure path $d_{\tau_{1:M_t}}$. This notation treats the prespecified network as encoding the direct spillover channels, so treatment paths outside
\(\{i\}\cup\mathcal N^1(i)\) do not enter the unit-level potential outcome
directly, while neighboring treatment assignments are represented through the
exposure path \(d_{\tau_{1:M_t}}\). 

For a target pair of own-treatment and spillover-exposure paths, $(x_{\tau_{1:M_t}},d_{\tau_{1:M_t}})$ and $(x'_{\tau_{1:M_t}},d'_{\tau_{1:M_t}})$, we assume that
the conditional mean potential outcomes are common across
units, although their distributions conditional on $\mathcal{G}_n$ may otherwise differ. Accordingly, the direct, spillover, and total causal effects are defined as
\begin{equation}
\begin{aligned}
\mathrm{DE}_t
&=
\e\left[
Y_{i,t}(x_{\tau_{1:M_t}},d_{\tau_{1:M_t}})
-
Y_{i,t}(x'_{\tau_{1:M_t}},d_{\tau_{1:M_t}})
\right],\\
\mathrm{SE}_t
&=
\e\left[
Y_{i,t}(x_{\tau_{1:M_t}},d_{\tau_{1:M_t}})
-
Y_{i,t}(x_{\tau_{1:M_t}},d'_{\tau_{1:M_t}})
\right],\\
\mathrm{TE}_t
&=
\e\left[
Y_{i,t}(x_{\tau_{1:M_t}},d_{\tau_{1:M_t}})
-
Y_{i,t}(x'_{\tau_{1:M_t}},d'_{\tau_{1:M_t}})
\right].
\end{aligned}
\label{eq:causal_effects}
\end{equation}
The direct effect isolates changes in own-treatment paths while holding
spillover exposure fixed; the spillover effect isolates changes in exposure paths while holding own treatment fixed; and the total effect captures simultaneous changes in both intervention paths. When useful, the total
effect can be decomposed after choosing an intermediate path, \(\mathrm{TE}_t\{(x,d),(x',d')\} = \mathrm{DE}_t(x,x';d)+\mathrm{SE}_t(x';d,d')\), with an analogous decomposition obtained by changing the exposure path first.

\subsection{Identification and sequential weighting representation}

To ensure identifiability of the path-indexed potential outcomes and causal effects, we impose the following consistency, sequential ignorability, and positivity assumptions, adapted to the spatiotemporal spillover setting.

\begin{assumption}[Consistency]\label{assump:consistency}
    The observed outcome agrees with the corresponding potential outcome under the realized own-treatment and spillover-exposure paths, i.e., $\left( X_{i,\tau_{1:M_t}},D_{i,\tau_{1:M_t}} \right) = \left( x_{\tau_{1:M_t}},d_{\tau_{1:M_t}} \right)$ implies $Y_{i,t} = Y_{i,t}(x_{\tau_{1:M_t}},d_{\tau_{1:M_t}})$, where $M_t=\max\{m:\tau_m \leq t\}$ indexes the most recent decision time at or before $t$.
\end{assumption}

\begin{assumption}[Sequential ignorability]\label{assump:ignorability}
The assignment of own treatment and spillover exposure is sequentially
independent of the potential outcomes conditional on the observed pre-decision history. That is, for any target intervention paths
\((x_{\tau_{1:m}},d_{\tau_{1:m}})\) and each decision time
\(\tau_{m'}\leq\tau_m\),
\begin{equation*}
    Y_{i,t}(x_{\tau_{1:m}},d_{\tau_{1:m}}) \perp\!\!\!\perp (X_{i,\tau_{m'}},D_{i,\tau_{m'}}) \mid H_{i,\tau_{m'}} .
\end{equation*}
\end{assumption}

\begin{assumption}[Positivity]\label{assump:positivity}
Let
\(\mathcal A_{\tau_m}(H_{i,\tau_m})\subseteq\mathcal X\times\mathcal D\)
denote the conditional support of
\((X_{i,\tau_m},D_{i,\tau_m})\mid H_{i,\tau_m}\). The generalized
propensity score \(p_{\tau_m}(x_{\tau_m},d_{\tau_m}\mid H_{i,\tau_m})\) is defined as the Radon--Nikodym density of this conditional
distribution. We assume
\begin{equation*}
p_{\tau_m}(x_{\tau_m},d_{\tau_m}\mid H_{i,\tau_m})>0
\end{equation*}
for all units \(i\), decision times \(\tau_m\), and intervention values
\((x_{\tau_m},d_{\tau_m})\in
\mathcal A_{\tau_m}(H_{i,\tau_m})\).
\end{assumption}

These assumptions extend standard longitudinal causal inference to the present setting with structured interference. Consistency follows the usual potential-outcome formulation \citep{rubin1974estimating}, while sequential ignorability extends longitudinal unconfoundedness \citep{robins1997causal,
robins2008estimation} to jointly account for own-treatment and spillover-exposure assignments, thereby allowing neighborhood treatments to influence outcomes through \(D_{i,\tau_{m}}\) while maintaining a localized interference structure defined by the prespecified network. Under these assumptions, a common representation follows by defining $K_h(u,d)=h^{-q}K\{(u-d)/h\}$ for continuous spillover exposure, where $K$ is a bounded kernel on $\mathbb R^q$, and
$K_h(u,d)=\mathbf{1}\{u=d\}$ for discrete spillover exposure.

\begin{proposition}
\label{prop:representation}
Under Assumptions~\ref{assump:consistency}--\ref{assump:positivity} and regularity conditions: the kernel and local-support conditions in Assumption~\SuppKernelSupportAssumption{}, and the mean-surface continuity condition in Assumption~\SuppMeanSurfaceAssumption{} of the Supplementary Material, the mean potential outcome under a joint own-treatment and spillover-exposure path admits the
sequential weighting representation
\[
\e\big[Y_{i,t}(x_{\tau_{1:M_t}},d_{\tau_{1:M_t}})\big]
=
\lim_{h\to0}\e\bigg[
Y_{i,t}\prod_{\ell=1}^{M_t}
\frac{\mathbf{1}\{X_{i,\tau_\ell}=x_{\tau_\ell}\}K_h(D_{i,\tau_\ell},d_{\tau_\ell})}
{p_{\tau_\ell}(x_{\tau_\ell},D_{i,\tau_\ell}\mid H_{i,\tau_\ell})}
\bigg].
\]
\end{proposition}

For discrete spillover exposure, Proposition~\ref{prop:representation} reduces to the ordinary sequential IPW representation. Yet in either case, estimation hinges on learning the sequential treatment assignment mechanism given the growing pre-decision history.

\section{Neural Representation Learning and Kernel-Weighted Causal Estimation}
\label{sec:estimation}



\subsection{Two-Scale Recurrent Encoding of Longitudinal Histories}
\label{subsec:two_scale_encoding}

Spatiotemporal causal adjustment requires conditioning on a growing pre-decision
history containing longitudinal information from both the focal unit and its
neighbors, creating a high-dimensional and multi-modal adjustment problem. To address the resulting complexity, we use an LSTM encoder as a flexible mechanism for compressing this growing history into fixed-dimensional recurrent states (Figure~\ref{fig:background}(b)).

To construct the encoder input on the process-time scale, we carry the most
recent own treatment and spillover exposure forward between decisions, both
of which change only at decision times. Thus, for
\(t=\tau_\ell+1,\ldots,\tau_{\ell+1}\), we form
\(O_{i,t}=(V_{i,t-1},Y_{i,t-1},X_{i,\tau_\ell},D_{i,\tau_\ell})\)
from the latest observed covariates and outcome, together with the current
intervention, and apply a GraphSAGE-mean aggregation layer:
\[
I_{i,t}
=
\sigma\Big(
W_{\mathrm{own}}O_{i,t}
+
W_{\mathrm{nbr}}
\frac{1}{|\mathcal N^1(i)|}
\sum_{j\in\mathcal N^1(i)}O_{j,t}
\Big),
\]
where the weight matrices \(W_{\mathrm{own}}\) and \(W_{\mathrm{nbr}}\) are
learned jointly with the LSTM and downstream assignment-model parameters, and
\(\sigma\) is the ReLU activation function. The resulting fixed-dimensional
representation \(I_{i,t}\) aggregates focal-unit and neighborhood information
and is then passed to a shared LSTM to update the recurrent states along the
process-time scale:
\[
(h_{i,t},c_{i,t})
=
\operatorname{LSTM}_{\theta}
\left(I_{i,t},h_{i,t-1},c_{i,t-1}\right),
\qquad t=1,\ldots,T,
\]
where \(h_{i,t}\) and \(c_{i,t}\) are the hidden and cell states, respectively,
and \(\theta\) is shared across units and process times. The recursion starts
from \((h_{i,0},c_{i,0})=(0, 0)\), with baseline
information entering through \(I_{i,1}\), formed from
\(O_{i,1}=(V_{i,0},Y_{i,0},0,0)\).
The two states jointly provide a compact summary of baseline information,
evolving focal-unit and graph-aggregated neighborhood histories, and the active
intervention regime for assignment modeling at subsequent decision times.






\subsection{Decision-Time Representations and Joint Assignment Models}
\label{subsec:decision_representation}

While the recurrent states constructed above summarize the evolving history on
the process-time scale, the intervention assignment mechanisms are modeled only
at decision times. Accordingly, at each decision time \(\tau_m\), we use the
focal-unit states \((h_{i,\tau_m},c_{i,\tau_m})\) together with the neighborhood
summaries \(h^s_{i,\tau_m}\) and \(c^s_{i,\tau_m}\), defined as the averages of
\(h_{j,\tau_m}\) and \(c_{j,\tau_m}\), respectively, over
\(j\in\mathcal N^1(i)\). An MLP \(g_Z\) then maps these pre-decision summaries
into three decision-time representations:
\[
\left(
Z^X_{i,\tau_m},
Z^D_{i,\tau_m},
Z^C_{i,\tau_m}
\right)
=
g_Z\left(
h_{i,\tau_m},
c_{i,\tau_m},
h^s_{i,\tau_m},
c^s_{i,\tau_m}
\right).
\]
Here, \(Z^X_{i,\tau_m}\), \(Z^D_{i,\tau_m}\), and \(Z^C_{i,\tau_m}\)
represent own-treatment-specific, spillover-specific, and shared pre-decision
information, respectively.

Reflecting this representation structure, \(Z^X_{i,\tau_m}\) and
\(Z^D_{i,\tau_m}\) enter the own-treatment and spillover-exposure models, respectively, whereas \(Z^C_{i,\tau_m}\) enters both to retain their shared pre-decision information. We specify the two assignment components as
\begin{align*}
    e_{i,\tau_m}
\triangleq
\Pr\left(
X_{i,\tau_m}=1\mid H_{i,\tau_m}
\right)
&=
\operatorname{expit}\Big\{
f_X\big(
Z^X_{i,\tau_m},
Z^C_{i,\tau_m}
\big)
\Big\},\\ 
p_{D,\tau_m}
\left(
d\mid H_{i,\tau_m},X_{i,\tau_m}=x
\right)
&\triangleq
f_D\Big(
d\mid
Z^D_{i,\tau_m},
Z^C_{i,\tau_m},
x
\Big),
\end{align*}
where we implement both \(f_X\) and \(f_D\) as feed-forward neural networks. Accordingly, the fitted joint generalized propensity score used for causal
weighting is
\[
\widehat p_{\tau_m}(x,d\mid H_{i,\tau_m})
=
\widehat e_{i,\tau_m}^{\,x}
(1-\widehat e_{i,\tau_m})^{1-x}
\widehat p_{D,\tau_m}
(d\mid H_{i,\tau_m},X_{i,\tau_m}=x).
\]


\subsection{Joint Training of Representations and Assignment Models}
\label{subsec:assignment_learning}

The designed encoder, representation map \(g_Z\), and assignment models
\(f_X\) and \(f_D\) form an integrated system for intervention prediction
and subsequent causal estimation. We therefore estimate their parameters
jointly from the observed spatiotemporal data using a training objective
that combines assignment likelihood losses, auxiliary reconstruction, and regularization terms.

Specifically, we use standard likelihood-based losses \(\mathcal L_X\) and
\(\mathcal L_D\) for own treatment and spillover exposure, respectively,
together with an auxiliary reconstruction loss \(\mathcal L_I\) that encourages
the recurrent states to retain recent process-time information. Let
\(\hat I_{i,t}\) denote the reconstruction of \(I_{i,t}\) from
\((h_{i,t},c_{i,t})\). Then
\begin{align*}
\mathcal L_X
&=
-\frac{1}{nL}
\sum_{i=1}^{n}\sum_{m=1}^{L}
\left[
X_{i,\tau_m}\log e_{i,\tau_m}
+
\left(1-X_{i,\tau_m}\right)
\log\left(1-e_{i,\tau_m}\right)
\right],\\
\mathcal L_D
&=
-\frac{1}{nL}
\sum_{i=1}^{n}\sum_{m=1}^{L}
\log
p_{D,\tau_m}
\left(
D_{i,\tau_m}
\mid
H_{i,\tau_m},X_{i,\tau_m}
\right),\\
\mathcal L_I
&=
\frac{1}{nT\sum_{r=1}^{q_I}a_r}
\sum_{i=1}^{n}\sum_{t=1}^{T}
\sum_{r=1}^{q_I}
a_r\,
\rho\Big(
\frac{
\hat I_{i,t,r}-I_{i,t,r}
}{
\sigma_r
}
\Big),
\end{align*}
where \(q_I\) is the dimension of the auxiliary target, while \(a_r\) and
\(\sigma_r\) are respectively the contribution weight and scale normalization
for its \(r\)th component, and \(\rho\) is the Huber function,
\(\rho(u)=u^2/2\) for \(|u|\leq1\) and \(|u|-1/2\) otherwise. The
normalizations \(\sigma_r\) and the target \(I_{i,t}\) are held fixed during
optimization.

Let \(\Theta\) collect the parameters of the GraphSAGE aggregation layer,
LSTM encoder, representation map \(g_Z\), assignment models \(f_X\) and
\(f_D\), and the reconstruction map used to obtain \(\hat I_{i,t}\).
Combining the three loss components, we jointly estimate these parameters
through
\[
\widehat\Theta
=
\arg\min_{\Theta}
\left\{
w_X\mathcal L_X
+
w_D\mathcal L_D
+
w_I\mathcal L_I
+
\lambda\|\Theta\|_2^2
\right\},
\]
where \(w_X\), \(w_D\), and \(w_I\) control the relative contributions of
the three loss components, and \(\lambda\) is the penalty tuning parameter.
The likelihood losses guide the learned representations toward information
relevant to the two assignment mechanisms, while \(\mathcal L_I\) helps
preserve recent process-time information.

During optimization, gradients are blocked through the neighborhood aggregation
branch when \(h^s_{i,\tau_m}\) and \(c^s_{i,\tau_m}\) enter a loss. This treats
the aggregated neighborhood summaries as contextual inputs and prevents
supervision from propagating back through multiple neighboring representations.

\subsection{Self-Normalized Kernel Weighting for Causal Effect Estimation}
\label{subsec:kernel_weighting}

The fundamental target of our causal analysis, i.e., the direct, spillover, and total effects defined in \eqref{eq:causal_effects}, all rely on estimating the mean potential outcome under arbitrary own-treatment and spillover-exposure paths. 


For an outcome time $t$, recall that
$M_t$ indexes the most recent decision time at or before $t$,
i.e., \(\tau_{M_t}\leq t<\tau_{M_t+1}\). Given a prespecified window length
\(M\leq M_t\), set \(m_0=M_t-M+1\), and write
\(x_{t;M}=(x_{\tau_{m_0}},\ldots,x_{\tau_{M_t}})\) and
\(d_{t;M}=(d_{\tau_{m_0}},\ldots,d_{\tau_{M_t}})\). Here, the pre-window history is left natural and absorbed into
$H_{i,\tau_{m_0}}$, so $Y_{i,t}(x_{t;M},d_{t;M})\triangleq Y_{i,t}(( X_{i,\tau_{1:m_0-1}},x_{t;M}),(D_{i,\tau_{1:m_0-1}},d_{t;M}))$ represents a
partial-window intervention over $\tau_{m_0},\ldots,\tau_{M_t}$. The corresponding common mean potential outcome is
\(\mu_{n,t}(x_{t;M},d_{t;M})
=
\e_{\mathcal G_n}\left[
Y_{i,t}(x_{t;M},d_{t;M})
\right]\).
Using the generalized propensity score
\(\widehat p_{\tau_\ell}
(x_{\tau_\ell},D_{i,\tau_\ell}\mid H_{i,\tau_\ell})\)
induced by the jointly fitted assignment mechanisms, which condition on the
pre-decision histories $H_{i,\tau_\ell}$, we define the localized weight
\[
\widehat{\omega}_{i,t;M}(h)
=
\prod_{\ell=m_0}^{M_t}
\frac{
\mathbf{1}\{X_{i,\tau_\ell}=x_{\tau_\ell}\}
K_h\left(D_{i,\tau_\ell},d_{\tau_\ell}\right)
}{
\widehat p_{\tau_\ell}
\left(
x_{\tau_\ell},D_{i,\tau_\ell}
\mid H_{i,\tau_\ell}
\right)
}.
\]
Correspondingly, the Horvitz--Thompson estimator and its self-normalized
counterpart, the H\'ajek estimator, of the mean potential outcome at time \(t\)
are
\begin{equation}
\begin{aligned}
\widehat{\mu}_{\mathrm{HT},t}^{\mathrm{K\text{-}IPW},M}
(x_{t;M},d_{t;M})
&=
\frac{1}{n}
\sum_{i=1}^{n}
\widehat{\omega}_{i,t;M}(h)Y_{i,t},
\\
\widehat{\mu}_{\mathrm{H},t}^{\mathrm{K\text{-}IPW},M}
(x_{t;M},d_{t;M})
&=
\frac{
\sum_{i=1}^{n}
\widehat{\omega}_{i,t;M}(h)Y_{i,t}
}{
\sum_{i=1}^{n}
\widehat{\omega}_{i,t;M}(h)
}.
\end{aligned}
\label{eq:estimator}
\end{equation}
Further, let
\(\mathcal T\subseteq\{1,\ldots,T\}\)
denote a prespecified finite collection of outcome times entering the
longitudinal estimand, and define $\mathcal T_M=\{t\in\mathcal T:M_t\geq M\}$ as the subset of outcome times for which an \(M\)-decision window is
available. The longitudinal estimator corresponding to the H\'ajek estimator is
\begin{equation}
\widehat{\Psi}_{\mathrm{H}}^{\mathrm{K\text{-}IPW},M}
=
\frac{1}{|\mathcal{T}_M|}
\sum_{t\in\mathcal{T}_M}
\widehat{\mu}_{\mathrm{H},t}^{\mathrm{K\text{-}IPW},M}
(x_{t;M},d_{t;M}), \label{eq:longitudinal estimator}
\end{equation}
and
\(\widehat{\Psi}_{\mathrm{HT}}^{\mathrm{K\text{-}IPW},M}\)
is defined analogously using
\(\widehat{\mu}_{\mathrm{HT},t}^{\mathrm{K\text{-}IPW},M}\).
For a given outcome time \(t\), setting \(M=M_t\) recovers the full-history
estimator, whereas fixing \(M\) across outcome times yields a rolling-window
estimator. Thus, the targeted causal effects in \eqref{eq:causal_effects} are estimated by taking the corresponding contrasts of these estimated mean potential outcomes across the specified own-treatment and spillover-exposure paths.

\section{Large-Sample Properties and Sensitivity Analysis}

We establish the large-sample properties of the kernel-weighted estimators for
continuous spillover exposure, with the corresponding discrete-exposure
results deferred to the Supplementary Material. Throughout, the exposure dimension $q$, the window length $M$, and the process- and decision-time horizons remain fixed, while $n\to\infty$ and $h_n\to0$. All probability limits and asymptotic distributions are understood conditionally on the observed network and shared baseline information, i.e., $\mathcal G_n$, with this conditioning suppressed in the notation below.

We first establish large-sample results for oracle versions of the estimators
in \eqref{eq:estimator}, constructed using the true generalized propensity score, before showing how these results transfer to their feasible counterparts. Define the oracle localized weight
\[
\omega^\circ_{i,t;M}(h_n)
=
\prod_{\ell=m_0}^{M_t}
\frac{
\mathbf{1}\{X_{i,\tau_\ell}=x_{\tau_\ell}\}
K_{h_n}(D_{i,\tau_\ell},d_{\tau_\ell})
}{
p_{\tau_\ell}
(x_{\tau_\ell},D_{i,\tau_\ell}\mid H_{i,\tau_\ell})
}.
\]
Replacing $\widehat\omega_{i,t;M}(h_n)$ in \eqref{eq:estimator} with
$\omega^\circ_{i,t;M}(h_n)$ yields the oracle Horvitz--Thompson and H\'ajek
estimators, denoted by
$\widehat{\mu}^{\circ,\mathrm{K\text{-}IPW},M}_{HT,t}$ and
$\widehat{\mu}^{\circ,\mathrm{K\text{-}IPW},M}_{H,t}$, respectively. For \(\zeta\in\{HT,H\}\), the corresponding oracle longitudinal estimator
\(\widehat{\Psi}^{\circ,\mathrm{K\text{-}IPW},M}_{\zeta}\) is defined
analogously to \eqref{eq:longitudinal estimator}, using \(\widehat{\mu}^{\circ,\mathrm{K\text{-}IPW},M}_{\zeta,t}\) in place of its feasible counterpart.

The formal regularity conditions are stated in the Supplementary Material.
Apart from standard conditions on the kernel, support, and smoothness
\citep[see, e.g.,][]{fan1996local}, overlap and moments \citep{kennedy2017non}, two conditions merit emphasis: a common conditional-mean structure and a uniformly bounded degree of the dependency graph. The former allows the mean potential outcomes, and hence the causal effects, to be represented by the common target $\mu_{n,t}(x_{t;M},d_{t;M})$; without it, the same
formulation can instead target the average of the unit-specific conditional
means. The latter ensures sparsity of the cross-unit dependence structure and is
practically plausible in infectious-disease applications, where dependence is
typically localized among a limited set of connected jurisdictions. Under these conditions, the oracle estimators are consistent, as stated in Theorem~\ref{thm:oracle-consistency}.



\begin{theorem}[Oracle consistency]
\label{thm:oracle-consistency}
For a given outcome time \(t\) and target intervention paths
\((x_{t;M},d_{t;M})\), suppose
Assumptions~\ref{assump:consistency}--\ref{assump:positivity}
and Assumptions~\SuppKernelSupportAssumption--\SuppDependenceAssumption{}
of the Supplementary Material hold. If \(h_n\to0\) and
\(nh_n^{qM}\to\infty\), then
\[
\widehat{\mu}^{\circ,\mathrm{K\text{-}IPW},M}_{\zeta,t}
(x_{t;M},d_{t;M})
-
\mu_{n,t}(x_{t;M},d_{t;M})
\overset{p}{\longrightarrow}
0,
\qquad
\zeta\in\{HT,H\}.
\]
\end{theorem}

For an $M$-decision window, the effective number of observations localized
around the target spillover-exposure path is of order $nh_n^{qM}$. Thus, a
longer intervention history enriches the longitudinal target but increases the
localization dimension and compounds the pathwise overlap requirement. Because \(\mathcal T_M\) is fixed and finite, the pointwise consistency in
Theorem~\ref{thm:oracle-consistency} further implies consistency of the
longitudinal mean estimators \eqref{eq:longitudinal estimator} and the
corresponding direct, spillover, and total effect contrasts, as stated in
Corollary~\SuppLongitudinalConsistencyCorollary{} of the Supplementary
Material.


Beyond consistency, asymptotic normality further requires the nondegenerate limiting-covariance condition stated in the Supplementary Material.


\begin{theorem}[Oracle asymptotic normality]
\label{thm:oracle-clt}
Suppose the conditions of Theorem~\ref{thm:oracle-consistency} and
Assumption~\SuppCovarianceAssumption{} of the Supplementary Material
hold. If \((nh_n^{qM})^{1/2}h_n^2\to0\), then, for
\(\zeta\in\{HT,H\}\),
\[
(nh_n^{qM})^{1/2}
\left\{
\widehat{\mu}^{\circ,\mathrm{K\text{-}IPW},M}_{\zeta,t}
(x_{t;M},d_{t;M})
-
\mu_{n,t}(x_{t;M},d_{t;M})
\right\}
\overset{d}{\longrightarrow}
\mathcal N(0,\sigma^2_{\zeta,t}),
\]
where $\sigma^2_{\zeta,t}\in(0,\infty)$ is the corresponding asymptotic
variance.
\end{theorem}

An analogous result holds for discrete spillover exposure. In this
case, the localization kernel reduces to an indicator, so no bandwidth sequence
or smoothing condition is required, and the normalization becomes the
usual $n^{1/2}$ rate.

The preceding pointwise results further extend to the longitudinal estimands
and their corresponding causal-effect contrasts.

\begin{corollary}[Asymptotic normality of longitudinal causal-effect estimators]
\label{cor:oracle-effect-clt}
Suppose the conditions of Theorem~\ref{thm:oracle-clt}, or their
discrete-exposure counterparts stated in the Supplementary Material, hold. Then, for
\(\zeta\in\{HT,H\}\), the oracle longitudinal mean estimators in
\eqref{eq:longitudinal estimator} and the resulting plug-in estimators of the
direct, spillover, and total effects in \eqref{eq:causal_effects} are
asymptotically normal at rate \((nh_n^{qM})^{1/2}\) for continuous spillover
exposure and \(n^{1/2}\) for discrete spillover exposure.
\end{corollary}

In practice, however, the generalized propensity scores entering the feasible
estimators are learned using LSTM-encoded histories and downstream MLPs, and
their convergence rates therefore depend on the network architecture and
training procedure. We thus formulate the oracle-to-feasible extension directly
in terms of propensity-score estimation error. Let
\[
\Delta_n
=
\max_{i,\ell,x}
\left|
\frac{
\widehat p_{\tau_\ell}
(x,D_{i,\tau_\ell}\mid H_{i,\tau_\ell})
}{
p_{\tau_\ell}
(x,D_{i,\tau_\ell}\mid H_{i,\tau_\ell})
}
-1
\right|,
\]
where the maximum is taken over the units, decision times, and target
own-treatment values entering the estimators under consideration. Then
$\Delta_n=o_p(1)$ suffices for feasible consistency, whereas
$\Delta_n=o_p\{(nh_n^{qM})^{-1/2}\}$ for continuous spillover exposure and
$\Delta_n=o_p(n^{-1/2})$ for discrete spillover exposure ensure that the
feasible estimators inherit their corresponding oracle asymptotic
distributions. Moreover, these conclusions extend to the longitudinal mean and
causal-effect estimators when the corresponding rate condition holds uniformly
over the fixed and finite collection of outcome times and intervention paths
entering their component estimators. Formal conditions and conclusions for these oracle-to-feasible transfers are stated in Proposition~\SuppOracleToFeasibleProposition{} of the Supplementary
Material.

Beyond the preceding large-sample analysis, we assess sensitivity to departures
from sequential ignorability induced by unmeasured focal-unit and neighborhood
history \(U_{i,1:\tau_m}\) relevant to the joint assignment process and
potential outcomes. Suppose that sequential ignorability would hold after
conditioning on \((H_{i,\tau_m},U_{i,1:\tau_m})\). Denote the corresponding joint conditional probability mass or density by
\(p^\ast_{\tau_m}(x,d\mid H_{i,\tau_m},U_{i,1:\tau_m})\). Following Rosenbaum-type hidden-bias models and their spatiotemporal extensions
\citep{rosenbaum1987sensitivity,
papadogeorgou2022causal}, we assume that, at the realized joint
assignment,
\[
\Gamma^{-1}
\leq
\rho_{i,\tau_m}
\triangleq
\frac{
p_{\tau_m}
(X_{i,\tau_m},D_{i,\tau_m}\mid H_{i,\tau_m})
}{
p^\ast_{\tau_m}
(X_{i,\tau_m},D_{i,\tau_m}
\mid H_{i,\tau_m},U_{i,1:\tau_m})
}
\leq
\Gamma,
\quad {\rm for\; some\;}
\Gamma\geq1.
\]
Here, larger values of \(\Gamma\) permit greater hidden bias, with
\(\Gamma=1\) recovering the original analysis. For the \(M\)-decision window
ending at \(t\), let \(r_i=\prod_{\ell=m_0}^{M_t}\rho_{i,\tau_\ell}\) denote unit \(i\)'s pathwise sensitivity multiplier, so that \(\Gamma^{-M}\leq r_i\leq\Gamma^M\). With \(\boldsymbol r=(r_1,\ldots,r_n)\), we then define the sensitivity-adjusted H\'ajek estimator by
\[
\widehat\mu_{\mathrm H,t}
(\boldsymbol r;x_{t;M},d_{t;M})
=
\frac{
\sum_{i=1}^{n}
\widehat\omega_{i,t;M}(h)r_iY_{i,t}
}{
\sum_{i=1}^{n}
\widehat\omega_{i,t;M}(h)r_i
}.
\]
Accordingly, varying \(\boldsymbol r\) over its admissible range yields the
sensitivity interval for the mean potential outcome $\mathcal I_{\mu,t}^{\Gamma}(x_{t;M},d_{t;M})
=
\Big[
\underline{\widehat\mu}_{\mathrm H,t}^{\,\Gamma}
(x_{t;M},d_{t;M}),
\,
\overline{\widehat\mu}_{\mathrm H,t}^{\,\Gamma}
(x_{t;M},d_{t;M})
\Big]$, where
\[
\begin{aligned}
\underline{\widehat\mu}_{\mathrm H,t}^{\,\Gamma}
(x_{t;M},d_{t;M})
&=
\min_{\boldsymbol r\in[\Gamma^{-M},\Gamma^M]^n}
\widehat\mu_{\mathrm H,t}
(\boldsymbol r;x_{t;M},d_{t;M}),\\
\overline{\widehat\mu}_{\mathrm H,t}^{\,\Gamma}
(x_{t;M},d_{t;M})
&=
\max_{\boldsymbol r\in[\Gamma^{-M},\Gamma^M]^n}
\widehat\mu_{\mathrm H,t}
(\boldsymbol r;x_{t;M},d_{t;M}).
\end{aligned}
\]
Equivalent linear-program formulations of these endpoint problems through the
Charnes Cooper transformation
\citep{papadogeorgou2022causal} are given in
Proposition~\SuppSensitivityLPProposition{} of the Supplementary Material.

For the longitudinal estimands, let
\(x_{\cdot;M}=\{x_{t;M}:t\in\mathcal T_M\}\) and
\(d_{\cdot;M}=\{d_{t;M}:t\in\mathcal T_M\}\) collect the corresponding
time-specific intervention paths. Averaging the time-specific endpoints over
\(\mathcal T_M\) yields lower and upper sensitivity bounds for the
corresponding longitudinal mean, denoted by
\(\underline{\widehat\Psi}_{\mathrm H}^{\,\Gamma}\) and
\(\overline{\widehat\Psi}_{\mathrm H}^{\,\Gamma}\), respectively. For a causal
contrast comparing \((x_{\cdot;M},d_{\cdot;M})\) with
\((x'_{\cdot;M},d'_{\cdot;M})\), the conservative sensitivity interval is
\[
\begin{aligned}
\mathcal I_{\mathrm{effect}}^{\Gamma}
=
\Big[
&
\underline{\widehat\Psi}_{\mathrm H}^{\,\Gamma}
(x_{\cdot;M},d_{\cdot;M})
-
\overline{\widehat\Psi}_{\mathrm H}^{\,\Gamma}
(x'_{\cdot;M},d'_{\cdot;M}),\;
\overline{\widehat\Psi}_{\mathrm H}^{\,\Gamma}
(x_{\cdot;M},d_{\cdot;M})
-
\underline{\widehat\Psi}_{\mathrm H}^{\,\Gamma}
(x'_{\cdot;M},d'_{\cdot;M})
\Big].
\end{aligned}
\]
Accordingly, the sensitivity intervals for direct, spillover, and total effects are obtained by setting \(d'_{\cdot;M}=d_{\cdot;M}\), setting
\(x'_{\cdot;M}=x_{\cdot;M}\), or allowing both paths to differ, respectively.

\section{Numerical Experiments}\label{simulation}

We conduct simulation studies to evaluate the proposed estimator for dynamic causal effects under spatiotemporal interference. The main goal is to assess whether the proposed estimator can recover path-indexed mean potential outcomes and the corresponding direct, spillover, and total effects, relative to the same estimator using oracle propensity values from the data-generating process and an outcome-regression benchmark, and to quantify its finite-sample performance.

\subsection{Simulation Design}

\noindent{\bf\emph{Time-varying confounding and spillover exposure:}} We give a concise sketch of the simulation design, with the complete specification of the data-generating process and parameter values deferred to the Supplementary Material. Longitudinal data are generated for \(n\) units connected through an undirected Erd\H{o}s--R\'enyi network. Covariates and outcomes are observed over \(T=20\) process times, whereas binary treatment is updated at the four decision times \(\tau_{1:4}=(1,6,11,16)\) and remains in effect until the next decision. Each unit has ten time-varying covariates. Meanwhile, to induce time-varying confounding, the earlier treatments are designed to affect subsequent covariates and outcomes, while later treatment assignments depend on the resulting histories through recent outcome changes and a weighted summary of lagged covariates, as well as on baseline risk and previous treatment. Spillover exposure is specified as the proportion of treated first-order neighbors, \(D_{i,\tau_m}=|\mathcal N^1(i)|^{-1}\sum_{j\in\mathcal N^1(i)}X_{j,\tau_m}\). Although this proportion takes finitely many values for a
fixed network, we apply Gaussian localization on its continuous
proportion scale to borrow information across nearby
spillover-exposure paths.
This smoothing choice is motivated by the smooth
spillover-response functions in the data-generating process.
The assignment model retains the discrete probability mass
function induced by treated-neighbor counts.
The numerical experiments therefore evaluate a continuous-scale
smoothing approximation, with its finite-sample performance
and sensitivity to bandwidth assessed empirically.

\noindent{\bf\emph{Network dependence, temporal carryover, and effect heterogeneity:}} Within the longitudinal dynamics, we incorporate network dependence through two channels: lagged neighborhood outcomes affect subsequent covariate updates, while neighborhood covariate summaries and accumulated own-treatment and spillover exposures affect a latent disease-activity state $C_{i,t}$. Because the observed outcome evolves only gradually toward $C_{i,t}$, this construction induces delayed responses and temporal carryover, making $Y_{i,T}$ depend on the full intervention paths. Treatment responsiveness is also allowed to vary with baseline burden, thereby evaluating the estimators under heterogeneous treatment effects. The resulting generation mechanism is summarized in Algorithm~\ref{alg:sim_dgp}.

\bigskip
\begin{algorithm}[!ht]
\caption{Schematic data-generating process for one simulation replication}
\label{alg:sim_dgp}
\small
\begin{algorithmic}[1]

\State Generate the Erd\H{o}s--R\'enyi network and, for any
unit-level variable (e.g., $Y_{i,t}$), define its neighborhood summary (e.g., $Y_{i,t}^s$) as the average of that corresponding variable among its first order neighbor.

\State Initialize the baseline for all \(i\): generate the
ten-dimensional covariate vector \(V_{i,0}\), latent disease-activity
state \(C_{i,0}\), and outcome \(Y_{i,0}\); define the baseline burden
\(\mathcal B_{i,0}\Leftarrow
\sum_{\ell=1}^{10}w_\ell V_{i,0}^{(\ell)}\); set \(X_{i,0}=0\); and
initialize the discounted cumulative own-treatment and spillover
exposures \(\widetilde X_{i,0}=\widetilde D_{i,0}=0\). Compute the baseline risk score \(R_{i,0}\) as a standardized weighted combination of
\((Y_{i,0},\mathcal B_{i,0},Y^s_{i,0})\).

\For{each $i\in\{1,\cdots,n\}$, together with its neighboring
units, and \(t=1,\ldots,T\)}

    \State Generate \(V_{i,t}\) from its mean-reverting transition
    conditional on
    \((V_{i,t-1},Y_{i,t-1},Y^s_{i,t-1})\), and set the time-varying burden score
    \(\mathcal B_{i,t}\Leftarrow
    \sum_{\ell=1}^{10}w_\ell V_{i,t}^{(\ell)}\).

    \If{\(t=\tau_m\) for some \(m\)}
        \State Compute
        \(\pi_{i,t}\) as a function of
        \((R_{i,0},X_{i,t-1},Y_{i,t-1},Y_{i,t-2},
        \mathcal B_{i,t-1})\), and generate
        \(X_{i,t}\sim\operatorname{Bernoulli}(\pi_{i,t})\).
    \Else \State Set \(X_{i,t}\Leftarrow X_{i,t-1}\).
    \EndIf

    \State Set \(D_{i,t}\Leftarrow N_i^{-1}\sum_{j\in\mathcal N^1(i)}X_{j,t}\) and update
    \(\widetilde X_{i,t}\Leftarrow
    \rho_X\widetilde X_{i,t-1}+X_{i,t}\) and
    \(\widetilde D_{i,t}\Leftarrow
    \rho_X\widetilde D_{i,t-1}+D_{i,t}\), where \(\rho_X\) is the
    common decay parameter.

    \State Generate \(C_{i,t}\) from its mean-reverting transition
    conditional on
    $(C_{i,t-1},\mathcal B_{i,t-1},\mathcal B^s_{i,t-1},$ $
    \widetilde X_{i,t},\widetilde D_{i,t},\mathcal B_{i,0})$,
    allowing treatment responsiveness to vary with baseline burden.

    \State Generate
    \(Y_{i,t}\Leftarrow
    (1-\kappa_Y)Y_{i,t-1}+\kappa_YC_{i,t}
    +\varepsilon^Y_{i,t}\), where \(\kappa_Y\) controls the rate at
    which the outcome approaches the latent state and
    \(\varepsilon^Y_{i,t}\sim N(0,\sigma_Y^2)\) is the outcome error.
\EndFor

\end{algorithmic}
\end{algorithm}


\noindent{\bf\emph{Target paths and causal estimands:}} Across the four decision times, we consider five representative own-treatment paths \(x^{\zeta'}\), indexed by $\zeta'\in\{$never, early, late, intermittent, frequent$\}$ (e.g., \(x^{\mathrm{intermittent}}=(1,0,1,0)\)), together with three reference spillover-exposure paths \(d^{\zeta''}\), \(\zeta''\in\{L,M,H\}\), representing low, mid, and high exposure patterns. Accordingly, mean potential outcome \(\mu_{n,T}(x,d)\) is evaluated for all 15 combinations. Following \eqref{eq:causal_effects}, we report direct effects comparing \((x,d^M)\) with \((x^{\mathrm{never}},d^M)\), spillover effects comparing \((x,d^H)\) with \((x,d^L)\), and total effects comparing \((x,d^H)\) with \((x^{\mathrm{never}},d^L)\), where \(x\) ranges over the four non-never paths for the direct and total effects and over all five paths for the spillover effects.

\noindent{\bf\emph{Monte Carlo evaluation:}} Ground-truth mean potential outcomes under the respective target regimes are approximated once using the two-hop local structural Monte Carlo procedure detailed in the Supplementary Material, and the resulting values are then held fixed across simulation replications. For each \(n\in\{5\times 10^3,\,5\times 10^4\}\), we generate \(R=100\) independent analysis data sets and report empirical bias, absolute relative bias (ARB), and mean-squared error (MSE) for the proposed, oracle-propensity, and OLS estimators. For interval evaluation of the proposed estimator, we report the average width and empirical coverage of 95\% percentile intervals constructed using \(B_{\mathrm{boot}}=100\) randomized disjoint network-block bootstrap resamples from each analysis network, holding the fitted model fixed and jointly recomputing all target means and contrasts within each resample.

\subsection{Simulation Results}\label{sim:results}

\begin{figure}[!htbp]
\centering
\includegraphics[width=\textwidth]{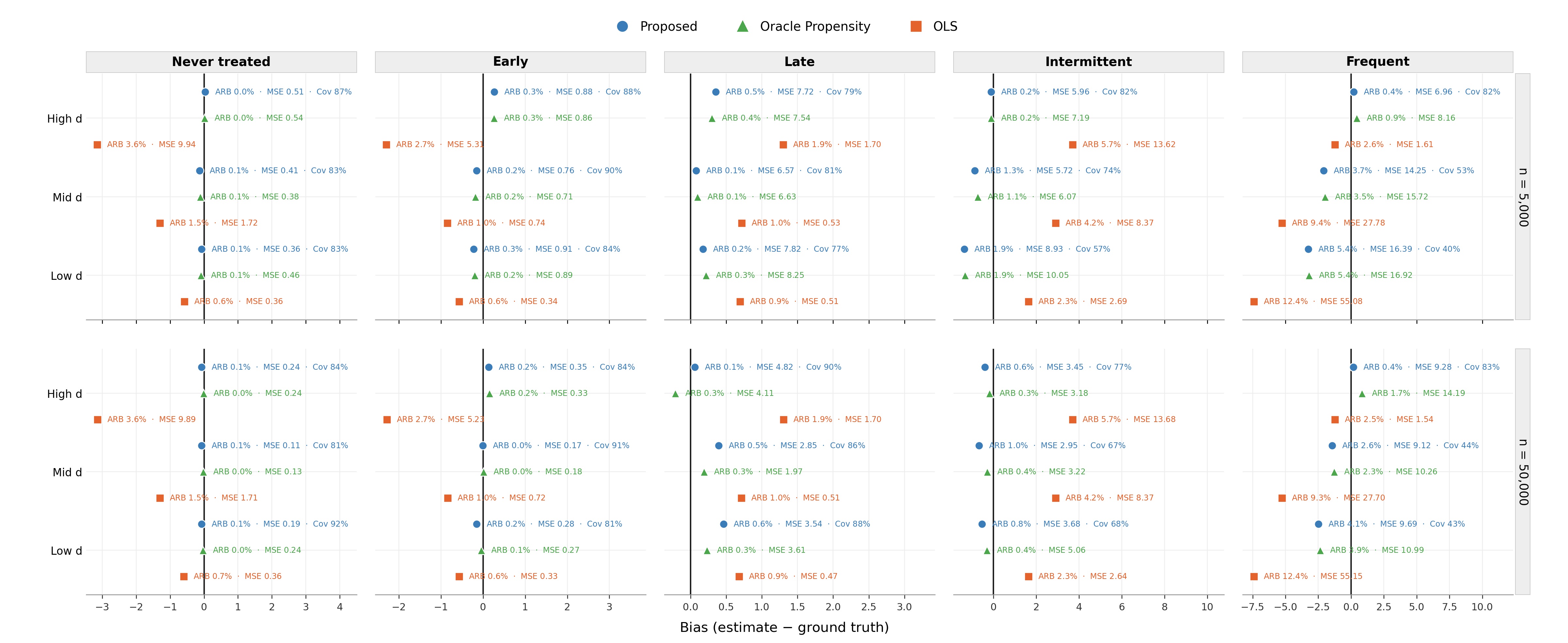}
\caption{Bias in estimating the 15 target mean potential outcomes. For each sample size, columns correspond to own-treatment paths and rows to spillover-exposure paths. The vertical reference lines mark zero bias; labels report ARB (absolute relative bias) and MSE for all estimators and empirical coverage for the proposed estimator.}
\label{fig:mean_potential_outcome}
\end{figure}

All results reported in this subsection use the complete four-decision window, \(M=M_T=4\), bandwidth \(h=0.03\), and are based on $R=100$ independent replications for each sample size. Figure~\ref{fig:mean_potential_outcome} summarizes estimation of the $15$ target mean potential outcomes, whereas Table~\ref{tab:13causal_effects} reports the corresponding causal effects. Stress-test results under alternative data-generating
processes and bandwidth sensitivity analyses are provided
in the Supplementary Material.


\noindent{\bf\emph{Mean potential outcomes:}} Figure~\ref{fig:mean_potential_outcome} shows that, across both sample sizes, the proposed estimator produced biases generally concentrated near zero across the $15$ target regimes and closely resembled those of its oracle-propensity counterpart. In contrast, OLS exhibited appreciable systematic bias across all five own-treatment paths and multiple spillover-exposure levels. Increasing the sample size generally reduced MSE and moved the proposed and oracle-propensity estimates closer to their targets, whereas the OLS pattern changed little.

\noindent{\bf\emph{Causal effects:}} Across the $13$ effects in Table~\ref{tab:13causal_effects}, the proposed estimator had smaller absolute empirical bias than OLS at both sample sizes. Its mean ARB decreased from $0.073$ to $0.048$ as the sample size $n$ increased, closely matching the corresponding oracle-propensity values of $0.074$ and $0.047$, while the OLS value remained approximately $0.303$; hence, propensity estimation introduced little additional bias in this design.

For interval evaluation, we exclude the three frequent-path
effects from the aggregate summaries of interval width and
coverage at both sample sizes.
This choice is motivated by the diagnostics of effective sample size (ESS): among the five own-treatment paths, the frequent path has the lowest mean ESS of the localized weights at each of the three reference spillover levels
(Supplementary Material). These frequent-path effects are nevertheless retained in Table~\ref{tab:13causal_effects} as a support stress test:
their absolute biases remain smaller than those of OLS,
but well-calibrated network-bootstrap intervals are
difficult to obtain. Across the remaining ten effects, the average interval width decreased from \(5.190\) to \(3.877\) as \(n\) increased, while mean empirical coverage increased from \(85.0\%\) to \(87.3\%\); although still below the nominal $95\%$ level, these intervals remain reasonably informative given the complex network dependence and pathwise localization. The remaining undercoverage may reflect finite-sample bias, uneven local support, and the inherent difficulty of calibrating block resampling under network dependence. A feasible estimator of the limiting variance, as available in simpler network-interference settings, may provide a more direct route to calibrated inference.

Cellwise MSE comparisons with OLS were less uniform than the bias comparisons. Nevertheless, the proposed estimator had lower average MSE at both sample sizes, decreasing from $8.335$ to $5.384$ as $n$ increased, whereas the OLS average remained approximately $8.88$. For the individual effects for which OLS attained lower MSE, its lower across-replication variability offset its larger bias in the MSE calculation, while the OLS bias pattern showed little attenuation as the sample size increased, rendering MSE alone insufficient as a measure of estimator performance in this setting.

\begin{table}[!htbp]
\centering
\caption{Estimation of causal-effect contrasts at bandwidth \(h=0.03\) and \(R=100\) replications.}
\label{tab:simulation_effects}
\scriptsize
\setlength{\tabcolsep}{2.4pt}
\begin{tabular}{lcrrrrrrrr}
\toprule
& & \multicolumn{4}{c}{Proposed}
& \multicolumn{2}{c}{Oracle propensity}
& \multicolumn{2}{c}{OLS} \\
\cmidrule(lr){3-6}
\cmidrule(lr){7-8}
\cmidrule(lr){9-10}
Contrast
& GT
& Bias (ARB)
& MSE
& Width
& Cov.
& Bias (ARB)
& MSE
& Bias (ARB)
& MSE \\
\midrule

\multicolumn{10}{l}{\textit{Panel A: \(n=5\times10^3\)}}\\
\addlinespace[0.3ex]

SE (never)
& -2.233
& +0.110 (0.049)
& \textbf{0.776}
& 2.374 & 87\%
& \textbf{+0.102 (0.046)}
& 0.872
& -2.566 (1.150)
& 6.622 \\

SE (early)
& -3.068
& +0.492 (0.160)
& 1.871
& 3.226 & 94\%
& \textbf{+0.460 (0.150)}
& \textbf{1.842}
& -1.731 (0.564)
& 3.033 \\

SE (late)
& -5.402
& +0.176 (0.033)
& 16.307
& 8.994 & 84\%
& \textbf{+0.082 (0.015)}
& 15.684
& +0.603 (0.112)
& \textbf{0.399} \\

SE (intermittent)
& -6.855
& +1.248 (0.182)
& 15.295
& 8.738 & 86\%
& \textbf{+1.231 (0.180)}
& 17.134
& +2.056 (0.300)
& \textbf{4.262} \\

SE (frequent)
& -10.966
& \textbf{+3.475 (0.317)}
& \textbf{22.975}
& 9.694 & 63\%
& +3.737 (0.341)
& 26.428
& +6.168 (0.562)
& 38.075 \\

DE (early)
& -2.370
& \textbf{-0.025 (0.010)}
& 1.098
& 2.677 & 90\%
& -0.072 (0.030)
& 1.019
& +0.455 (0.192)
& \textbf{0.219} \\

TE (early)
& -4.998
& \textbf{+0.342 (0.068)}
& \textbf{1.188}
& 2.714 & 89\%
& +0.344 (0.069)
& 1.203
& -1.716 (0.343)
& 3.002 \\

DE (late)
& -16.734
& \textbf{+0.204 (0.012)}
& 7.546
& 5.893 & 83\%
& +0.209 (0.013)
& 7.432
& +2.022 (0.121)
& \textbf{4.102} \\

TE (late)
& -21.392
& +0.428 (0.020)
& 8.047
& 6.117 & 78\%
& \textbf{+0.386 (0.018)}
& 8.188
& +1.881 (0.088)
& \textbf{3.591} \\

DE (intermittent)
& -20.474
& -0.751 (0.037)
& \textbf{5.915}
& 5.240 & 79\%
& \textbf{-0.632 (0.031)}
& 6.316
& +4.199 (0.205)
& 17.654 \\

TE (intermittent)
& -25.346
& -0.035 (0.001)
& \textbf{6.384}
& 5.930 & 80\%
& \textbf{-0.022 (0.001)}
& 7.943
& +4.272 (0.169)
& 18.314 \\

DE (frequent)
& -33.261
& -1.958 (0.059)
& \textbf{13.550}
& 6.533 & 63\%
& \textbf{-1.886 (0.057)}
& 15.394
& -3.959 (0.119)
& 15.736 \\

TE (frequent)
& -41.353
& \textbf{+0.271 (0.007)}
& 7.398
& 6.069 & 85\%
& +0.510 (0.012)
& 8.650
& -0.667 (0.016)
& \textbf{0.552} \\

\midrule
\multicolumn{10}{l}{\textit{Panel B: \(n=5\times10^4\)}}\\
\addlinespace[0.3ex]

SE (never)
& -2.233
& \textbf{+0.001 (0.000)}
& \textbf{0.457}
& 1.521 & 87\%
& +0.017 (0.008)
& 0.482
& -2.547 (1.141)
& 6.491 \\

SE (early)
& -3.068
& +0.291 (0.095)
& 0.639
& 2.139 & 88\%
& \textbf{+0.193 (0.063)}
& \textbf{0.564}
& -1.712 (0.558)
& 2.934 \\

SE (late)
& -5.402
& \textbf{-0.403 (0.075)}
& 9.085
& 7.159 & 89\%
& -0.441 (0.082)
& 8.560
& +0.622 (0.115)
& \textbf{0.391} \\

SE (intermittent)
& -6.855
& +0.144 (0.021)
& 7.018
& 6.676 & 91\%
& \textbf{+0.125 (0.018)}
& 8.766
& +2.075 (0.303)
& \textbf{4.310} \\

SE (frequent)
& -10.966
& \textbf{+2.665 (0.243)}
& \textbf{19.055}
& 8.245 & 70\%
& +3.165 (0.289)
& 27.375
& +6.187 (0.564)
& 38.283 \\

DE (early)
& -2.370
& +0.060 (0.025)
& 0.286
& 1.553 & 95\%
& \textbf{+0.038 (0.016)}
& 0.365
& +0.462 (0.195)
& \textbf{0.215} \\

TE (early)
& -4.998
& +0.211 (0.042)
& 0.528
& 1.953 & 90\%
& \textbf{+0.182 (0.036)}
& \textbf{0.483}
& -1.690 (0.338)
& 2.861 \\

DE (late)
& -16.734
& +0.462 (0.028)
& 3.028
& 4.206 & 88\%
& \textbf{+0.218 (0.013)}
& \textbf{2.167}
& +2.022 (0.121)
& 4.090 \\

TE (late)
& -21.392
& \textbf{+0.135 (0.006)}
& 4.800
& 5.158 & 93\%
& -0.177 (0.008)
& 4.009
& +1.901 (0.089)
& \textbf{3.618} \\

DE (intermittent)
& -20.474
& -0.615 (0.030)
& \textbf{2.938}
& 3.749 & 71\%
& \textbf{-0.272 (0.013)}
& 3.332
& +4.202 (0.205)
& 17.661 \\

TE (intermittent)
& -25.346
& -0.327 (0.013)
& 3.788
& 4.656 & 81\%
& \textbf{-0.159 (0.006)}
& \textbf{3.554}
& +4.295 (0.169)
& 18.454 \\

DE (frequent)
& -33.261
& -1.377 (0.041)
& \textbf{9.085}
& 5.126 & 45\%
& \textbf{-1.261 (0.038)}
& 10.259
& -3.954 (0.119)
& 15.640 \\

TE (frequent)
& -41.353
& \textbf{+0.256 (0.006)}
& 9.281
& 5.838 & 88\%
& +0.849 (0.021)
& 14.390
& -0.642 (0.016)
& \textbf{0.424} \\

\bottomrule
\end{tabular}

\vspace{0.6ex}
\begin{minipage}{0.98\textwidth}
\footnotesize
\textit{Note:} GT denotes the Monte Carlo ground-truth value. Width and coverage refer to the proposed estimator's $95\%$ percentile intervals obtained from the network-block bootstrap. The three frequent-path effects are retained as support stress tests but excluded from the aggregate summaries
of interval width and coverage because this regime has limited effective local support.\\
\end{minipage}
\label{tab:13causal_effects}
\end{table}

\section{Application: County-Level Analysis of COVID-19 Policies}\label{realdata}

We apply the proposed framework to estimate the effects of state-level
COVID-19 mitigation policies on subsequent county-level case rates while accounting for spillovers across geographically adjacent counties. We select Business and Economic Restrictions as a more direct transmission-prevention policy domain, targeting non-essential movement and commercial contacts. In contrast, Education and Childcare represents a more indirect domain, with effects potentially extending beyond schools to household exposure, caregiving, and family routines. We present exploratory findings here while deferring detailed data
sources, analytical procedures, exploratory analyses, and local-support diagnostics to the supplement.

\subsection{Data and Analysis}\label{application:design}

\noindent{\bf\emph{Data and network:}}
We analyze a longitudinal panel of 3,108 counties across 48 U.S.
states (excluding Alaska and Hawaii) and the District of Columbia,
covering April to October 2020. COVID-19 case and death data are
collected from the CDC, and policy data are collected from the
Boston University COVID-19 State Policy Database. These records
are linked to county covariates compiled from County Health
Rankings \& Roadmaps, county election records, and NOAA temperature data. Within this panel, policies are measured and assigned at the state level, whereas health outcomes are measured at the county level. To represent geographic
connections among these counties, we construct a network using
the U.S. Census Bureau's county adjacency file, retaining links
both within and across state borders. 

\noindent{\bf\emph{Policy exposures and target paths:}}
For each policy domain, state-month intensity is the average daily
proportion of recorded policy types that are active.
A threshold of \(0.2\) distinguishes high-intensity
(\(X=1\)) from low-intensity (\(X=0\)) policy months, with treatment
status fixed within each month. Correspondingly, spillover exposure is the proportion
of neighboring counties with \(X=1\).

Own treatment and spillover exposure were closely aligned in the
most commonly observed policy histories. We therefore consider target paths with
\(x_{\tau_m}=d_{\tau_m}\in\{0,1\}\) at each monthly decision
from April 2020 onward.
For Business and Economic Restrictions, September and October
endpoints are used to examine the longer-term implications of
earlier restrictions during reopening, while for Education and
Childcare, July and August endpoints focus on earlier outcomes
to limit complications from subsequent autumn changes in policies
and behaviour. Each selected path is compared with the joint
baseline \(x_{\tau_m}=d_{\tau_m}=0\) throughout the corresponding
decision history.

\noindent{\bf\emph{Outcomes, adjustment, and inference:}}
For each endpoint month, we report two summaries of new cases
per 10,000 residents: the last-week rate and the average weekly rate
within that month. Adjustment incorporates baseline
county characteristics, time-varying death rates and temperature,
and histories of case rates, own treatment, and spillover exposure,
together with neighborhood summaries.
We estimate effects using the full county network and bandwidth
\(h=0.0833\). Point estimates are averaged over \(100\) refits of the
assignment models using different random seeds. The \(95\%\) percentile
intervals are obtained by pooling \(1{,}000\) disjoint network-block
bootstrap estimates from each refit, with blocks formed by randomized
one-hop clustering. 

\subsection{Estimated Policy Effects}\label{application:results}

\begin{figure}[!htbp]
\centering
\includegraphics[width=0.9\textwidth]{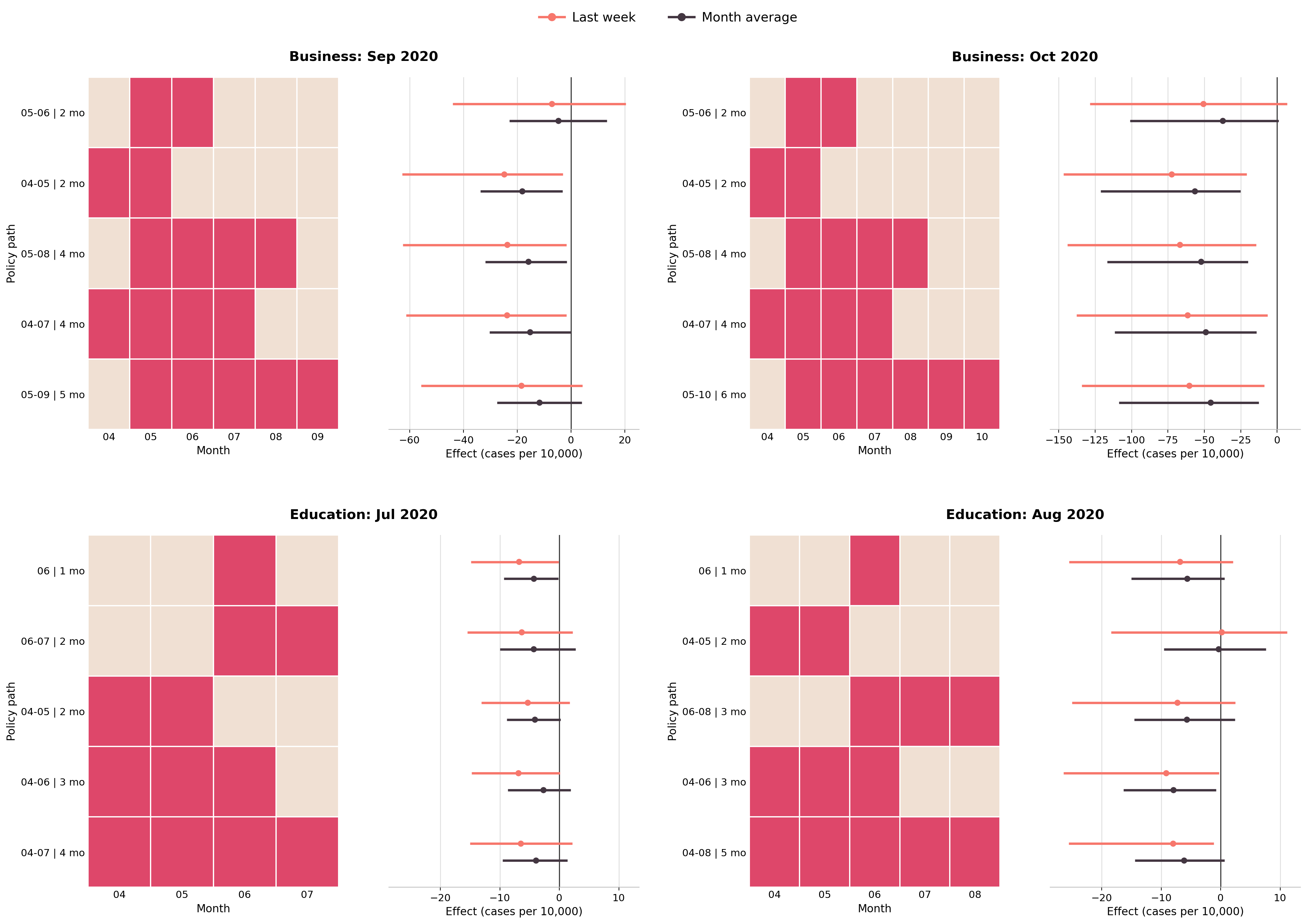}
\caption{Estimated effects of selected policy paths relative to
the joint low-intensity baseline, in new cases per 10,000 residents.
Red cells mark months with \(x_{\tau_m}=d_{\tau_m}=1\);
the remaining cells indicate \(x_{\tau_m}=d_{\tau_m}=0\). Red and black indicate the last-week and month-average weekly case rates,
respectively.}
\label{fig:policy_effects}
\end{figure}

Relative to the joint low-intensity baseline, the two policy domains
show different patterns of association with subsequent case rates
(Figure~\ref{fig:policy_effects}). Business and Economic Restrictions
are associated with lower subsequent case rates, and
Education and Childcare shows negative but smaller
associations.

\noindent{\bf\emph{Business and economic restrictions:}}
Negative associations with subsequent case rates were already evident
at the September endpoint and became more pronounced in October.
This pattern is consistent with early reductions in routine contacts
across workplaces, retail settings, and dining helping to lower the
initial transmission base and remaining relevant during reopening.
For policy design, these findings suggest the key point that
interventions targeting high-frequency community contacts may help
prevent later escalation when transmission risk is rising, with
implications extending beyond the immediate implementation period.

Besides, the nonmonotone pattern across treatment windows is also
policy-relevant: longer or earlier restrictions should not be
interpreted as a simple policy dose. Policy timing in 2020 was
endogenous to local epidemic conditions, with areas facing faster
early growth or greater underlying risk more likely to intervene
earlier and maintain restrictions longer. Policy evaluation should
therefore consider timing, underlying risk, and the transmission
channels targeted, alongside duration.

\noindent{\bf\emph{Education and childcare:}}
This indirect measure tells a different policy story.
The July and August endpoints showed broadly similar results,
with negative but more modest associations and no
comparably strong lagged pattern. This is consistent with their
likely mechanism: education-related policies do not primarily
suppress broad public mixing, but affect transmission
through more specific behavioural pathways involving schools,
households, caregiving, and family routines. For policy design,
these findings suggest that such measures may operate less as
broad epidemic-control tools and more as targeted interventions,
with effects depending on local institutional arrangements,
family constraints, and the availability of alternative support.


\section{Discussion}\label{discussion}

This work develops a unified framework for estimating spatiotemporal causal effects when treatments spill over across connected units and evolve over time. It links path-indexed causal targets to neural adjustment for high-dimensional, multimodal focal-unit and neighborhood histories and a self-normalized estimating procedure for spillover exposure, thereby providing a principled way to evaluate how policy effects unfold over time and extend across connected populations.

One practical implication of this construction is that its scientific scope extends beyond pandemic policy evaluation because application-specific information can enter the framework without making the causal target implicit. Across applications, domain knowledge can guide the specification of the network and exposure mapping, representation learning can summarize growing focal-unit and neighborhood histories, and the causal target can remain an explicit contrast between intervention paths. This organization allows richer or newly available data modalities to strengthen adjustment without turning the estimand into a latent or prediction-defined quantity. The same structure is relevant to regional pollution controls linked by atmospheric transport, health-care policies connected through referral networks, and local economic interventions transmitted through commuting or trade. Across these domains, the scientific inputs used to define connectivity and adjustment may change, while the causal contrast remains explicitly tied to the intervention paths of interest, thereby preserving its substantive interpretation.

The same division of roles also clarifies where scientific judgment and empirical support remain indispensable. External domain knowledge can often guide the network and exposure mapping, as geography and mobility do in pandemic studies, but it may not determine a unique specification. Because the estimand and resulting conclusions depend on these choices, sensitivity across scientifically plausible alternatives is important. Meanwhile, a more intrinsic limitation arises from pathwise support: longer intervention windows and higher-dimensional spillover summaries define richer causal questions but leave fewer observations near the target exposure path, with an effective local sample size of order ($nh_n^{qM}$) for continuous spillover exposures. Increasing the bandwidth enlarges the effective local sample by incorporating information from exposure paths farther from the target, but at the cost of greater localization bias; bandwidth selection must therefore balance the gain in support against the resulting bias, and flexible representation learning cannot eliminate this design-level trade-off. Finally, uncertainty quantification remains challenging under network dependence. The disjoint network-block bootstrap used here is a pragmatic approximation, and its finite-sample performance depends largely on how well the resampled neighborhoods reproduce the dependence structure relevant to the estimator.

An intriguing future direction for addressing pathwise support is to develop a data-driven, unit-specific bandwidth selection procedure calibrated to the local density of exposure paths, with smaller bandwidths in data-rich regions and larger bandwidths where support is sparse. Likewise, more reliable uncertainty quantification calls for resampling or variance-estimation procedures that better preserve dependence induced by overlapping neighborhoods. Beyond these methodological refinements, a broader scientific extension is to allow connectivity to evolve in response to prior interventions, with time-varying exposure mappings defined over the evolving network. Such an extension would permit effects operating through intervention-induced changes in connectivity to be distinguished from spillovers transmitted through a fixed network.




\section{Conflict of interest}
The authors declare no conflicts of interest.


\section{Data availability}
Sample data underlying this article are available in the
Causal\_spillover\_temporal repository at
\url{https://github.com/JuniperZhang/Causal_spillover_temporal}. The full data underlying this article will be shared upon reasonable request to the corresponding author.

\section{Acknowledgments and funding information}

The authors used ChatGPT to assist with manuscript polishing and with the development and refinement of plotting code. All resulting materials were reviewed and revised by the authors, who take full responsibility for the final manuscript and analyses.

\subsection*{Funding}

This work was supported by the Centers for Disease Control and Prevention
(grant number U01CK000674); the Eunice Kennedy Shriver National Institute
of Child Health and Human Development, National Institutes of Health
(grant number R01HD099348); and the Youth Science Fund Project (Youth Science Fund (Category C)) of the
National Natural Science Foundation of China
(grant number 12501403).



\bibliographystyle{plainnat}
\bibliography{references}

\begin{thebibliography}{34}
\providecommand{\natexlab}[1]{#1}
\providecommand{\url}[1]{\texttt{#1}}
\expandafter\ifx\csname urlstyle\endcsname\relax
  \providecommand{\doi}[1]{doi: #1}\else
  \providecommand{\doi}{doi: \begingroup \urlstyle{rm}\Url}\fi

\bibitem[Aronow and Samii(2017)]{aronow2017estimating}
Peter~M Aronow and Cyrus Samii.
\newblock Estimating average causal effects under general interference, with application to a social network experiment.
\newblock 2017.

\bibitem[Bica et~al.(2020)Bica, Alaa, Jordon, and Van Der~Schaar]{bica2020estimating}
Ioana Bica, Ahmed~M Alaa, James Jordon, and Mihaela Van Der~Schaar.
\newblock Estimating counterfactual treatment outcomes over time through adversarially balanced representations.
\newblock \emph{arXiv preprint arXiv:2002.04083}, 2020.

\bibitem[Chang et~al.(2021)Chang, Pierson, Koh, Gerardin, Redbird, Grusky, and Leskovec]{chang2021mobility}
Serina Chang, Emma Pierson, Pang~Wei Koh, Jaline Gerardin, Beth Redbird, David Grusky, and Jure Leskovec.
\newblock Mobility network models of covid-19 explain inequities and inform reopening.
\newblock \emph{Nature}, 589\penalty0 (7840):\penalty0 82--87, 2021.

\bibitem[Chang et~al.(2023)Chang, Vrabac, Leskovec, and Ugander]{chang2023estimating}
Serina Chang, Damir Vrabac, Jure Leskovec, and Johan Ugander.
\newblock Estimating geographic spillover effects of covid-19 policies from large-scale mobility networks.
\newblock In \emph{Proceedings of the AAAI Conference on Artificial Intelligence}, volume~37, pages 14161--14169, 2023.

\bibitem[Cole and Hern{\'a}n(2008)]{cole2008constructing}
Stephen~R Cole and Miguel~A Hern{\'a}n.
\newblock Constructing inverse probability weights for marginal structural models.
\newblock \emph{American journal of epidemiology}, 168\penalty0 (6):\penalty0 656--664, 2008.

\bibitem[Egami et~al.(2022)Egami, Fong, Grimmer, Roberts, and Stewart]{egami2022make}
Naoki Egami, Christian~J Fong, Justin Grimmer, Margaret~E Roberts, and Brandon~M Stewart.
\newblock How to make causal inferences using texts.
\newblock \emph{Science Advances}, 8\penalty0 (42):\penalty0 eabg2652, 2022.

\bibitem[Fan(2018)]{fan1996local}
Jianqing Fan.
\newblock \emph{Local polynomial modelling and its applications: monographs on statistics and applied probability 66}.
\newblock Routledge, 2018.

\bibitem[Flaxman et~al.(2020)Flaxman, Mishra, Gandy, Unwin, Mellan, Coupland, Whittaker, Zhu, Berah, Eaton, et~al.]{flaxman2020estimating}
Seth Flaxman, Swapnil Mishra, Axel Gandy, H~Juliette~T Unwin, Thomas~A Mellan, Helen Coupland, Charles Whittaker, Harrison Zhu, Tresnia Berah, Jeffrey~W Eaton, et~al.
\newblock Estimating the effects of non-pharmaceutical interventions on covid-19 in europe.
\newblock \emph{Nature}, 584\penalty0 (7820):\penalty0 257--261, 2020.

\bibitem[Forastiere et~al.(2021)Forastiere, Airoldi, and Mealli]{forastiere2021identification}
Laura Forastiere, Edoardo~M Airoldi, and Fabrizia Mealli.
\newblock Identification and estimation of treatment and interference effects in observational studies on networks.
\newblock \emph{Journal of the American Statistical Association}, 116\penalty0 (534):\penalty0 901--918, 2021.

\bibitem[Giffin et~al.(2022)Giffin, Gong, Majumder, Rappold, Reich, and Yang]{giffin2022estimating}
Andrew Giffin, Wenlong Gong, Suman Majumder, Ana~G Rappold, Brian~J Reich, and Shu Yang.
\newblock Estimating intervention effects on infectious disease control: The effect of community mobility reduction on coronavirus spread.
\newblock \emph{Spatial Statistics}, 52:\penalty0 100711, 2022.

\bibitem[Guo et~al.(2024)Guo, Shen, and Li]{guo2024estimating}
Suhan Guo, Furao Shen, and Ni~Li.
\newblock Estimating the treatment effect over time under general interference through deep learner integrated tmle.
\newblock \emph{arXiv preprint arXiv:2412.04799}, 2024.

\bibitem[Hale et~al.(2021)Hale, Angrist, Goldszmidt, Kira, Petherick, Phillips, Webster, Cameron-Blake, Hallas, Majumdar, et~al.]{hale2021global}
Thomas Hale, Noam Angrist, Rafael Goldszmidt, Beatriz Kira, Anna Petherick, Toby Phillips, Samuel Webster, Emily Cameron-Blake, Laura Hallas, Saptarshi Majumdar, et~al.
\newblock A global panel database of pandemic policies (oxford covid-19 government response tracker).
\newblock \emph{Nature human behaviour}, 5\penalty0 (4):\penalty0 529--538, 2021.

\bibitem[Halloran and Struchiner(1995)]{halloran1995causal}
M~Elizabeth Halloran and Claudio~J Struchiner.
\newblock Causal inference in infectious diseases.
\newblock \emph{Epidemiology}, 6\penalty0 (2):\penalty0 142--151, 1995.

\bibitem[Holtz et~al.(2020)Holtz, Zhao, Benzell, Cao, Rahimian, Yang, Allen, Collis, Moehring, Sowrirajan, et~al.]{holtz2020interdependence}
David Holtz, Michael Zhao, Seth~G Benzell, Cathy~Y Cao, Mohammad~Amin Rahimian, Jeremy Yang, Jennifer Allen, Avinash Collis, Alex Moehring, Tara Sowrirajan, et~al.
\newblock Interdependence and the cost of uncoordinated responses to covid-19.
\newblock \emph{Proceedings of the National Academy of Sciences}, 117\penalty0 (33):\penalty0 19837--19843, 2020.

\bibitem[Hong and Raudenbush(2006)]{hong2006evaluating}
Guanglei Hong and Stephen~W Raudenbush.
\newblock Evaluating kindergarten retention policy: A case study of causal inference for multilevel observational data.
\newblock \emph{Journal of the American Statistical Association}, 101\penalty0 (475):\penalty0 901--910, 2006.

\bibitem[Hudgens and Halloran(2008)]{hudgens2008toward}
Michael~G Hudgens and M~Elizabeth Halloran.
\newblock Toward causal inference with interference.
\newblock \emph{Journal of the american statistical association}, 103\penalty0 (482):\penalty0 832--842, 2008.

\bibitem[Imai and Ratkovic(2015)]{imai2015robust}
Kosuke Imai and Marc Ratkovic.
\newblock Robust estimation of inverse probability weights for marginal structural models.
\newblock \emph{Journal of the American Statistical Association}, 110\penalty0 (511):\penalty0 1013--1023, 2015.

\bibitem[Kennedy et~al.(2017)Kennedy, Ma, McHugh, and Small]{kennedy2017non}
Edward~H Kennedy, Zongming Ma, Matthew~D McHugh, and Dylan~S Small.
\newblock Non-parametric methods for doubly robust estimation of continuous treatment effects.
\newblock \emph{Journal of the Royal Statistical Society Series B: Statistical Methodology}, 79\penalty0 (4):\penalty0 1229--1245, 2017.

\bibitem[Li et~al.(2021)Li, Hu, Lu, Utsumi, Chakraborty, Sow, Madan, Li, Ghalwash, Shahn, et~al.]{li2021g}
Rui Li, Stephanie Hu, Mingyu Lu, Yuria Utsumi, Prithwish Chakraborty, Daby~M Sow, Piyush Madan, Jun Li, Mohamed Ghalwash, Zach Shahn, et~al.
\newblock G-net: a recurrent network approach to g-computation for counterfactual prediction under a dynamic treatment regime.
\newblock In \emph{Machine Learning for Health}, pages 282--299. PMLR, 2021.

\bibitem[Lim(2018)]{lim2018forecasting}
Bryan Lim.
\newblock Forecasting treatment responses over time using recurrent marginal structural networks.
\newblock \emph{Advances in neural information processing systems}, 31, 2018.

\bibitem[Liu and Wong(2026)]{liu2026ai}
Qiao Liu and Wing~Hung Wong.
\newblock An ai-powered bayesian generative modeling approach for causal inference in observational studies.
\newblock \emph{Journal of the American Statistical Association}, pages 1--12, 2026.

\bibitem[Liu et~al.(2024)Liu, Chen, and Wong]{liu2024encoding}
Qiao Liu, Zhongren Chen, and Wing~Hung Wong.
\newblock An encoding generative modeling approach to dimension reduction and covariate adjustment in causal inference with observational studies.
\newblock \emph{Proceedings of the National Academy of Sciences}, 121\penalty0 (23):\penalty0 e2322376121, 2024.

\bibitem[Louizos et~al.(2017)Louizos, Shalit, Mooij, Sontag, Zemel, and Welling]{louizos2017causal}
Christos Louizos, Uri Shalit, Joris~M Mooij, David Sontag, Richard Zemel, and Max Welling.
\newblock Causal effect inference with deep latent-variable models.
\newblock \emph{Advances in neural information processing systems}, 30, 2017.

\bibitem[Ma et~al.(2020)Ma, Wang, and Tresp]{ma2020causal}
Yunpu Ma, Yuyi Wang, and Volker Tresp.
\newblock Causal inference under networked interference.
\newblock \emph{arXiv preprint arXiv:2002.08506}, 2020.

\bibitem[Ogburn and VanderWeele(2014)]{ogburn2014causal}
Elizabeth~L Ogburn and Tyler~J VanderWeele.
\newblock Causal diagrams for interference.
\newblock \emph{Statistical science}, pages 559--578, 2014.

\bibitem[Papadogeorgou et~al.(2022)Papadogeorgou, Imai, Lyall, and Li]{papadogeorgou2022causal}
Georgia Papadogeorgou, Kosuke Imai, Jason Lyall, and Fan Li.
\newblock Causal inference with spatio-temporal data: Estimating the effects of airstrikes on insurgent violence in iraq.
\newblock \emph{Journal of the Royal Statistical Society: Series B (Statistical Methodology)}, 84\penalty0 (5):\penalty0 1969--1999, 2022.

\bibitem[Robins and Hernan(2008)]{robins2008estimation}
James Robins and Miguel Hernan.
\newblock Estimation of the causal effects of time-varying exposures.
\newblock \emph{Chapman \& Hall/CRC Handbooks of Modern Statistical Methods}, pages 553--599, 2008.

\bibitem[Robins(1997)]{robins1997causal}
James~M Robins.
\newblock Causal inference from complex longitudinal data.
\newblock In \emph{Latent variable modeling and applications to causality}, pages 69--117. Springer, 1997.

\bibitem[Rosenbaum(1987)]{rosenbaum1987sensitivity}
Paul~R Rosenbaum.
\newblock Sensitivity analysis for certain permutation inferences in matched observational studies.
\newblock \emph{Biometrika}, 74\penalty0 (1):\penalty0 13--26, 1987.

\bibitem[Rubin(1974)]{rubin1974estimating}
Donald~B Rubin.
\newblock Estimating causal effects of treatments in randomized and nonrandomized studies.
\newblock \emph{Journal of educational Psychology}, 66\penalty0 (5):\penalty0 688, 1974.

\bibitem[Veitch et~al.(2020)Veitch, Sridhar, and Blei]{veitch2020adapting}
Victor Veitch, Dhanya Sridhar, and David Blei.
\newblock Adapting text embeddings for causal inference.
\newblock In \emph{Conference on uncertainty in artificial intelligence}, pages 919--928. PMLR, 2020.

\bibitem[Wang and Jetsupphasuk(2021)]{wang2021causal}
Ye~Wang and Michael Jetsupphasuk.
\newblock Causal inference in longitudinal data under unknown interference.
\newblock \emph{arXiv preprint arXiv:2106.15074}, 2021.

\bibitem[Zhang et~al.(2023)Zhang, Kennard, Smith, McFarland, McCallum, and Keith]{zhang2023causal}
Raymond Zhang, Neha~Nayak Kennard, Daniel Smith, Dan McFarland, Andrew McCallum, and Katherine~A Keith.
\newblock Causal matching with text embeddings: A case study in estimating the causal effects of peer review policies.
\newblock In \emph{Findings of the Association for Computational Linguistics: ACL 2023}, pages 1284--1297, 2023.

\bibitem[Zhou et~al.(2024)Zhou, Imai, Lyall, and Papadogeorgou]{zhou2024estimating}
Lingxiao Zhou, Kosuke Imai, Jason Lyall, and Georgia Papadogeorgou.
\newblock Estimating heterogeneous treatment effects for spatio-temporal causal inference.
\newblock \emph{arXiv preprint arXiv:2412.15128}, 2024.

\end{thebibliography}



\section*{Supplementary material}

# S1 Sensitivity Analysis for Unmeasured Confounding

This section provides additional details for the sensitivity analysis summarized in Section 4 of the main text. The main analysis uses a single parameter $\Gamma$ to bound hidden bias in the joint own-treatment and spillover-exposure assignment mechanism. We first show how this joint bound can be interpreted through its two assignment components and then describe the resulting pathwise bounds, longitudinal sensitivity intervals, and robustness summaries.

## S1.1 Component-wise representation of hidden bias

At decision time $\tau_m$, let $U_{i,\tau_m}$ denote possibly vector-valued unmeasured focal-unit and neighborhood information relevant to the joint assignment process and potential outcomes, and let

$$U_{i,1:\tau_m} = (U_{i,\tau_1}, \dots, U_{i,\tau_m})$$

collect the corresponding latent pre-decision history through $\tau_m$. As in the main text, suppose that sequential ignorability would hold after conditioning on $(H_{i,\tau_m}, U_{i,1:\tau_m})$.

Write $p_{\tau_m}(x, d \mid H_{i,\tau_m})$ for the joint conditional probability mass or density given the observed history, and write $p^*_{\tau_m}(x, d \mid H_{i,\tau_m}, U_{i,1:\tau_m})$ for its counterpart given both the observed and latent histories. To distinguish hidden bias in the two components of the joint assignment mechanism, factor

$$p_{\tau_m}(x, d \mid H_{i,\tau_m}) = p^X_{\tau_m}(x \mid H_{i,\tau_m}) p^D_{\tau_m}(d \mid H_{i,\tau_m}, X_{i,\tau_m} = x),$$

and analogously,

$$p^*_{\tau_m}(x, d \mid H_{i,\tau_m}, U_{i,1:\tau_m}) = p^{X,*}_{\tau_m}(x \mid H_{i,\tau_m}, U_{i,1:\tau_m}) p^{D,*}_{\tau_m}(d \mid H_{i,\tau_m}, U_{i,1:\tau_m}, X_{i,\tau_m} = x).$$

At the realized assignment, define the component-wise density ratios

$$\rho^X_{i,\tau_m} = \frac{p^X_{\tau_m}(X_{i,\tau_m} \mid H_{i,\tau_m})}{p^{X,*}_{\tau_m}(X_{i,\tau_m} \mid H_{i,\tau_m}, U_{i,1:\tau_m})}, \quad \text{and} \quad \rho^D_{i,\tau_m} = \frac{p^D_{\tau_m}(D_{i,\tau_m} \mid H_{i,\tau_m}, X_{i,\tau_m})}{p^{D,*}_{\tau_m}(D_{i,\tau_m} \mid H_{i,\tau_m}, U_{i,1:\tau_m}, X_{i,\tau_m})}.$$

The joint ratio in the main text therefore satisfies $\rho_{i,\tau_m} = \rho^X_{i,\tau_m}\rho^D_{i,\tau_m}$.

Following Rosenbaum-type hidden-bias models and their spatiotemporal extensions (Rosenbaum, 1987; Papadogeorgou et al., 2022), one possible component-wise calibration imposes

$$\Gamma_X^{-1} \le \rho^X_{i,\tau_m} \le \Gamma_X, \qquad \Gamma_D^{-1} \le \rho^D_{i,\tau_m} \le \Gamma_D, \qquad \Gamma_X, \Gamma_D \ge 1.$$

It follows that

$$(\Gamma_X\Gamma_D)^{-1} \le \rho_{i,\tau_m} \le \Gamma_X\Gamma_D.$$

Thus, the joint sensitivity model in the main text is recovered by setting $\Gamma = \Gamma_X\Gamma_D$.

For the binary own-treatment component, the bound on $\rho^X_{i,\tau_m}$ can be calibrated through a bounded logit tilt:

$$\text{logit}\left\{p^{X,*}_{\tau_m}(1 \mid H_{i,\tau_m}, U_{i,1:\tau_m})\right\} = \text{logit}\left\{p^X_{\tau_m}(1 \mid H_{i,\tau_m})\right\} + \delta^X_{i,\tau_m}, \quad \text{where } \left|\delta^X_{i,\tau_m}\right| \le \log\Gamma_X.$$

For the spillover component, the density-ratio bound has the analogous interpretation that unmeasured history can tilt the local exposure density at the realized exposure by at most a factor $\Gamma_D$. This formulation is useful in the continuous-exposure setting because the kernel estimator weights units near the target exposure path but evaluates the generalized propensity score at each realized $D_{i,\tau_m}$.

## S1.2 Pathwise accumulation and sensitivity-adjusted weights

Fix an outcome time $t$, let $m_0 = M_t - M + 1$, and consider the $M$-decision target path $(x_{t;M}, d_{t;M})$. The decision-specific hidden bias accumulates through the pathwise multiplier

$$r_i = \prod_{\ell=m_0}^{M_t} \rho_{i,\tau_\ell}.$$

Under the joint-$\Gamma$ model, $\Gamma^{-M} \le r_i \le \Gamma^M$. Hence, for $\Gamma > 1$, the admissible range of pathwise weight distortions expands with the length of the intervention window.

Conditional on the fitted observed-history weights, an admissible latent assignment mechanism perturbs unit $i$'s weight from $\widehat{\omega}_{i,t;M}(h)$ to $\widehat{\omega}_{i,t;M}(h)r_i$. For $\boldsymbol{r} = (r_1, \ldots, r_n)$, the corresponding sensitivity-adjusted Hájek estimator is

$$\widehat{\mu}_{\mathrm{H},t}(\boldsymbol{r}; x_{t;M}, d_{t;M}) = \frac{\sum_{i=1}^{n} \widehat{\omega}_{i,t;M}(h) r_i Y_{i,t}}{\sum_{i=1}^{n} \widehat{\omega}_{i,t;M}(h) r_i}.$$

Accordingly, allowing the latent multipliers to vary over their admissible range yields the sensitivity interval for the mean potential outcome,

$$\mathcal{I}_{\mu,t}^{\Gamma}(x_{t;M}, d_{t;M}) = \left[\underline{\widehat{\mu}}_{\mathrm{H},t}^{\Gamma}(x_{t;M}, d_{t;M}),\, \overline{\widehat{\mu}}_{\mathrm{H},t}^{\Gamma}(x_{t;M}, d_{t;M})\right],$$

where

$$\underline{\widehat{\mu}}_{\mathrm{H},t}^{\Gamma}(x_{t;M}, d_{t;M}) = \min_{\boldsymbol{r} \in [\Gamma^{-M}, \Gamma^{M}]^n} \widehat{\mu}_{\mathrm{H},t}(\boldsymbol{r}; x_{t;M}, d_{t;M}),$$
$$\overline{\widehat{\mu}}_{\mathrm{H},t}^{\Gamma}(x_{t;M}, d_{t;M}) = \max_{\boldsymbol{r} \in [\Gamma^{-M}, \Gamma^{M}]^n} \widehat{\mu}_{\mathrm{H},t}(\boldsymbol{r}; x_{t;M}, d_{t;M}).$$

The two endpoint problems above admit equivalent linear-program representations through the Charnes–Cooper transformation (Papadogeorgou et al., 2022). The formal result and proof are given in Proposition S4 in Section S2.

## S1.3 Longitudinal causal contrasts and robustness summaries

Let

$$x_{\cdot;M} = \{x_{t;M} : t \in \mathcal{T}_M\}, \qquad d_{\cdot;M} = \{d_{t;M} : t \in \mathcal{T}_M\}$$

collect the time-specific intervention paths entering a longitudinal mean. The corresponding lower and upper sensitivity bounds for the longitudinal mean are

$$\underline{\widehat{\Psi}}_{\mathrm{H}}^{\Gamma}(x_{\cdot;M}, d_{\cdot;M}) = \frac{1}{|\mathcal{T}_M|} \sum_{t \in \mathcal{T}_M} \underline{\widehat{\mu}}_{\mathrm{H},t}^{\Gamma}(x_{t;M}, d_{t;M}),$$
$$\overline{\widehat{\Psi}}_{\mathrm{H}}^{\Gamma}(x_{\cdot;M}, d_{\cdot;M}) = \frac{1}{|\mathcal{T}_M|} \sum_{t \in \mathcal{T}_M} \overline{\widehat{\mu}}_{\mathrm{H},t}^{\Gamma}(x_{t;M}, d_{t;M}).$$

Consider a causal contrast comparing $(x_{\cdot;M}, d_{\cdot;M})$ with $(x'_{\cdot;M}, d'_{\cdot;M})$. Its sensitivity interval is

$$\mathcal{I}^{\Gamma}_{\text{effect}} = \Big[\underline{\widehat{\Psi}}^{\Gamma}_{\mathrm{H}}(x_{\cdot;M}, d_{\cdot;M}) - \overline{\widehat{\Psi}}^{\Gamma}_{\mathrm{H}}(x'_{\cdot;M}, d'_{\cdot;M}),\ \overline{\widehat{\Psi}}^{\Gamma}_{\mathrm{H}}(x_{\cdot;M}, d_{\cdot;M}) - \underline{\widehat{\Psi}}^{\Gamma}_{\mathrm{H}}(x'_{\cdot;M}, d'_{\cdot;M})\Big].$$

The sensitivity intervals for direct, spillover, and total effects follow by setting $d'_{\cdot;M} = d_{\cdot;M}$, setting $x'_{\cdot;M} = x_{\cdot;M}$, or allowing both paths to differ, respectively.

These intervals are conservative because the latent multipliers are allowed to vary independently across outcome times and across the two intervention paths entering the contrast. Tighter bounds could be obtained by imposing additional restrictions linking the multipliers through a shared latent assignment mechanism.

For numerical experiments, we evaluate the sensitivity intervals over a prespecified increasing grid $\mathcal{G}_\Gamma$ beginning at $\Gamma = 1$. Since the admissible set $[\Gamma^{-M}, \Gamma^{M}]^n$ expands with $\Gamma$, both the mean and causal-effect sensitivity intervals are nested: increasing $\Gamma$ can only weakly widen the corresponding interval.

For a reported causal effect, a concise robustness summary is

$$\Gamma^* = \min\left\{\Gamma \in \mathcal{G}_\Gamma : 0 \in \mathcal{I}^{\Gamma}_{\text{effect}}\right\},$$

whenever this set is nonempty. Thus, $\Gamma^*$ is the smallest value on the prespecified grid at which the corresponding sensitivity interval contains zero. When scientifically meaningful, hidden bias in own-treatment and spillover assignment may be calibrated separately through $(\Gamma_X, \Gamma_D)$, although, under the component-wise bounds above, the resulting intervals depend only on their product.

# S2 Additional Theoretical Results, Technical Conditions, and Proofs

This section first states additional theoretical results for longitudinal means, causal-effect contrasts, discrete spillover exposure, and feasible estimators; it then gives the technical conditions and proofs

supporting these results and those in Section 4 of the main text. Throughout, we condition on $\mathcal{G}_n$ and suppress this conditioning in probability, expectation, and convergence notation. The exposure dimension $q$, window length $M$, and process- and decision-time horizons are fixed.

For a given outcome time $t$, let $M_t = \max\{m : \tau_m \leq t\}$ and $m_0 = M_t - M + 1$, and fix target paths $(x_{t;M}, d_{t;M})$. At a decision time, assignment precedes the outcome at that time, while $H_{i,\tau_m}$ remains the pre-decision history. In addition to the oracle weight $\omega^{\circ}_{i,t;M}(h)$ defined in the main text, write

$$A^{\circ}_{i,t;M}(h) = h^{qM}\omega^{\circ}_{i,t;M}(h),$$

and define

$$N^{\circ}_{n,t}(h) = \frac{1}{n}\sum_{i=1}^{n}\omega^{\circ}_{i,t;M}(h)Y_{i,t}, \qquad D^{\circ}_{n,t}(h) = \frac{1}{n}\sum_{i=1}^{n}\omega^{\circ}_{i,t;M}(h).$$

Thus, $N^{\circ}_{n,t}(h)$ is the oracle Horvitz–Thompson estimator and $N^{\circ}_{n,t}(h)/D^{\circ}_{n,t}(h)$ is the oracle Hájek estimator.

## S2.1 Additional theoretical results

The first result extends the pointwise consistency in Theorem 1 to the longitudinal mean estimators and the causal-effect contrasts considered in the main text.

**Corollary S1** (Consistency of longitudinal means and causal-effect contrasts)**.** *Let $\mathcal{T}_M$ be fixed and finite, and suppose the conditions of Theorem 1 hold for every $t \in \mathcal{T}_M$ and its corresponding target path. Define*

$$\Psi_{M,n} = \frac{1}{|\mathcal{T}_M|}\sum_{t\in\mathcal{T}_M}\mu_{n,t}(x_{t;M}, d_{t;M}).$$

*Then, for $\zeta \in \{HT, H\}$,*

$$\widehat{\Psi}^{\circ,\text{K-IPW},M}_{\zeta} - \Psi_{M,n} \xrightarrow{p} 0. \tag{S2.1}$$

*More generally, let* $\boldsymbol{\mu}_n = (\mu_{n,1}, \ldots, \mu_{n,J})^\top$ *collect any fixed finite collection of time- and path-specific conditional mean potential outcomes satisfying the conditions of Theorem 1 , and let* $\widehat{\boldsymbol{\mu}}^{\circ}_{\zeta}$ *collect their oracle estimators. Then, for every fixed conformable matrix* $C$*,*

$$C\left(\widehat{\boldsymbol{\mu}}^{\circ}_{\zeta} - \boldsymbol{\mu}_n\right) \xrightarrow{p} 0.$$

*This includes consistency of the oracle direct, spillover, and total effect estimators for their corresponding* $n$*-indexed conditional causal-effect contrasts in (2.1).*

We next consider discrete spillover exposure. Set $K_h(u,d) = \mathbf{1}\{u = d\}$ and define the oracle weights and estimators as in the main text. The following result gives the discrete counterparts of Theorem 1 and Theorem 2 .

**Proposition S2** (Discrete-exposure counterpart)**.** *Suppose Assumptions 1–3, the common-target condition in Assumption S7, the conditional finite-range network-dependence condition in Assumption S8, and the uniform-overlap and second-moment conditions in Assumption S10 hold. Then, for* $\zeta \in \{HT, H\}$*,*

$$\widehat{\mu}^{\circ,\text{K-IPW},M}_{\zeta,t}(x_{t;M}, d_{t;M}) - \mu_{n,t}(x_{t;M}, d_{t;M}) \xrightarrow{p} 0.$$

*If, in addition, the third-moment and nondegenerate limiting-covariance conditions in Assumption S10 hold, then*

$$\sqrt{n}\left\{\widehat{\mu}^{\circ,\text{K-IPW},M}_{\zeta,t}(x_{t;M}, d_{t;M}) - \mu_{n,t}(x_{t;M}, d_{t;M})\right\} \xrightarrow{d} \mathcal{N}(0, \sigma^2_{\zeta,t}).$$

*The consistency conclusion extends to the oracle longitudinal mean estimators and the direct, spillover, and total effect estimators for their corresponding* $n$*-indexed conditional targets. Under the corresponding joint covariance condition, these estimators are also asymptotically normal with the usual* $\sqrt{n}$*-normalization.*

We next turn to the feasible estimators. Recall the maximal relative propensity-score error $\Delta_n$ defined in Section 4 of the main text. The following result gives sufficient conditions for the

oracle-to-feasible transfer invoked there.

**Proposition S3** (Oracle-to-feasible transfer)**.** *Suppose $M$ is fixed and the conditions of the corresponding oracle consistency result hold. If $\Delta_n = o_p(1)$, then, for $\zeta \in \{HT, H\}$,*

$$\widehat{\mu}_{\zeta,t}^{\text{K-IPW},M} - \widehat{\mu}_{\zeta,t}^{\circ,\text{K-IPW},M} = O_p(\Delta_n). \tag{S2.2}$$

*For any fixed finite collection of outcome times and intervention paths, the same bound holds uniformly over its component estimators, provided that $\Delta_n$ is defined uniformly over that collection. Consequently, oracle consistency carries over to the corresponding feasible pointwise, longitudinal-mean, and causal-effect estimators.*

*If the conditions of the corresponding oracle asymptotic-normality result also hold, the oracle limiting distributions carry over whenever $\Delta_n = o_p\{(nh_n^{qM})^{-1/2}\}$ for continuous spillover exposure, or whenever $\Delta_n = o_p(n^{-1/2})$ for discrete spillover exposure. These oracle limits are given in Theorem 2 and Corollary 1 for continuous exposure and in Proposition S2 for discrete exposure.*

For sensitivity analysis, we established that the sensitivity endpoints defined in Section S1 admit the following Charnes–Cooper linear-program representation (Papadogeorgou et al., 2022).

**Proposition S4** (Linear-program representation of the sensitivity endpoints)**.** *Fix an outcome time $t$, a window length $M$, and target paths $(x_{t;M}, d_{t;M})$. Suppose that $\widehat{\omega}_{i,t;M}(h) \geq 0$ for all $i$ and $\sum_{i=1}^n \widehat{\omega}_{i,t;M}(h) > 0$. For any $\Gamma \geq 1$, the lower sensitivity endpoint*

$$\underline{\widehat{\mu}}_{\mathrm{H},t}^{\Gamma}(x_{t;M}, d_{t;M}) = \min_{\boldsymbol{r} \in [\Gamma^{-M}, \Gamma^{M}]^n} \frac{\sum_{i=1}^n \widehat{\omega}_{i,t;M}(h) r_i Y_{i,t}}{\sum_{i=1}^n \widehat{\omega}_{i,t;M}(h) r_i}$$

*equals the optimal value of the linear program*

$$\min_{\boldsymbol{q},\kappa} \sum_{i=1}^n \widehat{\omega}_{i,t;M}(h) q_i Y_{i,t}$$

*subject to*

$$\sum_{i=1}^n \widehat{\omega}_{i,t;M}(h) q_i = 1, \qquad \Gamma^{-M}\kappa \leq q_i \leq \Gamma^{M}\kappa, \quad i = 1, \ldots, n, \qquad \kappa \geq 0.$$

*The upper sensitivity endpoint has the same representation with* min *function replaced by* max *function.*

## S2.2 Technical conditions

The conditions below formalize the regularity requirements summarized in Section 4 of the main text. The kernel, local-support, localized-moment, smoothness, and $h_n$-scaled covariance conditions are specific to continuous spillover exposure. The common-target and conditional finite-range network-dependence conditions are shared by the continuous and discrete settings. For discrete spillover exposure, the corresponding overlap, moment, and covariance conditions are stated separately in Assumption S10. When a fixed finite collection of outcome times and target paths is considered, all uniformity statements are understood to hold over that collection.

**Assumption S5** (Kernel and local support)**.** *The kernel $K : \mathbb{R}^q \to \mathbb{R}_+$ is bounded, has compact support $\mathcal{K}$, and satisfies $\int K(u)\,du = 1$. There exists $h_0 > 0$ such that, for every $n$, unit $i = 1, \ldots, n$, decision index $\ell = m_0, \ldots, M_t$, and $0 < h \le h_0$,*

$$d_{\tau_\ell} + h\mathcal{K} \subset \text{interior}\Big(\text{supp}\{D_{i,\tau_\ell} \mid H_{i,\tau_\ell}, X_{i,\tau_\ell} = x_{\tau_\ell}\}\Big)$$

*almost surely. For asymptotic normality, $K$ additionally satisfies the second-order conditions*

$$\int uK(u)\,du = 0, \qquad \int \|u\|^2 K(u)\,du < \infty.$$

**Assumption S6** (Uniform local overlap and moments)**.** *There exists a constant $c > 0$, independent of $n$, such that, almost surely,*

$$\inf_{i \le n} \inf_{\ell = m_0, \ldots, M_t} \inf_{u \in d_{\tau_\ell} + h_0\mathcal{K}} p_{\tau_\ell}(x_{\tau_\ell}, u \mid H_{i,\tau_\ell}) \ge c.$$

*Moreover, there exist constants $C_r < \infty$, $r = 1, 2, 3$, independent of $n$, such that, uniformly over $i \le n$ and $0 < h \le h_0$,*

$$\mathbb{E}\left[\left|A^{\circ}_{i,t;M}(h)\right|^r \,\Big|\, \mathcal{G}_n\right] + \mathbb{E}\left[\left|A^{\circ}_{i,t;M}(h)Y_{i,t}\right|^r \,\Big|\, \mathcal{G}_n\right] \le C_r h^{qM}, \qquad r = 1, 2, 3.$$

*The $r = 1$ bound for $A^{\circ}_{i,t;M}(h)$ follows from exact normalization, while its outcome-weighted counterpart is used for the oracle-to-feasible transfer. The $r = 2$ bounds are used for consistency, and the $r = 3$ bounds are used for asymptotic normality. These localized moment bounds are implied, for example, by uniform local overlap, boundedness of $K$, and uniformly bounded conditional moments of the relevant outcomes.*

To formulate the common-target and mean-surface conditions, for $u_{t;M} = (u_{\tau_{m_0}}, \ldots, u_{\tau_{M_t}})$, define the unit-specific conditional mean surface

$$m_{n,i,t}(x_{t;M}, u_{t;M}) = \mathbb{E}_{\mathcal{G}_n} \{Y_{i,t}(x_{t;M}, u_{t;M})\}. \tag{S2.3}$$

One thing worth mention is, the common-mean condition below requires these surfaces to share the same value only at the target spillover path $d_{t;M}$; For $u_{t;M} \neq d_{t;M}$, however, the mean surfaces may remain unit-specific. This distinction allows cross-unit heterogeneity near the target while providing the local smoothness needed for kernel localization.

**Assumption S7** (Common target and continuous-exposure smoothness)**.** *For both continuous and discrete spillover exposure, the conditional mean at the target path is common across units:*

$$m_{n,i,t}(x_{t;M}, d_{t;M}) = \mu_{n,t}(x_{t;M}, d_{t;M}), \qquad i = 1, \ldots, n.$$

*For continuous spillover exposure, consistency additionally requires the surfaces $m_{n,i,t}(x_{t;M}, \cdot)$ to be uniformly bounded in a fixed neighborhood of $d_{t;M}$ and continuous at $d_{t;M}$, uniformly over $n$ and $i \leq n$. For asymptotic normality, these surfaces are twice continuously differentiable on a fixed neighborhood of $d_{t;M}$, with first and second derivatives uniformly bounded over $n$ and $i \leq n$.*

**Assumption S8** (Conditional finite-range network dependence)**.** *Let* $\mathrm{dist}(i, j)$ *denote the graph distance between units $i$ and $j$ in the prespecified network, and, for two index sets $I, J \subseteq \{1, \ldots, n\}$, define*

$$\mathrm{dist}(I, J) = \min_{i \in I,\, j \in J} \mathrm{dist}(i, j).$$

*There exists a fixed integer $r_{\text{dep}} \geq 1$, independent of $n$, such that, for every $n$ and any disjoint index sets $I$ and $J$,*

$$\operatorname{dist}(I, J) > r_{\text{dep}} \;\Rightarrow\; \left\{\left(A^{\circ}_{i,t;M}(h), A^{\circ}_{i,t;M}(h)Y_{i,t}\right) : i \in I\right\} \perp\!\!\!\perp \left\{\left(A^{\circ}_{j,t;M}(h), A^{\circ}_{j,t;M}(h)Y_{j,t}\right) : j \in J\right\} \;\Big|\; \mathcal{G}_n$$

*for sufficiently small $h$. Moreover, the graph has a uniformly bounded degree, i.e.,*

$$\sup_n \max_{1 \leq i \leq n} |\{j : \operatorname{dist}(i, j) \leq r_{\text{dep}}\}| < \infty.$$

*For discrete spillover exposure, the same condition holds with $A^{\circ}_{i,t;M}(h)$ replaced by $\omega^{\circ}_{i,t;M}$. For any fixed finite collection of outcome times and target paths, the condition holds for the corresponding stacked random vectors.*

To formulate the limiting covariance conditions under the preceding finite-range dependence structure, define the centered rescaled summands

$$\begin{aligned}
\xi_{HT,n,i,t}(h) &= A^{\circ}_{i,t;M}(h)Y_{i,t} - \mathbb{E}\left[A^{\circ}_{i,t;M}(h)Y_{i,t} \;\Big|\; \mathcal{G}_n\right], \\
\xi_{H,n,i,t}(h) &= A^{\circ}_{i,t;M}(h)\left\{Y_{i,t} - \mu_{n,t}(x_{t;M}, d_{t;M})\right\} \\
&\qquad - \mathbb{E}\left[A^{\circ}_{i,t;M}(h)\left\{Y_{i,t} - \mu_{n,t}(x_{t;M}, d_{t;M})\right\} \;\Big|\; \mathcal{G}_n\right].
\end{aligned} \tag{S2.4}$$

**Assumption S9** (Limiting covariance structure)**.** *For each $\zeta \in \{HT, H\}$,*

$$\frac{1}{nh_n^{qM}} \operatorname{Var}\left\{\sum_{i=1}^{n} \xi_{\zeta,n,i,t}(h_n) \;\Big|\; \mathcal{G}_n\right\} \longrightarrow \sigma^2_{\zeta,t} \in (0, \infty), \quad a.s. \tag{S2.5}$$

*More generally, let $j = 1, \ldots, J$ index a fixed finite collection of component means, each based on the same fixed number $M$ of decision times, and let $\boldsymbol{\xi}_{\zeta,n,i}(h_n)$ denote the $J$-vector collecting their corresponding centered summands. Then*

$$\frac{1}{nh_n^{qM}} \operatorname{Var}\left\{\sum_{i=1}^{n} \boldsymbol{\xi}_{\zeta,n,i}(h_n) \;\Big|\; \mathcal{G}_n\right\} \longrightarrow \boldsymbol{\Sigma}_{\zeta}, \quad a.s. \tag{S2.6}$$

*where $\boldsymbol{\Sigma}_{\zeta}$ is a finite, positive definite $J \times J$ matrix.*

Under Assumption S8, summands associated with units separated by more than $r_{\text{dep}}$ network steps are conditionally independent and hence have zero conditional covariance. Therefore, for

each $\zeta \in \{HT, H\}$,

$$\mathrm{Var}\left\{\sum_{i=1}^{n} \xi_{\zeta,n,i,t}(h_n) \,\middle|\, \mathcal{G}_n\right\} = \sum_{i=1}^{n} \sum_{\mathrm{dist}(i,k)\le r_{\mathrm{dep}}} \mathrm{Cov}\left\{\xi_{\zeta,n,i,t}(h_n), \xi_{\zeta,n,k,t}(h_n) \,\middle|\, \mathcal{G}_n\right\}.$$

The same decomposition applies entrywise to the covariance matrix on the left-hand side of (S2.6). Thus, the limiting covariance structure retains all potentially nonzero covariance terms within the fixed network-dependence radius.

For discrete spillover exposure, let $\xi^d_{\zeta,n,i,t}$, $\zeta \in \{HT, H\}$, denote the discrete counterpart of $\xi_{\zeta,n,i,t}(h)$, obtained by replacing $A^{\circ}_{i,t;M}(h)$ with $\omega^{\circ}_{i,t;M}$.

**Assumption S10** (Discrete-exposure counterparts)**.** *There exists a constant $c > 0$, independent of $n$, such that, for every $n$, almost surely,*

$$\inf_{\substack{1\le i\le n \\ m_0\le \ell\le M_t}} p_{\tau_\ell}(x_{\tau_\ell}, d_{\tau_\ell} \mid H_{i,\tau_\ell}) \ge c.$$

*Moreover, for $r = 2, 3$, there exists $C_r < \infty$ such that, uniformly over $n$ and $1 \le i \le n$,*

$$\mathbb{E}\left[|\omega^{\circ}_{i,t;M}|^r + |\omega^{\circ}_{i,t;M} Y_{i,t}|^r \,\middle|\, \mathcal{G}_n\right] \le C_r.$$

*The pointwise and joint limiting-covariance conditions in Assumption S9 hold with $\xi_{\zeta,n,i,t}(h_n)$ replaced by $\xi^d_{\zeta,n,i,t}$ and the normalization $nh_n^{qM}$ replaced by $n$.*

## S2.3 Proofs

### S2.3.1 Identification and pathwise localization

For continuous spillover exposure and $\ell = m_0, \ldots, M_t$, define the one-step oracle weight

$$W^{\circ}_{i,\tau_\ell}(h) = \frac{\mathbf{1}\{X_{i,\tau_\ell} = x_{\tau_\ell}\} K_h(D_{i,\tau_\ell}, d_{\tau_\ell})}{p_{\tau_\ell}(x_{\tau_\ell}, D_{i,\tau_\ell} \mid H_{i,\tau_\ell})},$$

so that $\omega^{\circ}_{i,t;M}(h) = \prod_{\ell=m_0}^{M_t} W^{\circ}_{i,\tau_\ell}(h)$.

**Lemma S11** (Exact normalization and pathwise localization). *Under Assumptions 1–3 and Assumption S5, for* $0 < h \le h_0$*, whenever the following expectations exist, we have*

$$\mathbb{E}\{\omega^{\circ}_{i,t;M}(h) \mid \mathcal{G}_n\} = 1, \tag{S2.7}$$

$$\mathbb{E}\{\omega^{\circ}_{i,t;M}(h) Y_{i,t} \mid \mathcal{G}_n\} = \int_{\mathbb{R}^{qM}} m_{n,i,t}(x_{t;M}, u_{t;M}) \prod_{\ell=m_0}^{M_t} K_h(u_{\tau_\ell}, d_{\tau_\ell})\, du_{t;M}. \tag{S2.8}$$

*Consequently, under Assumption S7,*

$$\sup_{i \le n} \left| \mathbb{E}\{\omega^{\circ}_{i,t;M}(h) Y_{i,t} \mid \mathcal{G}_n\} - \mu_{n,t}(x_{t;M}, d_{t;M}) \right| = o(1).$$

*In addition, under Assumptions S5 and S7, we further have*

$$\sup_{i \le n} \left| \mathbb{E}\{\omega^{\circ}_{i,t;M}(h) Y_{i,t} \mid \mathcal{G}_n\} - \mu_{n,t}(x_{t;M}, d_{t;M}) \right| = O(h^2).$$

*Proof of Lemma S11.* First, direct conditioning at decision time $\tau_\ell$ gives

$$\begin{aligned}
\mathbb{E}_{\mathcal{G}_n}\{W^{\circ}_{i,\tau_\ell}(h) \mid H_{i,\tau_\ell}\} &= \sum_{x \in \{0,1\}} \int_{\mathbb{R}^q} \frac{\mathbf{1}\{x = x_{\tau_\ell}\} K_h(u, d_{\tau_\ell})}{p_{\tau_\ell}(x_{\tau_\ell}, u \mid H_{i,\tau_\ell})} p_{\tau_\ell}(x, u \mid H_{i,\tau_\ell})\, du \\
&= \int_{\mathbb{R}^q} K_h(u, d_{\tau_\ell})\, du = 1.
\end{aligned}$$

Because the preceding one-step weights are measurable with respect to the history at the next decision time, i.e., $W^{\circ}_{i,\tau_k}(h)$ is $H_{i,\tau_\ell}$-measurable for arbitrary $k < \ell$, thus,

$$\begin{aligned}
\mathbb{E}\{\omega^{\circ}_{i,t;M}(h) \mid \mathcal{G}_n\} &= \mathbb{E}\Big\{ \prod_{\ell=m_0}^{M_t} W^{\circ}_{i,\tau_\ell}(h) \mid \mathcal{G}_n \Big\} = \mathbb{E}\Big\{ \mathbb{E}_{\mathcal{G}_n}\Big\{ \prod_{\ell=m_0}^{M_t} W^{\circ}_{i,\tau_\ell}(h) \mid H_{i,\tau_{M_t}} \Big\} \mid \mathcal{G}_n \Big\} \\
= & \mathbb{E}\Big\{ \prod_{\ell=m_0}^{M_t - 1} W^{\circ}_{i,\tau_\ell}(h) \cdot \mathbb{E}_{\mathcal{G}_n}\Big\{ W^{\circ}_{i,\tau_{M_t}}(h) \mid H_{i,\tau_{M_t}} \Big\} \mid \mathcal{G}_n \Big\} = \mathbb{E}\Big\{ \prod_{\ell=m_0}^{M_t - 1} W^{\circ}_{i,\tau_\ell}(h) \mid \mathcal{G}_n \Big\}
\end{aligned}$$

backward iteration of this identity from $\tau_{M_t}$ to $\tau_{m_0}$ proves (S2.7).

For the outcome-weighted identity, define

$$R_h = \mathbb{E}\left[ Y_{i,t} \prod_{\ell=m_0}^{M_t} W^{\circ}_{i,\tau_\ell}(h) \,\Big|\, \mathcal{G}_n \right].$$

By the consistency condition in Assumption 1 and the partial-window potential-outcome definition

in Section 3.4 of the main text, on the event $X_{i,\tau_{m_0:M_t}} = x_{t;M}$,

$$Y_{i,t} = Y_{i,t}(x_{t;M}, D_{i,\tau_{m_0:M_t}}).$$

Because $\prod_{\ell=m_0}^{M_t} W^{\circ}_{i,\tau_\ell}(h) = 0$ outside this event, it follows that

$$R_h = \mathbb{E}\left[ Y_{i,t}(x_{t;M}, D_{i,\tau_{m_0:M_t}}) \prod_{\ell=m_0}^{M_t} W^{\circ}_{i,\tau_\ell}(h) \;\middle|\; \mathcal{G}_n \right].$$

Notice that, when condition on $H_{i,\tau_{M_t}}$, the past exposures $D_{i,\tau_{m_0:M_t-1}}$ are then fixed, thus

$$\begin{aligned}
&\mathbb{E}\left[ Y_{i,t}(x_{t;M}, D_{i,\tau_{m_0:M_t-1}}, D_{i,\tau_{M_t}}) W^{\circ}_{i,\tau_{M_t}}(h) \;\middle|\; H_{i,\tau_{M_t}} \right] \\
&\quad = \sum_{x\in\{0,1\}} \int_{\mathbb{R}^q} \mathbb{E}\left[ Y_{i,t}(x_{t;M}, D_{i,\tau_{m_0:M_t-1}}, u) \;\middle|\; X_{i,\tau_{M_t}} = x, D_{i,\tau_{M_t}} = u, H_{i,\tau_{M_t}} \right] \\
&\qquad\qquad \times \frac{\mathbf{1}\{x = x_{\tau_{M_t}}\} K_h(u, d_{\tau_{M_t}})}{p_{\tau_{M_t}}(x_{\tau_{M_t}}, u \mid H_{i,\tau_{M_t}})} p_{\tau_{M_t}}(x, u \mid H_{i,\tau_{M_t}})\, du. \\
&\quad = \int_{\mathbb{R}^q} \mathbb{E}\left[ Y_{i,t}(x_{t;M}, D_{i,\tau_{m_0:M_t-1}}, u) \;\middle|\; X_{i,\tau_{M_t}} = x_{\tau_{M_t}}, D_{i,\tau_{M_t}} = u, H_{i,\tau_{M_t}} \right] K_h(u, d_{\tau_{M_t}})\, du. \\
&\quad = \int_{\mathbb{R}^q} \mathbb{E}\left[ Y_{i,t}(x_{t;M}, D_{i,\tau_{m_0:M_t-1}}, u) \;\middle|\; H_{i,\tau_{M_t}} \right] K_h(u, d_{\tau_{M_t}})\, du,
\end{aligned}$$

where the last equality follows from the sequential ignorability Assumption 2. Thus, by tower property,

$$\begin{aligned}
R_h &= \mathbb{E}\Big\{ \Big[ \prod_{\ell=m_0}^{M_t-1} W^{\circ}_{i,\tau_\ell}(h) \Big] \cdot \mathbb{E}\Big[ Y_{i,t}(x_{t;M}, D_{i,\tau_{m_0:M_t-1}}, D_{i,\tau_{M_t}}) W^{\circ}_{i,\tau_{M_t}}(h) \;\Big|\; H_{i,\tau_{M_t}} \Big] \;\Big|\; \mathcal{G}_n \Big\} \\
&= \mathbb{E}\Big\{ \Big[ \prod_{\ell=m_0}^{M_t-1} W^{\circ}_{i,\tau_\ell}(h) \Big] \cdot \Big[ \int_{\mathbb{R}^q} \mathbb{E}\left[ Y_{i,t}(x_{t;M}, D_{i,\tau_{m_0:M_t-1}}, u) \;\middle|\; H_{i,\tau_{M_t}} \right] K_h(u, d_{\tau_{M_t}})\, du \Big] \;\Big|\; \mathcal{G}_n \Big\} \\
&= \int_{\mathbb{R}^q} \mathbb{E}\left[ Y_{i,t}(x_{t;M}, D_{i,\tau_{m_0:M_t-1}}, u) \prod_{\ell=m_0}^{M_t-1} W^{\circ}_{i,\tau_\ell}(h) \;\middle|\; \mathcal{G}_n \right] K_h(u, d_{\tau_{M_t}})\, du
\end{aligned}$$

Repeating the same argument backward for $\tau_{M_t-1}, \dots, \tau_{m_0}$ yields

$$R_h = \int_{\mathbb{R}^{qM}} \mathbb{E}\{ Y_{i,t}(x_{t;M}, u_{t;M}) \mid \mathcal{G}_n \} \prod_{\ell=m_0}^{M_t} K_h(u_{\tau_\ell}, d_{\tau_\ell})\, du_{t;M},$$

which proves (S2.8) according to the definition of $m_{n,i,t}$ in (S2.3).

For each $\ell = m_0, \ldots, M_t$, $u_{\tau_\ell}, d_{\tau_\ell}, v_{\tau_\ell} \in \mathbb{R}^q$. Under the change of variables $u_{\tau_\ell} = d_{\tau_\ell} + hv_{\tau_\ell}$, the Jacobian matrix is $\partial u_{\tau_\ell}/\partial v_{\tau_\ell} = hI_q$. Therefore, $du_{\tau_\ell} = |\det(hI_q)|\, dv_{\tau_\ell} = h^q\, dv_{\tau_\ell}$, and hence

$$K_h(u_{\tau_\ell}, d_{\tau_\ell})\, du_{\tau_\ell} = h^{-q} K(v_{\tau_\ell}) h^q\, dv_{\tau_\ell} = K(v_{\tau_\ell})\, dv_{\tau_\ell}.$$

It follows from (S2.8) that

$$\mathbb{E}\{\omega^{\circ}_{i,t;M}(h) Y_{i,t} \mid \mathcal{G}_n\} = \int_{\mathbb{R}^{qM}} m_{n,i,t}(x_{t;M}, d_{t;M} + hv_{t;M}) \prod_{\ell=m_0}^{M_t} K(v_{\tau_\ell})\, dv_{t;M}.$$

By the normalization of the kernel function $K$,

$$\int_{\mathbb{R}^{qM}} \prod_{\ell=m_0}^{M_t} K(v_{\tau_\ell})\, dv_{t;M} = \prod_{\ell=m_0}^{M_t} \int_{\mathbb{R}^q} K(v_{\tau_\ell})\, dv_{\tau_\ell} = 1.$$

Together with the common-target condition in Assumption S7, this gives

$$\begin{aligned}
&\mathbb{E}\{\omega^{\circ}_{i,t;M}(h) Y_{i,t} \mid \mathcal{G}_n\} - \mu_{n,t}(x_{t;M}, d_{t;M}) \\
&= \int_{\mathbb{R}^{qM}} \{m_{n,i,t}(x_{t;M}, d_{t;M} + hv_{t;M}) - m_{n,i,t}(x_{t;M}, d_{t;M})\} \prod_{\ell=m_0}^{M_t} K(v_{\tau_\ell})\, dv_{t;M}.
\end{aligned}$$

Since the product kernel is supported on $\mathcal{K}^M$, is nonnegative, and integrates to one,

$$\begin{aligned}
&\sup_{i\le n} \left|\mathbb{E}\{\omega^{\circ}_{i,t;M}(h) Y_{i,t} \mid \mathcal{G}_n\} - \mu_{n,t}(x_{t;M}, d_{t;M})\right| \\
&\le \sup_{i\le n} \sup_{v_{t;M}\in\mathcal{K}^M} |m_{n,i,t}(x_{t;M}, d_{t;M} + hv_{t;M}) - m_{n,i,t}(x_{t;M}, d_{t;M})| = o(1),
\end{aligned}$$

where the last equality follows from the uniform continuity condition in Assumption S7.

For the second-order result, the twice continuous differentiability in Assumption S7 gives

$$\begin{aligned}
&m_{n,i,t}(x_{t;M}, d_{t;M} + hv_{t;M}) - m_{n,i,t}(x_{t;M}, d_{t;M}) \\
&= h\nabla_{u_{t;M}} m_{n,i,t}(x_{t;M}, d_{t;M})^\top v_{t;M} \\
&\quad + h^2 \int_0^1 (1-s) v_{t;M}^\top \nabla^2_{u_{t;M}} m_{n,i,t}(x_{t;M}, d_{t;M} + shv_{t;M}) v_{t;M}\, ds.
\end{aligned}$$

For every $k = m_0, \dots, M_t$, the first-moment condition in Assumption S5 implies

$$\int_{\mathbb{R}^{qM}} v_{\tau_k} \prod_{\ell=m_0}^{M_t} K(v_{\tau_\ell})\, dv_{t;M}$$
$$= \left\{ \int_{\mathbb{R}^q} v_{\tau_k} K(v_{\tau_k})\, dv_{\tau_k} \right\} \prod_{\substack{\ell=m_0 \\ \ell \neq k}}^{M_t} \left\{ \int_{\mathbb{R}^q} K(v_{\tau_\ell})\, dv_{\tau_\ell} \right\} = 0.$$

Thus, the integrated first-order Taylor term equals zero.

By the uniform bound on the second derivatives in Assumption S7, for all sufficiently small $h$, there exists $C < \infty$, independent of $n$ and $i$, such that

$$\sup_{i \le n} \sup_{v_{t;M} \in \mathcal{K}^M} \sup_{0 \le s \le 1} \left\| \nabla^2_{u_{t;M}} m_{n,i,t}(x_{t;M}, d_{t;M} + shv_{t;M}) \right\|_{\text{op}} \le C.$$

Consequently,

$$\sup_{i \le n} \left| \mathbb{E}\{\omega^{\circ}_{i,t;M}(h) Y_{i,t} \mid \mathcal{G}_n\} - \mu_{n,t}(x_{t;M}, d_{t;M}) \right|$$
$$\le Ch^2 \int_0^1 (1-s)\, ds \int_{\mathbb{R}^{qM}} \|v_{t;M}\|^2 \prod_{\ell=m_0}^{M_t} K(v_{\tau_\ell})\, dv_{t;M}$$
$$= Ch^2 \int_0^1 (1-s)\, ds \int_{\mathbb{R}^{qM}} \left( \sum_{\ell=m_0}^{M_t} \|v_{\tau_\ell}\|^2 \right) \cdot \prod_{\ell=m_0}^{M_t} K(v_{\tau_\ell})\, dv_{t;M}$$
$$= \frac{CMh^2}{2} \int_{\mathbb{R}^q} \|v\|^2 K(v)\, dv = O(h^2),$$

where the final integral is finite by Assumption S5. □

*Proof of Proposition 1 .* Take $M = M_t$ and $m_0 = 1$ in Lemma S11. Equation (S2.8) gives

$$\mathbb{E}\left[ Y_{i,t} \prod_{\ell=1}^{M_t} \frac{\mathbf{1}\{X_{i,\tau_\ell} = x_{\tau_\ell}\} K_h(D_{i,\tau_\ell}, d_{\tau_\ell})}{p_{\tau_\ell}(x_{\tau_\ell}, D_{i,\tau_\ell} \mid H_{i,\tau_\ell})} \right]$$
$$= \int_{\mathbb{R}^{qM_t}} m_{n,i,t}(x_{\tau_{1:M_t}}, u_{\tau_{1:M_t}}) \prod_{\ell=1}^{M_t} K_h(u_{\tau_\ell}, d_{\tau_\ell})\, du_{\tau_{1:M_t}},$$

where the right-hand side is the kernel-smoothed potential-outcome mean.

By the compact support and normalization of $K$ in Assumption S5,

$$\left| \int_{\mathbb{R}^{qM_t}} m_{n,i,t}(x_{\tau_{1:M_t}}, d_{\tau_{1:M_t}} + hv_{\tau_{1:M_t}}) \prod_{\ell=1}^{M_t} K(v_{\tau_\ell})\, dv_{\tau_{1:M_t}} - m_{n,i,t}(x_{\tau_{1:M_t}}, d_{\tau_{1:M_t}}) \right|$$
$$\leq \sup_{v_{\tau_{1:M_t}} \in \mathcal{K}^{M_t}} \left| m_{n,i,t}(x_{\tau_{1:M_t}}, d_{\tau_{1:M_t}} + hv_{\tau_{1:M_t}}) - m_{n,i,t}(x_{\tau_{1:M_t}}, d_{\tau_{1:M_t}}) \right| \xrightarrow{h\to 0} 0,$$

where the convergence follows from the continuity condition in Assumption S7. Therefore,

$$\lim_{h\to 0} \mathbb{E}\left[ Y_{i,t} \prod_{\ell=1}^{M_t} \frac{\mathbf{1}\{X_{i,\tau_\ell} = x_{\tau_\ell}\} K_h(D_{i,\tau_\ell}, d_{\tau_\ell})}{p_{\tau_\ell}(x_{\tau_\ell}, D_{i,\tau_\ell} \mid H_{i,\tau_\ell})} \right] = m_{n,i,t}(x_{\tau_{1:M_t}}, d_{\tau_{1:M_t}}) = \mathbb{E}_{\mathcal{G}_n}\{Y_{i,t}(x_{\tau_{1:M_t}}, d_{\tau_{1:M_t}})\}.$$

For discrete spillover exposure, the kernel is $\mathbf{1}\{u = d\}$, so the same calculation gives the ordinary sequential IPW representation without taking a bandwidth limit. □

### S2.3.2 Oracle consistency

*Proof of Theorem 1 .* Fix an outcome time $t$ and target intervention paths $(x_{t;M}, d_{t;M})$. Write $h = h_n$ and abbreviate $\mu_{n,t}(x_{t;M}, d_{t;M})$ as $\mu_{n,t}$. Throughout the proof, all expectations, variances, and convergence statements are conditional on $\mathcal{G}_n$.

*Step 1:* We first derive conditional second-moment bounds for the oracle denominator and outcome-weighted numerator summands. Since

$$\omega^\circ_{i,t;M}(h) = h^{-qM} A^\circ_{i,t;M}(h),$$

the $r = 2$ moment bounds in Assumption S6 imply

$$\begin{aligned} \sup_{i\leq n} \mathbb{E}\{\omega^\circ_{i,t;M}(h)^2 \mid \mathcal{G}_n\} &= h^{-2qM} \sup_{i\leq n} \mathbb{E}\{A^\circ_{i,t;M}(h)^2 \mid \mathcal{G}_n\} = O(h^{-qM}), \\ \sup_{i\leq n} \mathbb{E}\{\omega^\circ_{i,t;M}(h)^2 Y_{i,t}^2 \mid \mathcal{G}_n\} &= h^{-2qM} \sup_{i\leq n} \mathbb{E}\{A^\circ_{i,t;M}(h)^2 Y_{i,t}^2 \mid \mathcal{G}_n\} = O(h^{-qM}). \end{aligned} \tag{S2.9}$$

*Step 2:* We next derive the conditional variance bounds using the finite-range network-dependence condition. We first consider the oracle numerator $N^\circ_{n,t}(h)$. Its conditional variance can be decom-

posed as

$$\operatorname{Var}\{N^\circ_{n,t}(h) \mid \mathcal{G}_n\} = \frac{1}{n^2}\sum_{i=1}^n \operatorname{Var}\{\omega^\circ_{i,t;M}(h)Y_{i,t} \mid \mathcal{G}_n\} + \frac{1}{n^2}\sum_{i\neq j} \operatorname{Cov}\{\omega^\circ_{i,t;M}(h)Y_{i,t}, \omega^\circ_{j,t;M}(h)Y_{j,t} \mid \mathcal{G}_n\}.$$

For the diagonal terms, (S2.9) gives

$$\sum_{i=1}^n \operatorname{Var}\{\omega^\circ_{i,t;M}(h)Y_{i,t} \mid \mathcal{G}_n\} \leq \sum_{i=1}^n \mathbb{E}\{\omega^\circ_{i,t;M}(h)^2 Y_{i,t}^2 \mid \mathcal{G}_n\} \leq n \sup_{i\leq n} \mathbb{E}\{\omega^\circ_{i,t;M}(h)^2 Y_{i,t}^2 \mid \mathcal{G}_n\} = O(nh^{-qM}).$$

According to the conditional finite-range dependence in Assumption S8, we have

$$\operatorname{Cov}\{\omega^\circ_{i,t;M}(h)Y_{i,t}, \omega^\circ_{j,t;M}(h)Y_{j,t} \mid \mathcal{G}_n\} = 0$$

whenever $\operatorname{dist}(i,j) > r_{\text{dep}}$. Moreover, the uniformly bounded cardinality condition in Assumption S8 implies

$$\sum_{i=1}^n |\{j \neq i : \operatorname{dist}(i,j) \leq r_{\text{dep}}\}| = O(n).$$

Thus, only $O(n)$ ordered pairs can contribute nonzero conditional covariances. While by Cauchy–Schwarz inequality, we have

$$\begin{aligned}\left|\operatorname{Cov}\{\omega^\circ_{i,t;M}(h)Y_{i,t}, \omega^\circ_{j,t;M}(h)Y_{j,t} \mid \mathcal{G}_n\}\right| &\leq \left[\operatorname{Var}\{\omega^\circ_{i,t;M}(h)Y_{i,t} \mid \mathcal{G}_n\}\operatorname{Var}\{\omega^\circ_{j,t;M}(h)Y_{j,t} \mid \mathcal{G}_n\}\right]^{1/2} \\ &\leq \left[\mathbb{E}\{\omega^\circ_{i,t;M}(h)^2 Y_{i,t}^2 \mid \mathcal{G}_n\}\mathbb{E}\{\omega^\circ_{j,t;M}(h)^2 Y_{j,t}^2 \mid \mathcal{G}_n\}\right]^{1/2} = O(h^{-qM})\end{aligned}$$

uniformly over $i$ and $j$. Therefore,

$$\sum_{i\neq j} \left|\operatorname{Cov}\{\omega^\circ_{i,t;M}(h)Y_{i,t}, \omega^\circ_{j,t;M}(h)Y_{j,t} \mid \mathcal{G}_n\}\right| = O(nh^{-qM}).$$

Combining the diagonal and covariance bounds yields

$$\operatorname{Var}\{N^\circ_{n,t}(h) \mid \mathcal{G}_n\} = O\{(nh^{qM})^{-1}\}.$$

Replacing $\omega^{\circ}_{i,t;M}(h)Y_{i,t}$ by $\omega^{\circ}_{i,t;M}(h)$, the same argument gives

$$\mathrm{Var}\{D^{\circ}_{n,t}(h) \mid \mathcal{G}_n\} = O\{(nh^{qM})^{-1}\}.$$

*Step 3:* For any $\epsilon > 0$, Chebyshev inequality gives

$$\Pr\left(\left|N^{\circ}_{n,t}(h_n) - \mathbb{E}\{N^{\circ}_{n,t}(h_n) \mid \mathcal{G}_n\}\right| > \epsilon \;\middle|\; \mathcal{G}_n\right) \leq \frac{\mathrm{Var}\{N^{\circ}_{n,t}(h_n) \mid \mathcal{G}_n\}}{\epsilon^2} = O\{(nh_n^{qM})^{-1}\} \to 0,$$

where the limits follow from $nh_n^{qM} \to \infty$, in other words,

$$N^{\circ}_{n,t}(h_n) - \mathbb{E}\{N^{\circ}_{n,t}(h_n) \mid \mathcal{G}_n\} \xrightarrow{p} 0.$$

Similarly, we have

$$D^{\circ}_{n,t}(h_n) - \mathbb{E}\{D^{\circ}_{n,t}(h_n) \mid \mathcal{G}_n\} \xrightarrow{p} 0.$$

Meanwhile, according to Lemma S11, we have

$$\mathbb{E}\{D^{\circ}_{n,t}(h_n) \mid \mathcal{G}_n\} = \frac{1}{n}\sum_{i=1}^{n} \mathbb{E}\{\omega^{\circ}_{i,t;M}(h_n) \mid \mathcal{G}_n\} = 1,$$
$$\mathbb{E}\{N^{\circ}_{n,t}(h_n) \mid \mathcal{G}_n\} = \frac{1}{n}\sum_{i=1}^{n} \mathbb{E}\{\omega^{\circ}_{i,t;M}(h_n)Y_{i,t} \mid \mathcal{G}_n\} = \mu_{n,t} + o(1),$$

hence

$$D^{\circ}_{n,t}(h_n) \xrightarrow{p} 1, \qquad N^{\circ}_{n,t}(h_n) - \mu_{n,t} \xrightarrow{p} 0. \tag{S2.10}$$

Since the oracle Horvitz–Thompson estimator is $N^{\circ}_{n,t}(h_n)$, the second limit directly gives

$$\widehat{\mu}^{\circ,\text{K-IPW},M}_{HT,t}(x_{t;M}, d_{t;M}) - \mu_{n,t} = N^{\circ}_{n,t}(h_n) - \mu_{n,t} \xrightarrow{p} 0.$$

While for the oracle Hájek estimator, Assumption S7 implies that $\mu_{n,t}$ is uniformly bounded over $n$. Therefore, by Slutsky's theorem,

$$\widehat{\mu}^{\circ,\text{K-IPW},M}_{H,t}(x_{t;M}, d_{t;M}) - \mu_{n,t} = \frac{N^{\circ}_{n,t}(h_n)}{D^{\circ}_{n,t}(h_n)} - \mu_{n,t} = \frac{N^{\circ}_{n,t}(h_n) - \mu_{n,t} + \mu_{n,t}\{1 - D^{\circ}_{n,t}(h_n)\}}{D^{\circ}_{n,t}(h_n)} \xrightarrow{p} 0.$$

□

*Proof of Corollary S1.* For each fixed $t \in \mathcal{T}_M$, Theorem 1 gives that for $\zeta \in \{HT, H\}$,

$$\widehat{\mu}^{\circ,\text{K-IPW},M}_{\zeta,t}(x_{t;M}, d_{t;M}) - \mu_{n,t}(x_{t;M}, d_{t;M}) \xrightarrow{p} 0.$$

Since $\mathcal{T}_M$ is fixed and finite, averaging over $t \in \mathcal{T}_M$ preserves convergence in probability and proves the longitudinal result (S2.1). More generally, because the collection is fixed and finite, componentwise convergence implies

$$\widehat{\boldsymbol{\mu}}^{\circ}_{\zeta} - \boldsymbol{\mu}_n \xrightarrow{p} \mathbf{0}.$$

The continuous mapping theorem applied to the fixed linear map $\boldsymbol{v} \mapsto C\boldsymbol{v}$ then gives

$$C\left(\widehat{\boldsymbol{\mu}}^{\circ}_{\zeta} - \boldsymbol{\mu}_n\right) \xrightarrow{p} \mathbf{0}.$$

□

#### S2.3.3 Oracle asymptotic normality

*Proof of Theorem 2 .* Fix an outcome time $t$ and target intervention paths $(x_{t;M}, d_{t;M})$. Write $h = h_n$ and abbreviate $\mu_{n,t}(x_{t;M}, d_{t;M})$ as $\mu_{n,t}$. Throughout the proof, all expectations, variances, and convergence statements are conditional on $\mathcal{G}_n$.

For the Horvitz–Thompson estimator,

$$\begin{aligned}\sqrt{nh^{qM}}\left\{\widehat{\mu}^{\circ,\text{K-IPW},M}_{HT,t}(x_{t;M}, d_{t;M}) - \mu_{n,t}\right\} &= \sqrt{nh^{qM}}\{N^{\circ}_{n,t}(h) - \mu_{n,t}\} \\ &= \frac{1}{\sqrt{nh^{qM}}}\sum_{i=1}^{n}\xi_{HT,n,i,t}(h) + \sqrt{nh^{qM}}\left[\frac{1}{n}\sum_{i=1}^{n}\mathbb{E}\{\omega^{\circ}_{i,t;M}(h)Y_{i,t} \mid \mathcal{G}_n\} - \mu_{n,t}\right],\end{aligned} \tag{S2.11}$$

where $\xi_{HT,n,i,t}(h)$ is defined in (S2.4). By Lemma S11, the expression in square brackets is $O(h^2)$. Therefore, the second term in (S2.11) is $O\{\sqrt{nh^{qM}}h^2\} = o(1)$. Therefore, it remains to establish a central limit theorem for

$$\frac{1}{\sqrt{nh^{qM}}}\sum_{i=1}^{n}\xi_{HT,n,i,t}(h).$$

Let

$$s^2_{HT,n,t} = \mathrm{Var}\left\{ \sum_{i=1}^{n} \xi_{HT,n,i,t}(h) \;\middle|\; \mathcal{G}_n \right\}.$$

Assumption S9 gives

$$\frac{s^2_{HT,n,t}}{nh^{qM}} \longrightarrow \sigma^2_{HT,t} \in (0,\infty), \tag{S2.12}$$

Meanwhile, by the definition of $\xi_{HT,n,i,t}(h)$,

$$\begin{aligned}\mathbb{E}\{|\xi_{HT,n,i,t}(h)|^3 \mid \mathcal{G}_n\} &\leq 4\mathbb{E}\{|A^{\circ}_{i,t;M}(h)Y_{i,t}|^3 \mid \mathcal{G}_n\} + 4\left|\mathbb{E}\{A^{\circ}_{i,t;M}(h)Y_{i,t} \mid \mathcal{G}_n\}\right|^3 \\ &\leq 8\mathbb{E}\{|A^{\circ}_{i,t;M}(h)Y_{i,t}|^3 \mid \mathcal{G}_n\},\end{aligned}$$

thus, the $r = 3$ moment bound in Assumption S6 implies

$$\max_{i\leq n} \mathbb{E}\{|\xi_{HT,n,i,t}(h)|^3 \mid \mathcal{G}_n\} = O(h^{qM}).$$

For each $n$, define a graph

$$\mathcal{D}_n = \big(\{1,\ldots,n\}, E_n\big), \qquad E_n = \{\{i,j\} : 1 \leq i < j \leq n,\ \mathrm{dist}(i,j) \leq r_{\mathrm{dep}}\}.$$

If two disjoint index sets $I$ and $J$ have no edge between them in $\mathcal{D}_n$, then

$$\mathrm{dist}(I,J) > r_{\mathrm{dep}}.$$

Assumption S8 therefore implies

$$\{\xi_{HT,n,i,t}(h) : i \in I\} \perp\!\!\!\perp \{\xi_{HT,n,j,t}(h) : j \in J\} \;\Big|\; \mathcal{G}_n.$$

Thus, conditionally on $\mathcal{G}_n$, $\mathcal{D}_n$ is a dependency graph for $\{\xi_{HT,n,i,t}(h), 1 \leq i \leq n\}$. Its maximum degree is

$$D_n = \max_{i\leq n} |\{j \neq i : \mathrm{dist}(i,j) \leq r_{\mathrm{dep}}\}|,$$

and Assumption S8 ensures that $\sup_n D_n < \infty$.

For almost every realization of $\mathcal{G}_n$, apply Theorem 2.7 of Chen and Shao (2004) with $p = 3$ to the normalized summands

$$\left\{ \frac{\xi_{HT,n,i,t}(h)}{s_{HT,n,t}} : i = 1, \ldots, n \right\}.$$

The resulting Berry–Esseen bound satisfies

$$\sup_{z \in \mathbb{R}} \left| \Pr\left\{ \frac{\sum_{i=1}^n \xi_{HT,n,i,t}(h)}{s_{HT,n,t}} \leq z \,\middle|\, \mathcal{G}_n \right\} - \Phi(z) \right|$$
$$\leq C(D_n + 1)^{10} \frac{n \max_{i \leq n} \mathbb{E}\{|\xi_{HT,n,i,t}(h)|^3 \mid \mathcal{G}_n\}}{s_{HT,n,t}^3} = O\{(nh^{qM})^{-1/2}\} \to 0,$$

where $C$ is a universal constant and $\Phi(\cdot)$ is the cumulative distribution function of standard normal random variable. Hence,

$$\frac{\sum_{i=1}^n \xi_{HT,n,i,t}(h)}{s_{HT,n,t}} \xrightarrow{d} \mathcal{N}(0, 1).$$

Combining (S2.12) and Slutsky's theorem gives

$$\frac{1}{\sqrt{nh^{qM}}} \sum_{i=1}^n \xi_{HT,n,i,t}(h) \xrightarrow{d} \mathcal{N}(0, \sigma_{HT,t}^2).$$

Together with the fact that the second term in (S2.11) is $o(1)$, this proves the result for the oracle Horvitz–Thompson estimator.

For the Hájek estimator,

$$\begin{aligned} \sqrt{nh^{qM}} \left\{ \widehat{\mu}_{H,t}^{\circ,\text{K-IPW},M}(x_{t;M}, d_{t;M}) - \mu_{n,t} \right\} &= \sqrt{nh^{qM}} \left\{ \frac{N_{n,t}^\circ(h)}{D_{n,t}^\circ(h)} - \mu_{n,t} \right\} \\ &= \frac{\sqrt{nh^{qM}} \{N_{n,t}^\circ(h) - \mu_{n,t} D_{n,t}^\circ(h)\}}{D_{n,t}^\circ(h)}. \end{aligned} \quad \text{(S2.13)}$$

The proof of Theorem 1 gives $D_{n,t}^\circ(h) \xrightarrow{p} 1$. While the centered numerator in (S2.13) admits the decomposition

$$\sqrt{nh^{qM}} \{N_{n,t}^\circ(h) - \mu_{n,t} D_{n,t}^\circ(h)\} = \frac{1}{\sqrt{nh^{qM}}} \sum_{i=1}^n \xi_{H,n,i,t}(h) + \frac{\sqrt{nh^{qM}}}{n} \sum_{i=1}^n \mathbb{E}\left[ \omega_{i,t;M}^\circ(h)\{Y_{i,t} - \mu_{n,t}\} \,\middle|\, \mathcal{G}_n \right],$$

where $\xi_{H,n,i,t}(h)$ is defined in (S2.4). By (S2.7),

$$\frac{1}{n}\sum_{i=1}^{n}\mathbb{E}\left[\omega^{\circ}_{i,t;M}(h)\{Y_{i,t}-\mu_{n,t}\}\;\middle|\;\mathcal{G}_n\right]=\frac{1}{n}\sum_{i=1}^{n}\left[\mathbb{E}\{\omega^{\circ}_{i,t;M}(h)Y_{i,t}\mid\mathcal{G}_n\}-\mu_{n,t}\right].$$

Lemma S11 therefore implies that this quantity is $O(h^2)$. Hence, the second term contribution to the preceding decomposition is

$$\frac{\sqrt{nh^{qM}}}{n}\sum_{i=1}^{n}\mathbb{E}\left[\omega^{\circ}_{i,t;M}(h)\{Y_{i,t}-\mu_{n,t}\}\;\middle|\;\mathcal{G}_n\right]=O\{\sqrt{nh^{qM}}h^2\}=o(1). \tag{S2.14}$$

Next, let

$$s^2_{H,n,t}=\mathrm{Var}\left\{\sum_{i=1}^{n}\xi_{H,n,i,t}(h)\;\middle|\;\mathcal{G}_n\right\}.$$

Assumption S9 gives

$$\frac{s^2_{H,n,t}}{nh^{qM}}\longrightarrow\sigma^2_{H,t}\in(0,\infty).$$

The uniform boundedness condition in Assumption S7 gives $\sup_n|\mu_{n,t}|<\infty$. Consequently,

$$\begin{aligned}\mathbb{E}\{|\xi_{H,n,i,t}(h)|^3\mid\mathcal{G}_n\}&\le 8\mathbb{E}\left[\left|A^{\circ}_{i,t;M}(h)\{Y_{i,t}-\mu_{n,t}\}\right|^3\;\middle|\;\mathcal{G}_n\right]\\&\le 32\mathbb{E}\{|A^{\circ}_{i,t;M}(h)Y_{i,t}|^3\mid\mathcal{G}_n\}+32|\mu_{n,t}|^3\mathbb{E}\{|A^{\circ}_{i,t;M}(h)|^3\mid\mathcal{G}_n\}.\end{aligned}$$

The $r=3$ bounds in Assumption S6 thus imply

$$\max_{i\le n}\mathbb{E}\{|\xi_{H,n,i,t}(h)|^3\mid\mathcal{G}_n\}=O(h^{qM}).$$

The graph $\mathcal{D}_n$ constructed above is also a conditional dependency graph for $\{\xi_{H,n,i,t}(h):i=1,\ldots,n\}$, because each $\xi_{H,n,i,t}(h)$ is a measurable transformation of $(A^{\circ}_{i,t;M}(h),A^{\circ}_{i,t;M}(h)Y_{i,t})$ after conditioning on $\mathcal{G}_n$. Applying Theorem 2.7 of Chen and Shao (2004) with $p=3$ gives

$$\begin{aligned}&\sup_{z\in\mathbb{R}}\left|\Pr\left\{\frac{\sum_{i=1}^{n}\xi_{H,n,i,t}(h)}{s_{H,n,t}}\le z\;\middle|\;\mathcal{G}_n\right\}-\Phi(z)\right|\\&\quad\le C(D_n+1)^{10}\frac{n\max_{i\le n}\mathbb{E}\{|\xi_{H,n,i,t}(h)|^3\mid\mathcal{G}_n\}}{s^3_{H,n,t}}=O\{(nh^{qM})^{-1/2}\}\to 0.\end{aligned}$$

It follows that

$$\frac{1}{\sqrt{nh^{qM}}}\sum_{i=1}^{n}\xi_{H,n,i,t}(h) \xrightarrow{d} \mathcal{N}(0,\sigma_{H,t}^2).$$

Together with the $o(1)$ bias term in (S2.14),

$$\sqrt{nh^{qM}}\{N_{n,t}^{\circ}(h) - \mu_{n,t}D_{n,t}^{\circ}(h)\} \xrightarrow{d} \mathcal{N}(0,\sigma_{H,t}^2).$$

Finally, combining this result with $D_{n,t}^{\circ}(h) \xrightarrow{p} 1$ in (S2.10), Slutsky's theorem gives

$$\sqrt{nh^{qM}}\left\{\widehat{\mu}_{H,t}^{\circ,\text{K-IPW},M}(x_{t;M},d_{t;M}) - \mu_{n,t}\right\} \xrightarrow{d} \mathcal{N}(0,\sigma_{H,t}^2).$$

□

*Proof of Corollary 1 .* We first consider continuous spillover exposure. Let

$$\boldsymbol{\mu}_n = (\mu_{n,1},\ldots,\mu_{n,J})^{\top}$$

collect the fixed finite set of component means entering the longitudinal estimand and causal-effect contrasts, and let $\widehat{\boldsymbol{\mu}}_{\zeta}^{\circ}$ collect their corresponding oracle estimators, where $\zeta \in \{HT, H\}$. Recall that $\boldsymbol{\xi}_{\zeta,n,i}(h_n)$ denotes the $J$-vector of centered summands defined in Assumption S9.

Fix any nonzero vector $a \in \mathbb{R}^J$, and let

$$s_{\zeta,n}^2(a) = \operatorname{Var}\left\{\sum_{i=1}^{n} a^{\top}\boldsymbol{\xi}_{\zeta,n,i}(h_n) \;\middle|\; \mathcal{G}_n\right\}.$$

The joint limiting-covariance condition in Assumption S9 gives

$$\frac{s_{\zeta,n}^2(a)}{nh_n^{qM}} \longrightarrow a^{\top}\boldsymbol{\Sigma}_{\zeta}a > 0.$$

Because $J$ and $a$ are fixed, the componentwise third-moment bounds established in the proof of Theorem 2 imply

$$\max_{i\le n}\mathbb{E}\left\{\left|a^{\top}\boldsymbol{\xi}_{\zeta,n,i}(h_n)\right|^3 \;\middle|\; \mathcal{G}_n\right\} \le J^2\sum_{j=1}^{J}|a_j|^3\max_{i\le n}\mathbb{E}\left\{|\xi_{\zeta,n,i,j}(h_n)|^3 \;\middle|\; \mathcal{G}_n\right\} = O(h_n^{qM}),$$

where $\xi_{\zeta,n,i,j}(h_n)$ is the $j$th component of $\boldsymbol{\xi}_{\zeta,n,i}(h_n)$.

Because the scalar array

$$\left\{a^\top \boldsymbol{\xi}_{\zeta,n,i}(h_n) : i = 1, \ldots, n\right\}$$

has the same conditional dependency graph $\mathcal{D}_n$ constructed in the proof of Theorem 2 . Applying Theorem 2.7 of Chen and Shao (2004) with $p = 3$, its Berry–Esseen bound is of order

$$(D_n + 1)^{10} \frac{n \max_{i \le n} \mathbb{E}\left\{|a^\top \boldsymbol{\xi}_{\zeta,n,i}(h_n)|^3 \mid \mathcal{G}_n\right\}}{s^3_{\zeta,n}(a)} = O\{(nh_n^{qM})^{-1/2}\} \to 0.$$

Consequently,

$$\frac{1}{\sqrt{nh_n^{qM}}} \sum_{i=1}^{n} a^\top \boldsymbol{\xi}_{\zeta,n,i}(h_n) \xrightarrow{d} \mathcal{N}\left(0, a^\top \boldsymbol{\Sigma}_\zeta a\right).$$

Since this holds for every fixed $a \in \mathbb{R}^J$, the Cramér–Wold device gives

$$\frac{1}{\sqrt{nh_n^{qM}}} \sum_{i=1}^{n} \boldsymbol{\xi}_{\zeta,n,i}(h_n) \xrightarrow{d} \mathcal{N}(\mathbf{0}, \boldsymbol{\Sigma}_\zeta).$$

For the Horvitz–Thompson estimators, the componentwise decompositions in the proof of Theorem 2 , together with the fixed dimension $J$, give

$$\sqrt{nh_n^{qM}}\,(\widehat{\boldsymbol{\mu}}^\circ_{HT} - \boldsymbol{\mu}_n) = \frac{1}{\sqrt{nh_n^{qM}}} \sum_{i=1}^{n} \boldsymbol{\xi}_{HT,n,i}(h_n) + o_p(1).$$

It follows that

$$\sqrt{nh_n^{qM}}\,(\widehat{\boldsymbol{\mu}}^\circ_{HT} - \boldsymbol{\mu}_n) \xrightarrow{d} \mathcal{N}(\mathbf{0}, \boldsymbol{\Sigma}_{HT}).$$

For the Hájek estimators, let $D^\circ_{n,j}(h_n)$ denote the oracle weighted denominator for the $j$th component mean. The proof of Theorem 1 gives

$$\left(D^\circ_{n,1}(h_n), \ldots, D^\circ_{n,J}(h_n)\right)^\top \xrightarrow{p} (1, \ldots, 1)^\top,$$

where joint convergence follows because $J$ is fixed. The componentwise decompositions in the

proof of Theorem 2 give

$$\sqrt{nh_n^{qM}}\,(\widehat{\boldsymbol{\mu}}_H^\circ - \boldsymbol{\mu}_n) = \operatorname{diag}\left\{\frac{1}{D_{n,1}^\circ(h_n)}, \ldots, \frac{1}{D_{n,J}^\circ(h_n)}\right\}\left\{\frac{1}{\sqrt{nh_n^{qM}}}\sum_{i=1}^n \boldsymbol{\xi}_{H,n,i}(h_n) + o_p(1)\right\}.$$

Therefore, Slutsky's theorem yields

$$\sqrt{nh_n^{qM}}\,(\widehat{\boldsymbol{\mu}}_H^\circ - \boldsymbol{\mu}_n) \xrightarrow{d} \mathcal{N}(\mathbf{0}, \boldsymbol{\Sigma}_H).$$

Thus, combining the two results above, for any fixed contrast matrix $C$, the continuous mapping theorem gives

$$\sqrt{nh_n^{qM}}C\left(\widehat{\boldsymbol{\mu}}_\zeta^\circ - \boldsymbol{\mu}_n\right) \xrightarrow{d} \mathcal{N}\left(\mathbf{0}, C\boldsymbol{\Sigma}_\zeta C^\top\right).$$

The longitudinal mean corresponds to the fixed averaging map, while the direct, spillover, and total effects correspond to the appropriate fixed contrast matrices.

For discrete spillover exposure, Assumption S10 gives the corresponding joint limiting-covariance and third-moment conditions, with

$$\max_{i\le n}\mathbb{E}\left\{|a^\top \boldsymbol{\xi}_{\zeta,n,i}^d|^3 \mid \mathcal{G}_n\right\} = O(1).$$

The same dependency-graph Berry–Esseen bound is therefore $O(n^{-1/2})$. Repeating the preceding arguments gives the asserted joint asymptotic distributions with the $\sqrt{n}$ normalization. □

#### S2.3.4 Discrete spillover exposure

*Proof of Proposition S2.* Fix an outcome time $t$ and target intervention paths $(x_{t;M}, d_{t;M})$, and abbreviate $\mu_{n,t}(x_{t;M}, d_{t;M})$ as $\mu_{n,t}$. Throughout the proof, all expectations, variances, and convergence statements are conditional on $\mathcal{G}_n$. We follow the arguments used for the continuous-exposure results exactly, highlighting the changes caused by replacing kernel localization with exact matching on the discrete exposure path.

For discrete exposure, at decision time $\tau_\ell$, the one-step oracle weight becomes

$$W^\circ_{i,\tau_\ell} = \frac{\mathbf{1}\{X_{i,\tau_\ell} = x_{\tau_\ell}\}\mathbf{1}\{D_{i,\tau_\ell} = d_{\tau_\ell}\}}{p_{\tau_\ell}(x_{\tau_\ell}, d_{\tau_\ell} \mid H_{i,\tau_\ell})}.$$

Summing over the conditional support of $(X_{i,\tau_\ell}, D_{i,\tau_\ell})$ gives

$$\mathbb{E}\{W^\circ_{i,\tau_\ell} \mid H_{i,\tau_\ell}, \mathcal{G}_n\} = \sum_{x\in\{0,1\}} \sum_{u} \frac{\mathbf{1}\{x = x_{\tau_\ell}\}\mathbf{1}\{u = d_{\tau_\ell}\}}{p_{\tau_\ell}(x_{\tau_\ell}, d_{\tau_\ell} \mid H_{i,\tau_\ell})} p_{\tau_\ell}(x, u \mid H_{i,\tau_\ell}) = 1.$$

Repeating the backward-conditioning argument in Lemma S11 therefore gives

$$\mathbb{E}\{\omega^\circ_{i,t;M} \mid \mathcal{G}_n\} = 1.$$

For the outcome-weighted identity, the product weight is zero unless $X_{i,\tau_{m_0:M_t}} = x_{t;M}$ and $D_{i,\tau_{m_0:M_t}} = d_{t;M}$. On this event, the consistency condition in Assumption 1 and the partial-window potential-outcome definition in the main text give $Y_{i,t} = Y_{i,t}(x_{t;M}, d_{t;M})$. Consequently,

$$\mathbb{E}\{\omega^\circ_{i,t;M} Y_{i,t} \mid \mathcal{G}_n\} = \mathbb{E}\left[ Y_{i,t}(x_{t;M}, d_{t;M}) \prod_{\ell=m_0}^{M_t} W^\circ_{i,\tau_\ell} \;\middle|\; \mathcal{G}_n \right].$$

Notice that, at the last decision time, direct conditioning gives

$$\begin{aligned}
&\mathbb{E}\left[ Y_{i,t}(x_{t;M}, d_{t;M}) W^\circ_{i,\tau_{M_t}} \;\middle|\; H_{i,\tau_{M_t}}, \mathcal{G}_n \right] \\
&\quad = \sum_{x\in\{0,1\}} \sum_{u} \mathbb{E}\left[ Y_{i,t}(x_{t;M}, d_{t;M}) \;\middle|\; X_{i,\tau_{M_t}} = x, D_{i,\tau_{M_t}} = u, H_{i,\tau_{M_t}}, \mathcal{G}_n \right] \\
&\qquad \times \frac{\mathbf{1}\{x = x_{\tau_{M_t}}\}\mathbf{1}\{u = d_{\tau_{M_t}}\}}{p_{\tau_{M_t}}(x_{\tau_{M_t}}, d_{\tau_{M_t}} \mid H_{i,\tau_{M_t}})} p_{\tau_{M_t}}(x, u \mid H_{i,\tau_{M_t}}) \\
&\quad = \mathbb{E}\left[ Y_{i,t}(x_{t;M}, d_{t;M}) \;\middle|\; X_{i,\tau_{M_t}} = x_{\tau_{M_t}}, D_{i,\tau_{M_t}} = d_{\tau_{M_t}}, H_{i,\tau_{M_t}}, \mathcal{G}_n \right] \\
&\quad = \mathbb{E}\left[ Y_{i,t}(x_{t;M}, d_{t;M}) \;\middle|\; H_{i,\tau_{M_t}}, \mathcal{G}_n \right],
\end{aligned}$$

where the last equality follows from Assumption 2.

Because the preceding one-step weights are $H_{i,\tau_{M_t}}$-measurable, the tower property removes $W^\circ_{i,\tau_{M_t}}$. Repeating the same argument backward for $\tau_{M_t-1}, \dots, \tau_{m_0}$ yields

$$\mathbb{E}\{\omega^\circ_{i,t;M} Y_{i,t} \mid \mathcal{G}_n\} = \mathbb{E}\left[ Y_{i,t}(x_{t;M}, d_{t;M}) \;\middle|\; \mathcal{G}_n \right] = m_{n,i,t}(x_{t;M}, d_{t;M}) = \mu_{n,t},$$

where the final equality follows from the common-target condition in Assumption S7. So unlike the continuous case, the indicator kernel selects the target exposure path exactly, resulting the fact that there is no localization bias or bandwidth condition.

For variance terms, notice that, the $r = 2$ moment condition in Assumption S10 gives

$$\sup_{i \le n} \mathbb{E}\{|\omega^{\circ}_{i,t;M}|^2 \mid \mathcal{G}_n\} = O(1), \qquad \sup_{i \le n} \mathbb{E}\{|\omega^{\circ}_{i,t;M} Y_{i,t}|^2 \mid \mathcal{G}_n\} = O(1).$$

Consequently,

$$\sum_{i=1}^{n} \mathrm{Var}\{\omega^{\circ}_{i,t;M} Y_{i,t} \mid \mathcal{G}_n\} \le n \sup_{i \le n} \mathbb{E}\{|\omega^{\circ}_{i,t;M} Y_{i,t}|^2 \mid \mathcal{G}_n\} = O(n).$$

Under Assumption S8, only $O(n)$ ordered pairs can have nonzero conditional covariance. For each such pair, conditional Cauchy–Schwarz and the preceding second-moment bound give

$$\left|\mathrm{Cov}\{\omega^{\circ}_{i,t;M} Y_{i,t}, \omega^{\circ}_{j,t;M} Y_{j,t} \mid \mathcal{G}_n\}\right| = O(1).$$

Hence,

$$\sum_{i \ne j} \left|\mathrm{Cov}\{\omega^{\circ}_{i,t;M} Y_{i,t}, \omega^{\circ}_{j,t;M} Y_{j,t} \mid \mathcal{G}_n\}\right| = O(n),$$

and therefore

$$\mathrm{Var}\{N^{\circ}_{n,t} \mid \mathcal{G}_n\} = O(n^{-1}).$$

Replacing $\omega^{\circ}_{i,t;M} Y_{i,t}$ by $\omega^{\circ}_{i,t;M}$ gives, by the same argument,

$$\mathrm{Var}\{D^{\circ}_{n,t} \mid \mathcal{G}_n\} = O(n^{-1}).$$

Thus, Chebyshev inequalities leads to

$$D^{\circ}_{n,t} \xrightarrow{p} 1, \qquad N^{\circ}_{n,t} - \mu_{n,t} \xrightarrow{p} 0.$$

The first two consistency conclusions of the proposition then follow by the same Horvitz–Thompson and Hájek arguments used in the proof of Theorem 1 .

For asymptotic normality, the $r = 3$ moment condition in Assumption S10, together with the centering inequality used in the proof of Theorem 2 , gives

$$\max_{i \le n} \mathbb{E}\{|\xi^d_{HT,n,i,t}|^3 \mid \mathcal{G}_n\} = O(1).$$

Because $\mu_{n,t}$ is uniformly bounded, the same argument gives

$$\max_{i \le n} \mathbb{E}\{|\xi^d_{H,n,i,t}|^3 \mid \mathcal{G}_n\} = O(1).$$

For $\zeta \in \{HT, H\}$, let

$$(s^d_{\zeta,n,t})^2 = \mathrm{Var}\left\{ \sum_{i=1}^n \xi^d_{\zeta,n,i,t} \;\middle|\; \mathcal{G}_n \right\}.$$

The discrete limiting-covariance condition in Assumption S10 gives

$$\frac{(s^d_{\zeta,n,t})^2}{n} \to \sigma^2_{\zeta,t} \in (0, \infty).$$

The graph $\mathcal{D}_n$ constructed in the proof of Theorem 2 is also a conditional dependency graph for the discrete centered summands and has uniformly bounded maximum degree. Theorem 2.7 of Chen and Shao (2004), applied with $p = 3$, therefore gives the Berry–Esseen bound

$$C(D_n + 1)^{10} \frac{n \max_{i \le n} \mathbb{E}\{|\xi^d_{\zeta,n,i,t}|^3 \mid \mathcal{G}_n\}}{(s^d_{\zeta,n,t})^3} = O(n^{-1/2}) \to 0.$$

It follows that

$$\frac{1}{\sqrt{n}} \sum_{i=1}^n \xi^d_{\zeta,n,i,t} \xrightarrow{d} \mathcal{N}(0, \sigma^2_{\zeta,t}).$$

Because the weighting identities are exact,

$$\sqrt{n}\{N^\circ_{n,t} - \mu_{n,t}\} = \frac{1}{\sqrt{n}} \sum_{i=1}^n \xi^d_{HT,n,i,t},$$

and

$$\sqrt{n}\{N^\circ_{n,t} - \mu_{n,t} D^\circ_{n,t}\} = \frac{1}{\sqrt{n}} \sum_{i=1}^n \xi^d_{H,n,i,t}.$$

The pointwise Gaussian limits now follow directly for the Horvitz–Thompson estimator and, using $D^{\circ}_{n,t} \xrightarrow{p} 1$, by Slutsky's theorem for the Hájek estimator.

For any fixed finite collection of component means, the joint result follows from the same Cramér–Wold argument used in the proof of Corollary 1 . Applying the corresponding fixed averaging and contrast maps then gives the stated longitudinal-mean and causal-effect conclusions.

□

#### S2.3.5 Transfer from oracle to feasible estimators

*Proof of Proposition* S3. We first consider one component estimator for a fixed outcome time $t$ and target paths $(x_{t;M}, d_{t;M})$. For $\ell = m_0, \dots, M_t$, abbreviate

$$p_{i,\ell} = p_{\tau_\ell}(x_{\tau_\ell}, D_{i,\tau_\ell} \mid H_{i,\tau_\ell}), \qquad \widehat{p}_{i,\ell} = \widehat{p}_{\tau_\ell}(x_{\tau_\ell}, D_{i,\tau_\ell} \mid H_{i,\tau_\ell}).$$

By the definition of $\Delta_n$,

$$\widehat{p}_{i,\ell} = p_{i,\ell}(1 + \varepsilon_{i,\ell}), \qquad |\varepsilon_{i,\ell}| \le \Delta_n.$$

Because $\Delta_n = o_p(1)$, so with probability tending to one, we have,

$$\big|(1 + \varepsilon_{i,\ell})^{-1}\big| \le 2, \qquad \big|(1 + \varepsilon_{i,\ell})^{-1} - 1\big| = \frac{|\varepsilon_{i,\ell}|}{|1 + \varepsilon_{i,\ell}|} \le 2\Delta_n.$$

Therefore,

$$\begin{aligned} \left|\prod_{\ell=m_0}^{M_t} (1 + \varepsilon_{i,\ell})^{-1} - 1\right| &= \left|\sum_{k=m_0}^{M_t} \big\{(1 + \varepsilon_{i,k})^{-1} - 1\big\} \prod_{\ell=m_0}^{k-1} (1 + \varepsilon_{i,\ell})^{-1}\right| \\ &\le \sum_{k=m_0}^{M_t} \big|(1 + \varepsilon_{i,k})^{-1} - 1\big| \prod_{\ell=m_0}^{k-1} \big|(1 + \varepsilon_{i,\ell})^{-1}\big| \le \sum_{k=m_0}^{M_t} 2\Delta_n\, 2^{k-m_0} \le C_M \Delta_n \end{aligned}$$

where $C_M < \infty$ because $M$ is fixed. Since

$$\widehat{\omega}_{i,t;M}(h_n) = \omega^{\circ}_{i,t;M}(h_n) \prod_{\ell=m_0}^{M_t} (1 + \varepsilon_{i,\ell})^{-1},$$

it follows that

$$\left|\widehat{\omega}_{i,t;M}(h_n) - \omega^{\circ}_{i,t;M}(h_n)\right| \leq C_M \Delta_n \omega^{\circ}_{i,t;M}(h_n).$$

Let $\widehat{N}_{n,t}(h_n)$ and $\widehat{D}_{n,t}(h_n)$ denote the feasible numerator and denominator. For the Horvitz–Thompson estimator,

$$\left|\widehat{\mu}^{\text{K-IPW},M}_{HT,t} - \widehat{\mu}^{\circ,\text{K-IPW},M}_{HT,t}\right| = \left|\widehat{N}_{n,t}(h_n) - N^{\circ}_{n,t}(h_n)\right| \leq C_M \Delta_n \frac{1}{n}\sum_{i=1}^{n} \omega^{\circ}_{i,t;M}(h_n)|Y_{i,t}|.$$

For continuous spillover exposure, the $r = 1$ bound in Assumption S6 gives

$$\sup_{i\leq n} \mathbb{E}\left[\omega^{\circ}_{i,t;M}(h_n)|Y_{i,t}| \;\middle|\; \mathcal{G}_n\right] = h_n^{-qM} \sup_{i\leq n} \mathbb{E}\left[|A^{\circ}_{i,t;M}(h_n)Y_{i,t}| \;\middle|\; \mathcal{G}_n\right] = O(1).$$

Consequently,

$$\mathbb{E}\left[\frac{1}{n}\sum_{i=1}^{n} \omega^{\circ}_{i,t;M}(h_n)|Y_{i,t}| \;\middle|\; \mathcal{G}_n\right] = O(1),$$

and Chebyshev's inequality implies

$$\frac{1}{n}\sum_{i=1}^{n} \omega^{\circ}_{i,t;M}(h_n)|Y_{i,t}| = O_p(1).$$

For discrete spillover exposure, the same conclusion follows from Assumption S10 because

$$\mathbb{E}\{\omega^{\circ}_{i,t;M}|Y_{i,t}| \mid \mathcal{G}_n\} \leq \left[\mathbb{E}\{(\omega^{\circ}_{i,t;M})^2 Y_{i,t}^2 \mid \mathcal{G}_n\}\right]^{1/2} = O(1).$$

Therefore, in either setting,

$$\widehat{\mu}^{\text{K-IPW},M}_{HT,t} - \widehat{\mu}^{\circ,\text{K-IPW},M}_{HT,t} = \widehat{N}_{n,t}(h_n) - N^{\circ}_{n,t}(h_n) = O_p(\Delta_n).$$

By (S2.10) and the uniform boundedness of $\mu_{n,t}$ in Assumption S7, $N^{\circ}_{n,t}(h_n) = \mu_{n,t} + o_p(1) = O_p(1)$. Similarly, again by utilizing (S2.10), we have

$$\left|\widehat{D}_{n,t}(h_n) - D^{\circ}_{n,t}(h_n)\right| \leq C_M \Delta_n D^{\circ}_{n,t}(h_n) = O_p(\Delta_n),$$

Slutsky's theorem therefore gives $\widehat{D}_{n,t}(h_n) = D^{\circ}_{n,t}(h_n) + O_p(\Delta_n) \xrightarrow{p} 1$. Hence,

$$\begin{aligned}\left|\widehat{\mu}^{\text{K-IPW},M}_{H,t} - \widehat{\mu}^{\circ,\text{K-IPW},M}_{H,t}\right| &= \left|\frac{\widehat{N}_{n,t}(h_n)}{\widehat{D}_{n,t}(h_n)} - \frac{N^{\circ}_{n,t}(h_n)}{D^{\circ}_{n,t}(h_n)}\right| \\ &\le \frac{|\widehat{N}_{n,t}(h_n) - N^{\circ}_{n,t}(h_n)|}{|\widehat{D}_{n,t}(h_n)|} + \frac{|N^{\circ}_{n,t}(h_n)||\widehat{D}_{n,t}(h_n) - D^{\circ}_{n,t}(h_n)|}{|\widehat{D}_{n,t}(h_n)D^{\circ}_{n,t}(h_n)|} = O_p(\Delta_n).\end{aligned}$$

This proves (S2.2) for the Hájek estimator.

Because the collection of component estimators is fixed and finite, the preceding $O_p(\Delta_n)$ bound holds uniformly over its components. Hence, when $\Delta_n = o_p(1)$, feasible consistency follows immediately from oracle consistency. For continuous spillover exposure,

$$\sqrt{nh_n^{qM}}\left(\widehat{\mu}^{\text{K-IPW},M}_{\zeta,t} - \widehat{\mu}^{\circ,\text{K-IPW},M}_{\zeta,t}\right) = O_p\{\sqrt{nh_n^{qM}}\Delta_n\} = o_p(1)$$

when $\Delta_n = o_p\{(nh_n^{qM})^{-1/2}\}$. For discrete spillover exposure, the corresponding normalized difference is $o_p(1)$ when $\Delta_n = o_p(n^{-1/2})$. Slutsky's theorem then transfers the oracle pointwise, longitudinal-mean, and causal-effect asymptotic distributions to their feasible counterparts. □

#### S2.3.6 Linear-program representation of the sensitivity endpoints

*Proof of Proposition S4.* We prove the representation for the lower endpoint; the upper-endpoint argument is identical. For any feasible $\boldsymbol{r} = (r_1, \ldots, r_n) \in [\Gamma^{-M}, \Gamma^{M}]^n$, the assumptions imply

$$\sum_{i=1}^{n} \widehat{\omega}_{i,t;M}(h) r_i \ge \Gamma^{-M} \sum_{i=1}^{n} \widehat{\omega}_{i,t;M}(h) > 0.$$

Define

$$\kappa = \Big[\sum_{i=1}^{n} \widehat{\omega}_{i,t;M}(h) r_i\Big]^{-1}, \qquad q_i = \kappa r_i, \quad i = 1, \ldots, n.$$

Then

$$\sum_{i=1}^{n} \widehat{\omega}_{i,t;M}(h) q_i = \kappa \sum_{i=1}^{n} \widehat{\omega}_{i,t;M}(h) r_i = 1,$$

and the bounds on $r_i$ imply

$$\Gamma^{-M}\kappa \le q_i \le \Gamma^{M}\kappa.$$

Thus, $(\boldsymbol{q}, \kappa)$ is feasible for the linear program. Moreover,

$$\sum_{i=1}^{n} \widehat{\omega}_{i,t;M}(h) q_i Y_{i,t} = \kappa \sum_{i=1}^{n} \widehat{\omega}_{i,t;M}(h) r_i Y_{i,t} = \frac{\sum_{i=1}^{n} \widehat{\omega}_{i,t;M}(h) r_i Y_{i,t}}{\sum_{i=1}^{n} \widehat{\omega}_{i,t;M}(h) r_i},$$

so the transformation preserves the objective value.

Conversely, let $(\boldsymbol{q}, \kappa)$ be feasible for the linear program. If $\kappa = 0$, the constraints on $q_i$ imply $q_i = 0$ for every $i$, contradicting

$$\sum_{i=1}^{n} \widehat{\omega}_{i,t;M}(h) q_i = 1.$$

Hence, $\kappa > 0$. Define

$$r_i = \frac{q_i}{\kappa}, \quad i = 1, \dots, n.$$

The constraints on $q_i$ imply

$$\Gamma^{-M} \le r_i \le \Gamma^{M},$$

so $\boldsymbol{r}$ is feasible for the original endpoint problem. Furthermore,

$$\sum_{i=1}^{n} \widehat{\omega}_{i,t;M}(h) r_i = \frac{1}{\kappa} \sum_{i=1}^{n} \widehat{\omega}_{i,t;M}(h) q_i = \frac{1}{\kappa},$$

and therefore

$$\frac{\sum_{i=1}^{n} \widehat{\omega}_{i,t;M}(h) r_i Y_{i,t}}{\sum_{i=1}^{n} \widehat{\omega}_{i,t;M}(h) r_i} = \sum_{i=1}^{n} \widehat{\omega}_{i,t;M}(h) q_i Y_{i,t}.$$

The transformation maps feasible solutions in both directions while preserving their objective values. Hence, the two minimization problems have the same optimal value, and the maximization case follows identically. □

# S3 Additional Numerical Experiments and COVID-19 Policy Analyses

This section provides implementation details and additional diagnostic analyses supporting the numerical experiments and county-level COVID-19 policy application. For the numerical experiments, Section S3.1 specifies the detailed data-generating process and estimation procedures, examines propensity overlap and the effective sample sizes of the target paths, assesses the sensitivity of estimation accuracy and interval coverage to bandwidth choice, reports the hidden-bias sensitivity analysis, and evaluates the estimators under alternative data-generating processes. For the application, Section S3.2 documents data construction and missingness, examines temporal patterns of own and neighboring policy exposure alongside outcome and covariate distributions, and presents propensity-overlap diagnostics and implementation details for network-block bootstrap inference.

## S3.1 Numerical Experiments: Design, Diagnostics, and Bandwidth Sensitivity

### S3.1.1 Network Data Generation with Temporal Feedback and Spillover

**Network structure and time scales.** For each replication, we generate longitudinal observations for $n \in \{5000, 50000\}$ units connected through an undirected Erdős–Rényi network. Let $A = (A_{ij})$ denote its adjacency matrix, with independently generated edges

$$A_{ij} = A_{ji} \sim \text{Bernoulli}\left(\frac{\bar{d}}{n-1}\right), \qquad i < j,$$

where $\bar{d}$ is the targeted average degree and $A_{ii} = 0$. Neighborhood summaries are constructed using the row-normalized adjacency matrix $\widetilde{A} = \mathcal{D}_A^{-1} A$, where $\mathcal{D}_A = \text{diag}(N_1, \dots, N_n)$ and $N_i = \sum_{j=1}^{n} A_{ij} = |\mathcal{N}^1(i)|$. Inverse-degree weights are taken to be zero for isolated units, so the corre-

sponding rows of $\widetilde{A}$ are zero.

For each unit-level variable, the neighborhood summary is defined as the weighted average over first-order neighbors. Specifically, at process time $t$,

$$V_{i,t}^s = \sum_{j=1}^n \widetilde{A}_{ij} V_{j,t}, \qquad Y_{i,t}^s = \sum_{j=1}^n \widetilde{A}_{ij} Y_{j,t}.$$

Here the superscript $s$ denotes an aggregated neighborhood summary, and the covariate average is computed coordinatewise. The corresponding average of neighboring treatments is the spillover exposure $D_{i,t}$ defined below.

Covariates and outcomes evolve over $T = 20$ process times. Each unit has $Q = 10$ time-varying covariates, whereas binary treatment is updated only at $\tau_{1:4} = (1, 6, 11, 16)$ and remains in effect between decisions. The network is held fixed throughout each generated trajectory.

**Baseline burden and heterogeneous treatment response.** Let $V_{i,0}$, $C_{i,0}$, and $Y_{i,0}$ denote the baseline covariate vector, latent disease-activity state, and observed outcome.

Baseline covariate values are generated independently across units and coordinates by drawing from $N(\mu_{V0,\ell}, \sigma^2_{V0,\ell})$, $\ell = 1, \dots, Q$, and clipping the resulting draws to $[0, V_{\max,\ell}]$. The coordinate-specific parameters are given in Table S3.1.

Independently of the baseline covariates, baseline outcomes are generated from independent $N(50, 5^2)$ draws and clipped to $[0, 130]$. We initialize the latent disease-activity state as $C_{i,0} = Y_{i,0}$, and set $X_{i,0} = \widetilde{X}_{i,0} = \widetilde{D}_{i,0} = 0$, where $\widetilde{X}_{i,t}$ and $\widetilde{D}_{i,t}$ are the cumulative exposure summaries defined below.

Let $V_{i,t}^{(\ell)}$ denote the $\ell$-th coordinate of the covariate vector $V_{i,t}$. The covariate burden and its neighborhood counterpart are defined by

$$\mathcal{B}_{i,t} = \sum_{\ell=1}^Q w_\ell V_{i,t}^{(\ell)}, \qquad \mathcal{B}_{i,t}^s = \sum_{j=1}^n \widetilde{A}_{ij} \mathcal{B}_{j,t}, \qquad w_\ell \geq 0, \quad \sum_{\ell=1}^Q w_\ell = 1.$$

The weights $w_\ell$ are listed in Table S3.1. Units are ranked by $\mathcal{B}_{i,0}$ and divided into five equal-size groups, indexed by $G_i \in \{1, \dots, 5\}$ in ascending order of baseline burden. Group membership is

Table S3.1: Parameters for baseline covariate generation and covariate-burden weights.

| Coordinate $\ell$ | $\mu_{V0,\ell}$ | $\sigma_{V0,\ell}$ | $V_{\max,\ell}$ | $w_\ell$ |
|---|---|---|---|---|
| 1 | 52.3 | 8.0 | 95.0 | 1.06/9.76 |
| 2 | 37.7 | 6.5 | 80.0 | 0.92/9.76 |
| 3 | 47.4 | 7.0 | 85.0 | 1.16/9.76 |
| 4 | 34.2 | 5.5 | 70.0 | 0.66/9.76 |
| 5 | 58.5 | 7.5 | 90.0 | 1.36/9.76 |
| 6 | 51.0 | 6.0 | 78.0 | 1.14/9.76 |
| 7 | 31.2 | 5.0 | 65.0 | 0.60/9.76 |
| 8 | 41.2 | 5.5 | 72.0 | 0.94/9.76 |
| 9 | 32.5 | 4.5 | 60.0 | 1.02/9.76 |
| 10 | 46.3 | 6.5 | 88.0 | 0.90/9.76 |

therefore a fixed function of baseline covariates alone and remains unchanged throughout follow-up. It does not enter the treatment-assignment model or the covariate-evolution process as a separate predictor. Its only role is to scale the accumulated own-treatment benefit in the latent-state equation below through the group-specific multiplier $m_{G_i}$, thereby allowing treatment responsiveness to vary with baseline burden. These baseline-burden groups are not defined by realized treatment histories and are distinct from the five target own-treatment paths labelled *never*, *early*, *late*, *intermittent*, and *frequent*.

Besides, treatment assignment depends on the standardized baseline risk score

$$R_{i,0} = \frac{r_{i,0} - \bar{r}_0}{s_{r,0}}, \qquad r_{i,0} = \omega_{Y0} Y_{i,0} + \omega_{V0} \mathcal{B}_{i,0} + \omega_{YS0} Y^s_{i,0},$$

where $\bar{r}_0$ and $s_{r,0}$ are the empirical mean and standard deviation of the baseline scores.

**Covariate dynamics and temporal feedback.** For $t = 1, \ldots, T$, each covariate coordinate evolves according to

$$\begin{aligned} V^{(\ell)}_{i,t} &= (1 - \kappa_\ell) V^{(\ell)}_{i,t-1} + \mu_\ell + \gamma_{Y,\ell}(Y_{i,t-1} - c_{Y,\ell}) \\ &\qquad + \gamma_{YS,\ell}(Y^s_{i,t-1} - c_{Y,\ell}) + \varepsilon^{V,\ell}_{i,t}, \qquad \varepsilon^{V,\ell}_{i,t} \sim N(0, \sigma^2_{V,\ell}), \end{aligned}$$

followed by clipping to $[0, V_{\max,\ell}]$. The neighborhood-outcome term mirrors the own-outcome term,

using the same reference level $c_{Y,\ell}$ and a separate coefficient $\gamma_{YS,\ell}$, without introducing an additional neighborhood-specific reference level. The own-outcome term allows previous outcomes to affect subsequent covariates, while the neighborhood-outcome term transmits changes in neighboring units' outcomes into the focal unit's covariate trajectory. Because neighboring outcomes themselves respond to treatment, the latter provides an indirect spillover channel through covariate evolution, complementing the neighborhood-burden and spillover-exposure terms in the latent-state equation below. After updating the covariates across the network, we recompute $\mathcal{B}_{i,t}$ and $\mathcal{B}^s_{i,t}$. Together with the treatment and outcome mechanisms below, these updates create feedback from earlier treatments to later covariates and treatment decisions.

The coordinate-specific parameters $\kappa_\ell$, $\mu_\ell$, $\gamma_{Y,\ell}$, $c_{Y,\ell}$, and $\sigma_{V,\ell}$ are fixed across replications, with their values provided in the accompanying simulation configuration.

**History-dependent treatment assignment.** At each decision time $\tau_m$, treatment is generated from a Bernoulli distribution with success probability $\pi_{i,\tau_m}$, where

$$\begin{aligned}\operatorname{logit}(\pi_{i,\tau_m}) = \theta_0 + \theta_R R_{i,0} + \theta_p X_{i,\tau_m-1} + \theta_w W_{i,\tau_m} + \theta_i I_{i,\tau_m} \\ + \theta_v \mathcal{B}_{i,\tau_m-1} + \theta_{W2} W^2_{i,\tau_m} + \varepsilon^X_{i,\tau_m}, \qquad \varepsilon^X_{i,\tau_m} \sim N(0, \sigma^2_X).\end{aligned}$$

For $m \geq 2$, recent worsening and improvement are defined by

$$\begin{aligned}W_{i,\tau_m} &= \max\{Y_{i,\tau_m-1} - Y_{i,\tau_m-2}, 0\}, \\ I_{i,\tau_m} &= \max\{Y_{i,\tau_m-2} - Y_{i,\tau_m-1}, 0\}.\end{aligned}$$

At the first decision time, both quantities are set to zero.

The assignment model combines three sources of treatment variation. First, the component $\theta_0 + \theta_R R_{i,0}$ captures persistent differences in treatment propensity across units through the standardized baseline risk score. There is no additional unit-specific random intercept.

Second, treatment depends on recent treatment and outcome history. The term $X_{i,\tau_m-1}$ allows previous treatment status to influence the current decision, thereby permitting treatment to persist over time. Meanwhile, $W_{i,\tau_m}$ and $I_{i,\tau_m}$ allow treatment decisions to respond to recent worsening

and improvement, with separate coefficients permitting asymmetric responses to the two types of change. The quadratic term $\theta_{W2} W_{i,\tau_m}^2$ additionally allows treatment log-odds to increase more sharply during larger deteriorations, representing a nonlinear response to recent worsening.

Third, the term $\theta_v \mathcal{B}_{i,\tau_m-1}$ links treatment assignment to the most recent covariate burden, which also influences disease activity. Because earlier treatments affect subsequent outcomes and covariates, later treatment decisions depend on histories that have themselves been affected by prior treatment. This feedback induces the time-varying confounding that the numerical experiments are designed to assess.

At non-decision times, we set $X_{i,t} = X_{i,t-1}$. After updating treatment across all units, spillover exposure is computed as

$$D_{i,t} = \sum_{j=1}^{n} \widetilde{A}_{ij} X_{j,t}.$$

Thus, $D_{i,t}$ is the proportion of treated first-order neighbors for nonisolated units and is set to zero for isolated units.

**Accumulated exposure and covariate-burden effects.** To allow treatment effects to persist over process time, define

$$\widetilde{X}_{i,t} = \rho_X \widetilde{X}_{i,t-1} + X_{i,t}, \qquad \widetilde{D}_{i,t} = \rho_X \widetilde{D}_{i,t-1} + D_{i,t},$$

with $\widetilde{X}_{i,0} = \widetilde{D}_{i,0} = 0$. Both summaries use the same decay parameter $\rho_X$, so that earlier exposure contributes progressively less to the current latent state.

Covariate-burden effects are measured relative to a deterministic reference trajectory, so that burden is assessed against a time-specific benchmark that follows the mean-reversion component of the covariate process. For each coordinate, we introduce

$$\bar{v}_{\ell,t} = (1 - \kappa_\ell)\bar{v}_{\ell,t-1} + \mu_\ell, \qquad \bar{\mathcal{B}}_t = \sum_{\ell=1}^{Q} w_\ell \bar{v}_{\ell,t},$$

where $\bar{v}_{\ell,0} = \mu_{V0,\ell}$ is the location parameter used in the Gaussian baseline generator. Here $\bar{v}_{\ell,t}$

describes the reference evolution of the $\ell$-th covariate, and $\bar{\mathcal{B}}_t$ combines these coordinatewise trajectories using the same weights as the covariate burden. This reference calculation retains only the mean-reversion component, omitting outcome feedback and clipping. It is therefore determined by the DGP parameters and held fixed across replications.

The deviation $\mathcal{B}_{i,t-1} - \bar{\mathcal{B}}_{t-1}$ indicates whether the focal unit's lagged covariate burden lies above or below the reference level at the same process time. Using a time-varying reference allows the burden response to account for the temporal evolution built into the covariate process. To translate this deviation into a bounded contribution to latent disease activity, we use the sigmoid transformation $s(u) = (1 + e^{-u})^{-1}$ and define

$$q_{1,i,t} = s(\mathcal{B}_{i,t-1} - \bar{\mathcal{B}}_{t-1}), \qquad q_{S,i,t} = s(\mathcal{B}^s_{i,t-1} - \bar{\mathcal{B}}_{t-1}).$$

For the focal unit, $q_{1,i,t}$ equals $1/2$ at the reference level and increases as covariate burden rises above that level. The sigmoid transformation makes this response saturate for large deviations, thereby preventing extreme covariate burdens from producing arbitrarily large contributions to disease activity. The analogous quantity $q_{S,i,t}$ applies the same transformation and reference level to the neighborhood-averaged covariate burden. This allows the covariate conditions of neighboring units to affect the focal unit's latent disease activity, providing a further channel of network dependence alongside the neighborhood-outcome feedback in the covariate process.

**Latent disease activity and delayed outcome response.** The latent disease-activity state evolves as

$$\begin{aligned} C_{i,t} = \mu_C &+ \rho_C(C_{i,t-1} - \mu_C) + \lambda_1 q_{1,i,t} + \lambda_S q_{S,i,t} \\ &- m_{G_i}\beta_X \left(1 - e^{-a_X \widetilde{X}_{i,t}}\right) - \beta_{XS}\left(1 - e^{-a_X \widetilde{D}_{i,t}}\right) \\ &- \beta_{XD}\left(1 - e^{-a_X \widetilde{X}_{i,t}}\right)\left(1 - e^{-a_X \widetilde{D}_{i,t}}\right) + \varepsilon^C_{i,t}, \qquad \varepsilon^C_{i,t} \sim N(0, \sigma^2_C). \end{aligned}$$

The autoregressive component $\mu_C + \rho_C(C_{i,t-1} - \mu_C)$ combines persistence with mean reversion, retaining part of the previous latent state while pulling it toward the reference level $\mu_C$. The terms

$\lambda_1 q_{1,i,t}$ and $\lambda_S q_{S,i,t}$ then contribute increases in disease activity associated with the focal unit's and its neighbors' covariate burdens. These contributions become larger as the respective burdens rise relative to the common reference trajectory $\bar{\mathcal{B}}_{t-1}$. Using the same transformation and reference level gives the two burden effects a common construction, while their separate coefficients allow different magnitudes of influence.

Accumulated own treatment and spillover exposure reduce latent disease activity through the saturating response functions $1 - e^{-a_X \widetilde{X}_{i,t}}$ and $1 - e^{-a_X \widetilde{D}_{i,t}}$. These functions allow additional exposure to yield progressively smaller marginal reductions as accumulated exposure increases. The common shape parameter $a_X$ controls the rate of saturation, while $\beta_X$ and $\beta_{XS}$ determine the magnitudes of the corresponding treatment contributions.

The own-treatment contribution is further scaled by $m_{G_i}$, allowing treatment responsiveness to vary across baseline-burden groups. The additive spillover-treatment contribution uses a common coefficient $\beta_{XS}$ across these groups and is not multiplied by $m_{G_i}$. Thus, group-specific responsiveness is introduced through the own-treatment component, while the additive spillover component retains a common parameterization. As specified above, $G_i$ does not enter the treatment-assignment model as a separate predictor.

The final product term introduces an interaction between accumulated own treatment and spillover exposure. For $\beta_{XD} > 0$, greater accumulated spillover exposure strengthens the reduction in latent disease activity associated with own treatment, and greater accumulated own treatment analogously strengthens the spillover contribution. The interaction reuses the same saturating response functions and shape parameter $a_X$. This creates a nonadditive exposure-response structure that the OLS benchmark, which includes treatment, spillover exposure, and baseline covariates additively, does not explicitly represent.

Setting $\lambda_S = \beta_{XS} = \beta_{XD} = 0$ would remove the explicit neighborhood-burden, spillover-treatment, and interaction terms from the latent-state equation.

The observed outcome then moves gradually toward the latent state:

$$Y_{i,t} = (1 - \kappa_Y) Y_{i,t-1} + \kappa_Y C_{i,t} + \varepsilon^Y_{i,t}, \qquad \varepsilon^Y_{i,t} \sim N(0, \sigma^2_Y),$$

followed by clipping to $[y_{\min}, Y_{\max}] = [0, 130]$. The parameter $\kappa_Y$ controls the speed of this adjustment. Consequently, treatment effects can emerge with delay and persist through both accumulated exposure and the latent-state dynamics.

The Gaussian innovations in the longitudinal updates are mutually independent across units, times, equations, and covariate coordinates where applicable, and are independent of the network and baseline draws.

At each process time, the updates are performed across the network in the following order: covariates, treatment assignments, spillover and cumulative exposures, latent states, and observed outcomes. The resulting outcomes and neighborhood summaries then enter subsequent covariate updates and treatment decisions. The specified parameter values are summarized in Table S3.2.

### S3.1.2 Intervention Paths and Monte Carlo Approximation of Counterfactual Means

**Target intervention paths.** At the four decision times $\tau_{1:4} = (1, 6, 11, 16)$, the $2^4 = 16$ possible own-treatment sequences are grouped into five families according to their treatment-switching patterns:

$$\begin{aligned}
\text{never} &= \{(0,0,0,0)\}, \\
\text{early} &= \{(1,0,0,0), (1,1,0,0), (1,1,1,0)\}, \\
\text{late} &= \{(0,0,0,1), (0,0,1,1), (0,1,1,1)\}, \\
\text{intermittent} &= \{(0,0,1,0), (0,1,0,0), (0,1,0,1), (0,1,1,0), \\
&\qquad (1,0,0,1), (1,0,1,0), (1,0,1,1), (1,1,0,1)\}, \\
\text{frequent} &= \{(1,1,1,1)\}.
\end{aligned}$$

Table S3.2: Specified parameters of the data-generating process.

| Category | Parameter(s) | Value |
|---|---|---|
| **Design** | Sample size $n$ | 5000, 50000 |
| | Process-time horizon $T$; covariate dimension $Q$ | 20; 10 |
| | Decision times $\tau_{1:4}$ | $(1, 6, 11, 16)$ |
| **Network** | Expected network degree $\bar{d}$ | 6 |
| **Baseline Groups** | Group-specific multipliers $(m_1, \ldots, m_5)$ | $(0.50, 1.16, 0.92, 1.20, 1.10)$ |
| **Treatment** | Assignment coefficients $(\theta_0, \theta_R, \theta_p, \theta_w, \theta_i, \theta_v)$ | $(-3.85, 1.00, 0.28, 0.19, 0.08, 0.06)$ |
| | Quadratic worsening coefficient $\theta_{W2}$ | 0.05 |
| | Baseline-risk weights $(\omega_{Y0}, \omega_{V0}, \omega_{YS0})$ | $(0.10, 0.18, 0.08)$ |
| | Treatment-logit noise standard deviation $\sigma_X$ | 0.45 |
| **Dynamics** | Exposure decay $\rho_X$ | 0.63 |
| | Neighborhood-outcome feedback $\gamma_{YS,\ell}$ | $\gamma_{Y,\ell}/2$, $\ell = 1, \ldots, Q$ |
| | Latent-state location and persistence $(\mu_C, \rho_C)$ | $(72.0, 0.65)$ |
| | Own and neighborhood burden coefficients $(\lambda_1, \lambda_S)$ | $(5.2, 2.5)$ |
| | Own and spillover treatment coefficients $(\beta_X, \beta_{XS})$ | $(8.25, 3.0)$ |
| | Own-by-spillover interaction coefficient $\beta_{XD}$ | 10.0 |
| | Treatment-response shape $a_X$ | 0.85 |
| | Latent-state noise standard deviation $\sigma_C$ | 0.80 |
| **Outcome** | Outcome adjustment and noise $(\kappa_Y, \sigma_Y)$ | $(0.28, 0.45)$ |

We report results at one representative path from each family. Using the notation of the main text, these paths are

$$x^{\text{never}} = (0,0,0,0),\ \ x^{\text{early}} = (1,0,0,0),\ \ x^{\text{late}} = (0,0,0,1),$$
$$x^{\text{intermittent}} = (1,0,1,0),\ \ x^{\text{frequent}} = (1,1,1,1).$$

Therefore, each target specifies a particular treatment sequence, with treatment carried forward between decision times.

We combine each own-treatment path with three fixed reference spillover-exposure paths, representing low, mid, and high exposure patterns. These reference paths are selected once using a fixed reference data set of $n = 5000$ units. With entries displayed to two decimal places, these paths are

$$d^L = (0.00, 0.33, 0.20, 0.22),$$
$$d^M = (0.30, 0.33, 0.36, 0.22),$$
$$d^H = (0.43, 0.57, 0.50, 0.40).$$

The same three reference paths are shared across the five own-treatment paths and held fixed across analysis sample sizes, replications, and bandwidth comparisons. The resulting 15 target mean potential outcomes are combined into the 13 direct, spillover, and total effects specified in the main text.

**Two-hop local structural Monte Carlo approximation.** For each target intervention regime $(x, d)$, we approximate the counterfactual mean outcome by a two-hop local structural Monte Carlo simulation rather than a full $n$-unit network simulation. Each draw contains one focal unit, its first-hop neighbors, and a pool of second-hop neighbors for each first-hop unit. Neighbor counts are constructed using Poisson($\bar{d}$) draws, motivated by the large-$n$ limit of the analysis network's Binomial$(n-1, \bar{d}/(n-1))$ degree distribution. The local trajectories evolve according to the DGP in Section S3.1.1, with feedback truncated beyond the second hop.

At each decision time $\tau_m$, the focal unit's treatment is set to $x_{\tau_m}$. Because the requested spillover exposure $d_{\tau_m}$ may not be attainable on the sampled neighbor set, we round $N_i d_{\tau_m}$ to the nearest feasible treated-neighbor count, where $N_i$ is the focal unit's number of first-hop neighbors. A subset of that size is selected uniformly at random from the first-hop neighbors and assigned treatment; the remaining first-hop neighbors are untreated, so that the treated share equals the rounded target. The achieved exposure is recorded as a diagnostic. Selecting the treated subset instead from the natural assignment model, conditional on its size, changes the 15 reference means by at most 0.06 (mean absolute change 0.02), within Monte Carlo error and well below the estimation biases reported for the proposed estimator.

The counterfactual mean is estimated by its Monte Carlo counterpart using the average of the focal unit's final outcome over $B_{\text{MC}}$ such draws, with $B_{\text{MC}} = 10{,}000$ in the numerical experiments:

$$\widehat{\mu}^{\text{MC}}_{n,T}(x, d) = \frac{1}{B_{\text{MC}}} \sum_{b=1}^{B_{\text{MC}}} Y^{(b)}_{i,T}(x, d).$$

#### S3.1.3 Estimator Implementation and Network-Block Bootstrap

**Estimators and bandwidth.** We compare the proposed self-normalized estimator described in Section 3.4 of the main text with its oracle-propensity counterpart and an OLS outcome-regression benchmark. All estimators target the final outcome $Y_{i,T}$ under the intervention paths specified in Section S3.1.2, using the complete four-decision window $M = M_T = 4$.

The oracle-propensity estimator replaces the fitted assignment probabilities with their DGP counterparts. For own treatment, the oracle probability is $\mathrm{E}\{\mathrm{expit}(\eta_{i,\tau_m} + \varepsilon^X_{i,\tau_m}) \mid H_{i,\tau_m}\}$, where $\eta_{i,\tau_m}$ is the treatment-assignment logit in Section S3.1.1 without the noise term, including $\theta_{W2} W^2_{i,\tau_m}$; the expectation over $\varepsilon^X_{i,\tau_m}$ is evaluated by Gauss–Hermite quadrature. For spillover exposure, the treated-neighbor count follows a Poisson–binomial distribution constructed from the neighbors' own-treatment probabilities; dividing this count by the focal unit's degree gives the corresponding spillover exposure. The proposed and oracle-propensity estimators use the same Gaussian localization kernel and bandwidth at each decision time. The primary analyses use $h = 0.03$, and the

final localized weights are not truncated. The bandwidth comparison with $h = 0.02, 0.04$, and $0.05$ is reported in the bandwidth-sensitivity subsection below.

The OLS benchmark includes own-treatment history, spillover-exposure history, and baseline covariates additively, without interaction or nonlinear terms. Specifically, we regress $Y_{i,T}$ on the four decision-time own-treatment indicators, the four corresponding spillover exposures, and $V_{i,0}$, including an intercept. For each target regime, we replace the observed treatment and spillover paths with the target values, retain each unit's baseline covariates, and average the resulting predictions over all $n$ units.

**Neural architecture and training settings.** The proposed estimator uses the GraphSAGE-mean aggregation, LSTM encoder, and decision-time representation and assignment models introduced in Sections 3.1–3.3 of the main text. Training includes the auxiliary reconstruction loss $\mathcal{L}_I$ introduced in Section 3.3 of the main text, which encourages the recurrent states to retain process-time input information. Table S3.3 summarizes the architecture, optimization settings, and training and validation sample sizes.

For each replication, the validation data are generated separately from the analysis data under the same DGP. Validation loss determines early stopping and learning-rate scheduling, and the model parameters attaining the lowest validation loss are retained for estimation.

**Neighborhood-averaged recurrent states.** The neighborhood averages $h^s_{i,\tau_m}$ and $c^s_{i,\tau_m}$ are computed once for all units at the start of each training epoch and held fixed over that epoch, which is also how the gradient blocking of Section 3.3 is applied.

**Monte Carlo replications and evaluation.** For each $n \in \{5000, 50000\}$, we conduct $R = 100$ independent Monte Carlo replications. Each replication regenerates the network and longitudinal data, retrains the neural representation and assignment models, and recomputes all three estimators. Thus, variation across replications includes variation from both data generation and model fitting.

Point-estimation performance is assessed using empirical bias, absolute relative bias, and mean

Table S3.3: Neural architecture and training settings for the numerical experiments.

| Category | Parameter(s) | Value |
|---|---|---|
| **Architecture** | LSTM hidden dimension; number of layers | 128; 2 |
| | Dimensions of $(Z^X, Z^D, Z^C)$ | $(16, 16, 32)$ |
| | GraphSAGE-mean encoding dimension | 128 |
| | Optional outcome-prediction head | Disabled |
| | Auxiliary reconstruction head for $\mathcal{L}_I$ | Enabled |
| **Optimization** | Optimizer | AdamW |
| | Initial learning rate | $10^{-3}$ |
| | $L_2$ penalty $\lambda$ (AdamW's native weight decay) | $10^{-4}$ |
| | Batch size; maximum number of epochs | 64; 300 |
| | Dropout | 0.1 |
| **Scheduling** | Early-stopping patience | 20 epochs |
| | Learning-rate scheduler patience | 10 epochs |
| **Refinement** | K-IPW weight clipping (upper percentile of the localized weights) | None (100th percentile) |
| **Loss Weights** | Own-treatment $w_X$; spillover $w_D$; reconstruction $w_I$ | 1.0; 1.0; 1.0 |
| | Outcome $w_Y$ (inactive: outcome-prediction head disabled) | 0.1 |
| **Sample Sizes** | Training / validation size, $n = 50000$ | 50000/5000 |
| | Training / validation size, $n = 5000$ | 5000/500 |

squared error relative to the Monte Carlo reference values in Section S3.1.2. Effective sample sizes are examined separately for each own-treatment and reference spillover-path combination, as reported in the overlap and effective-sample-size subsection below.

**Network-block bootstrap inference.** Within each Monte Carlo replication, we construct 95% percentile intervals for the proposed estimator using $B_{\text{boot}} = 100$ network-block bootstrap resamples. For the oracle-propensity and OLS benchmarks, we report point-estimation performance only.

For the proposed estimator, the fitted neural model is held fixed throughout the bootstrap resamples. The intervals therefore describe variability of the final weighted estimator conditional on that replication's fitted assignment models, without adding uncertainty from refitting those models. Model retraining instead takes place across the outer Monte Carlo replications, in which

the models are trained anew. An alternative that retrains the neural model within bootstrap resamples is available but substantially more computationally expensive.

Under this fixed-model procedure, the observed network is partitioned into disjoint blocks using randomized 2-hop-max clustering. Blocks are sampled with replacement, and the self-normalized weighted means are recomputed using the resulting block multiplicities. The same block resample is used jointly for all 15 target means and their 13 causal contrasts. Consequently, each contrast is recomputed from its two component means within the same resample, retaining their covariance.

The interval endpoints are the 2.5th and 97.5th percentiles of the corresponding bootstrap estimates. We summarize interval performance by average width and empirical coverage across Monte Carlo replications.

#### S3.1.4 Simulated Temporal Dynamics and Treatment-Exposure Distributions

**Covariate distributions and temporal dynamics.** Figures S3.1 and S3.2 summarize the baseline distributions and temporal evolution of the ten covariates specified in Section S3.1.1. The baseline distributions differ across coordinates in both location and dispersion, while the trajectories exhibit different temporal patterns of decline and recovery over process time.

**Treatment-path frequencies and spillover exposure.** Using the five treatment-path families defined in Section S3.1.2, we classify units according to their realized own-treatment histories at $\tau_{1:4} = (1, 6, 11, 16)$. Figure S3.3 shows the share of units in each family. The *intermittent* family is the largest, accounting for 48.5% of units, whereas the *frequent* family is the smallest, accounting for 1.8%, consistent with the limited support for frequent-path targets noted in the main text.

Figure S3.4 shows the distribution of spillover exposure $D_{i,t}$, pooled across all $T = 20$ process times and stratified by realized own-treatment family. Because the Erdős–Rényi network is formed independently of baseline covariates, exposure differs little across families: the median exposure is 0.333 in every family, mean exposure is slightly higher in the frequent family (0.360) and lowest in the never family (0.318), and the distributions overlap substantially across all five families.

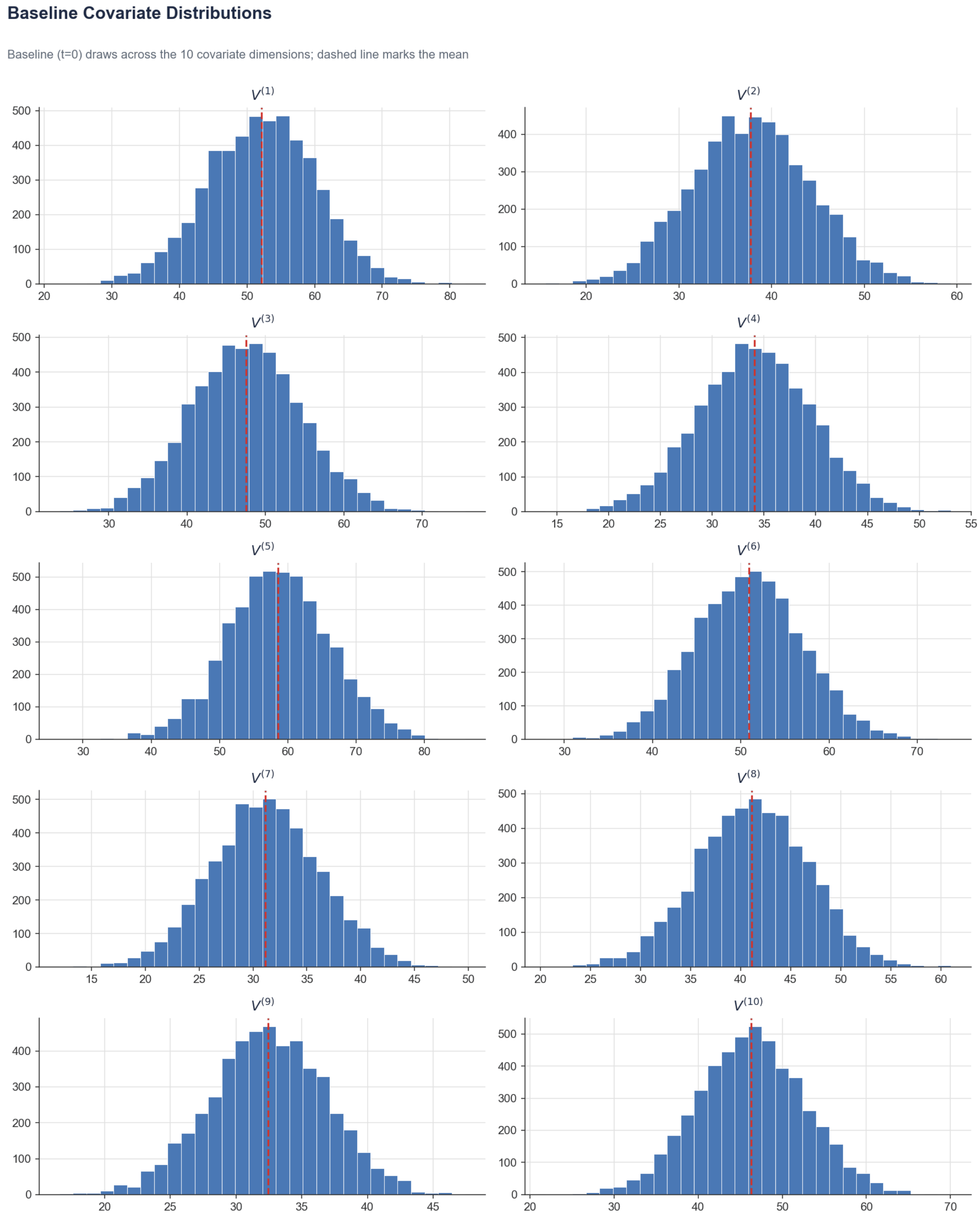


Figure S3.1: Baseline distributions of the simulated covariates. Different coordinates exhibit distinct locations and dispersions. Dashed lines indicate the corresponding means.

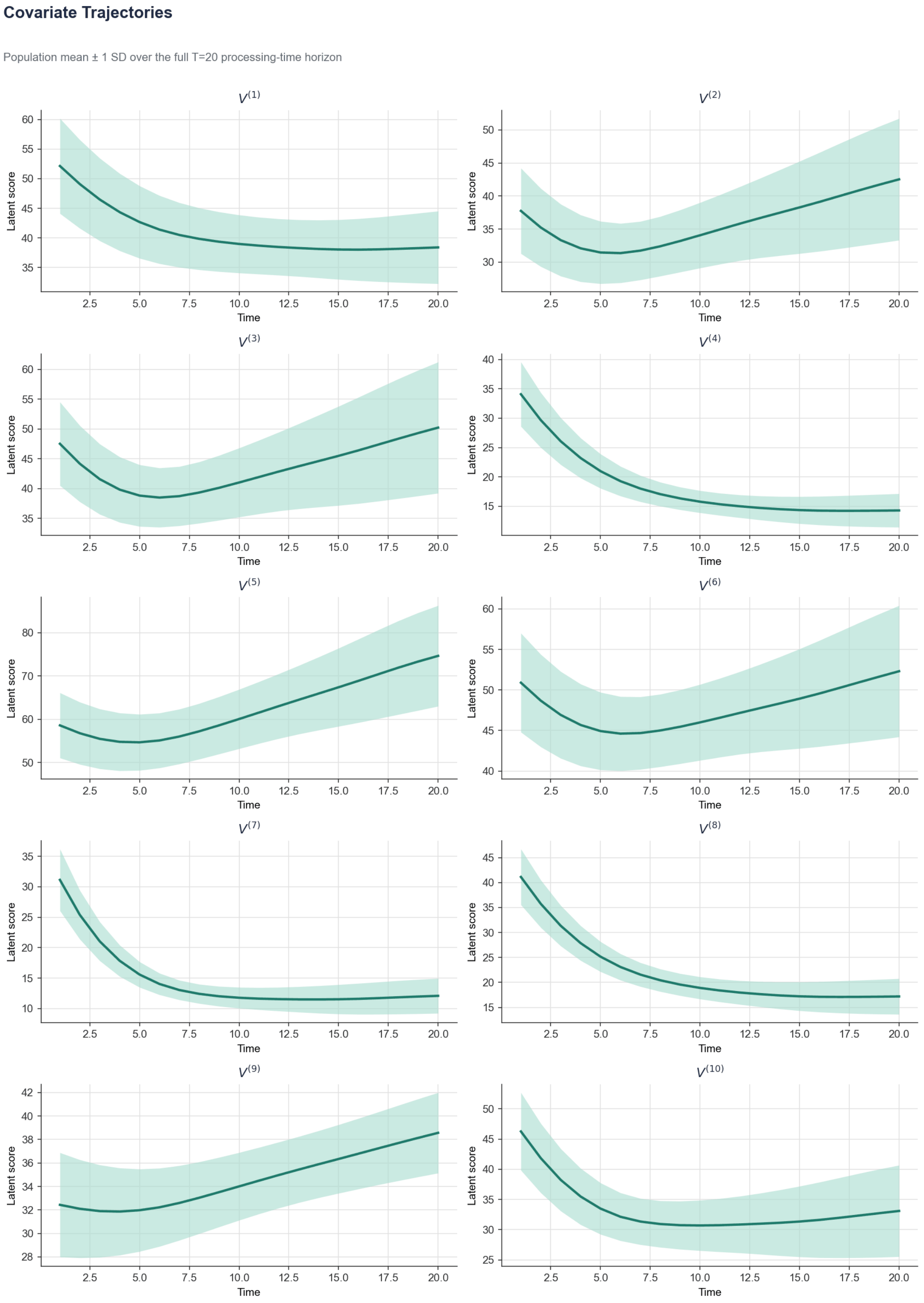


Figure S3.2: Temporal trajectories of the simulated covariates. The covariate dimensions follow distinct dynamic patterns over time, including decline followed by a plateau and decline followed by recovery.

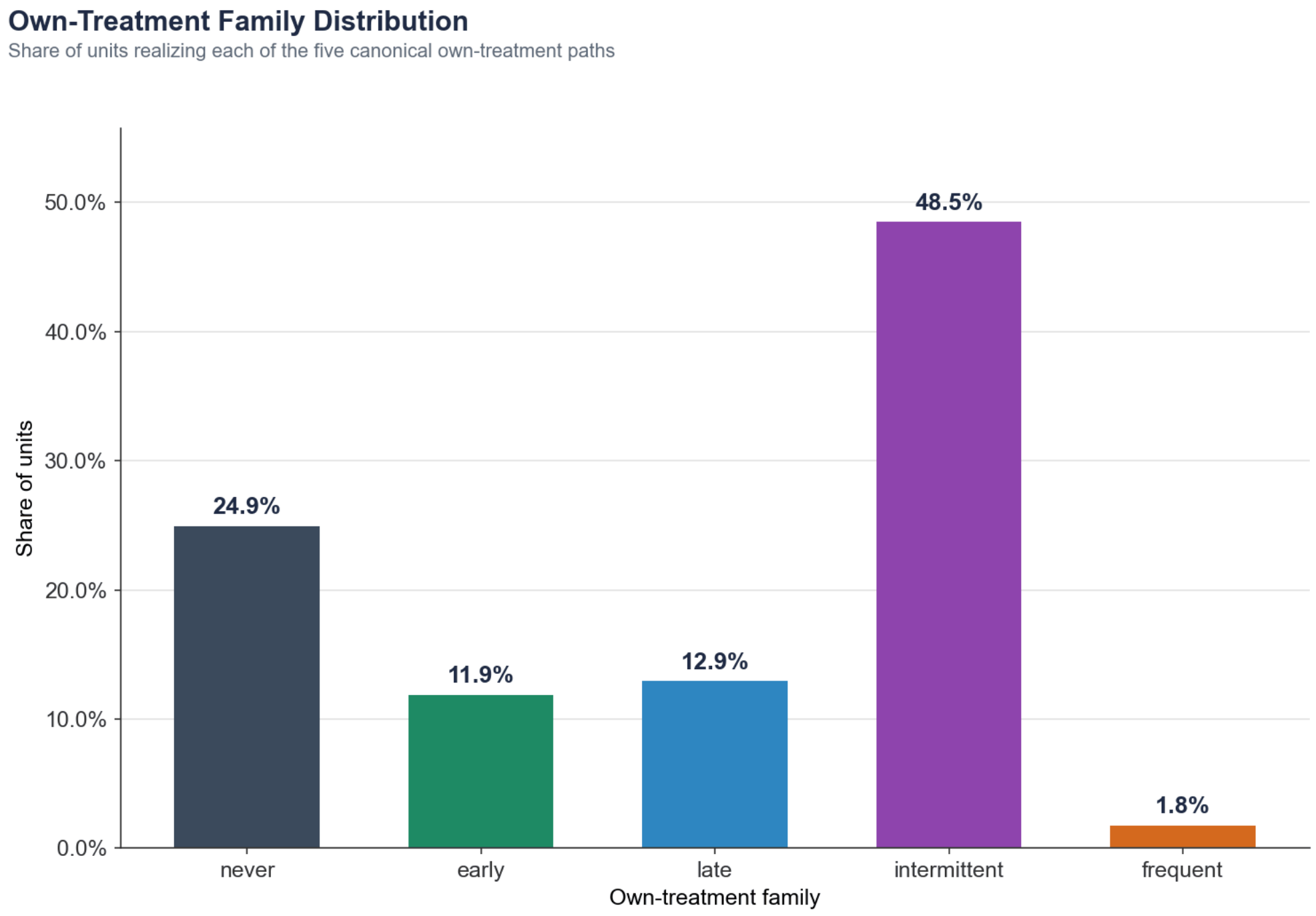


Figure S3.3: Distribution of realized own-treatment histories across the five treatment-path families defined in Section S3.1.2.

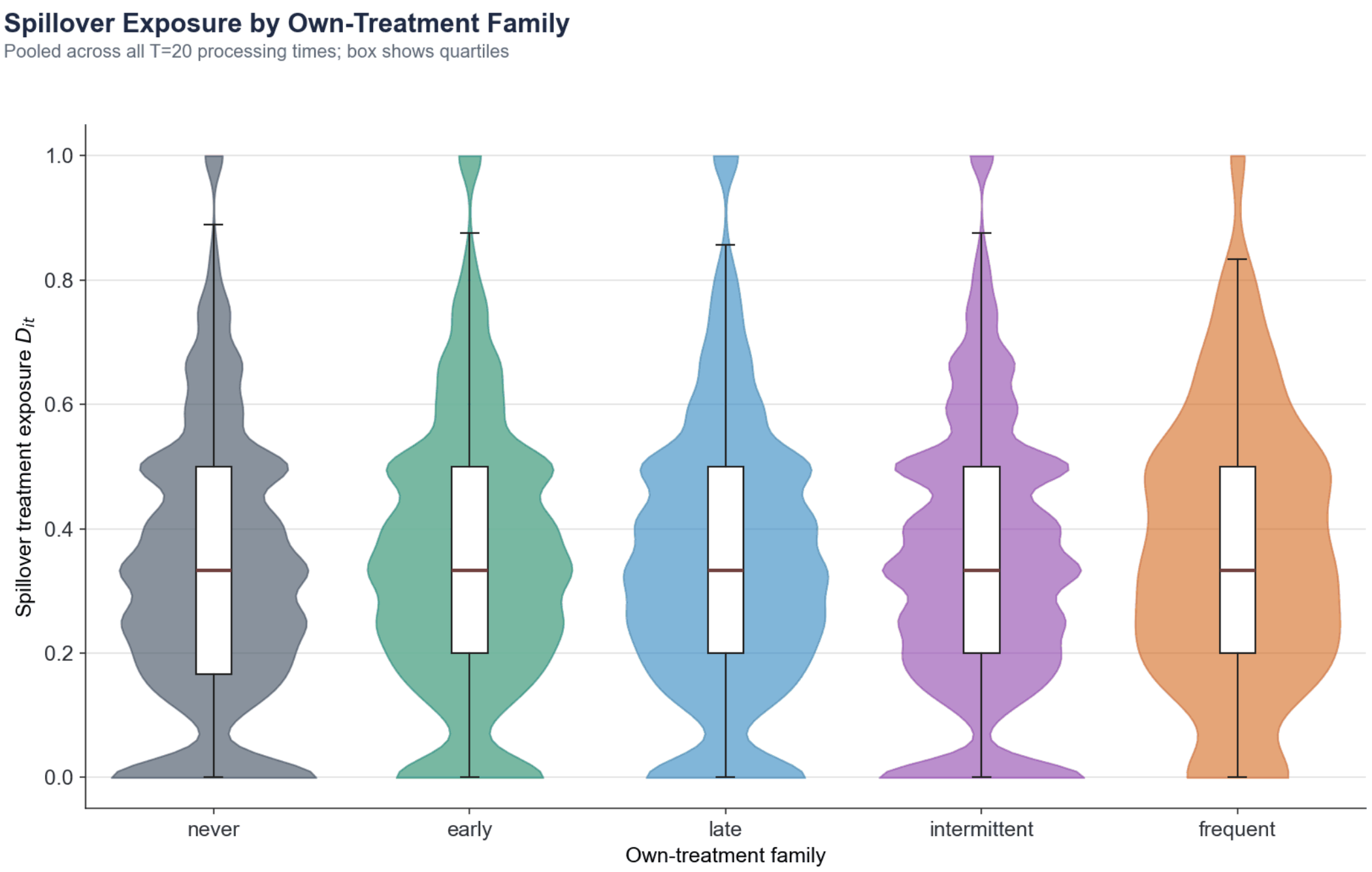


Figure S3.4: Distribution of spillover exposure $D_{i,t}$, pooled across the $T = 20$ process times, by realized own-treatment family. Spillover exposure differs little across families and shows substantial overlap.

**Outcome trajectories.** Figure S3.5 presents the overall outcome trajectory and the mean trajectories stratified by realized own-treatment family. The overall population mean rises over the observed horizon, steeply at first and more slowly after about $t = 6$, reflecting the mean-reverting component of the latent-state dynamics and the gradual adjustment of observed outcomes to the latent disease state (Section S3.1.1). Higher outcomes indicate greater disease activity. By the final process time $T$, the never and early families have the highest mean outcomes, while the frequent family receives the most treatment and plateaus at the lowest mean outcome. Because treatment is assigned according to baseline risk, lagged covariate burden, and recent outcome changes (Section S3.1.1), frequent membership also reflects greater underlying risk and treatment need: the frequent family starts with the highest mean outcome at $t = 1$ and falls below the other families only after repeated treatment. Selection into treatment therefore works against the observed ordering at $T$. The observed differences across families combine treatment responses with selection into different treatment histories and cannot directly be interpreted as causal effects. This confounding by indication is exactly the identification challenge that the proposed estimator is designed to address by reweighting based on the fitted assignment models.

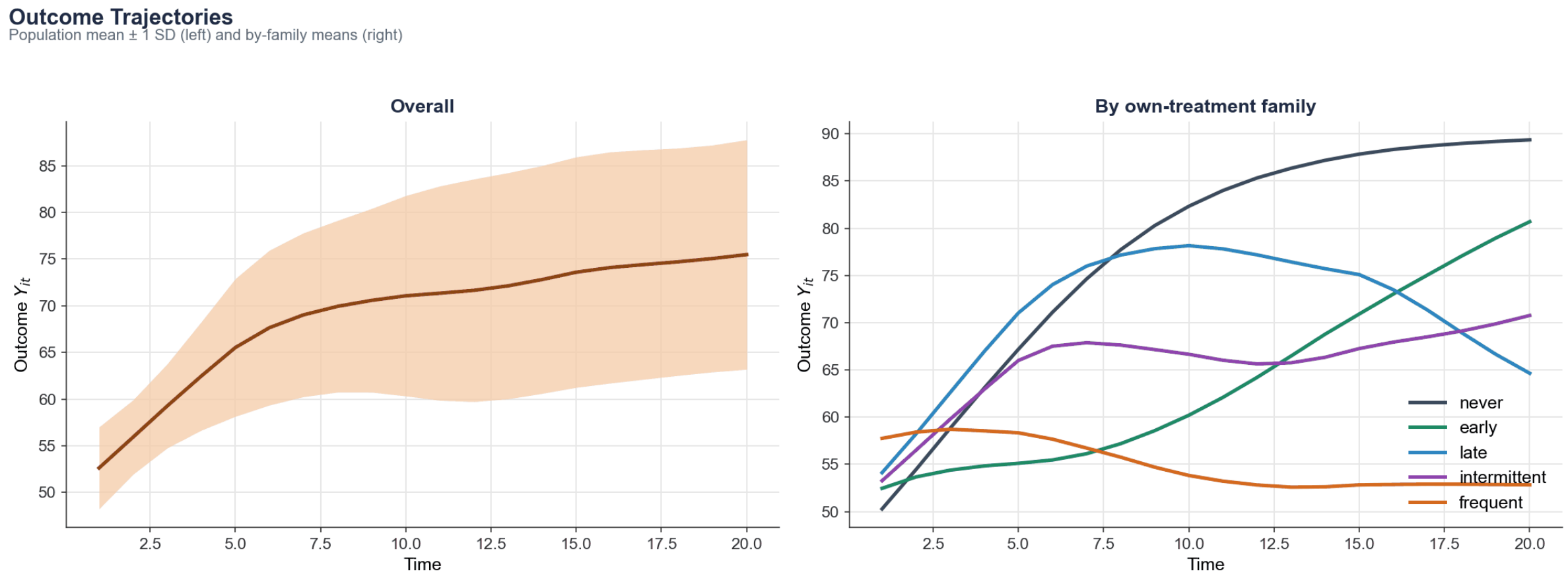


Figure S3.5: Outcome trajectories in the simulated data: the overall mean with a shaded band indicating one standard deviation above and below the mean (left), and mean trajectories by realized own-treatment family (right).

#### S3.1.5 Propensity Overlap and Local Support for the Frequent Treatment Path

**Own-treatment propensity overlap.** Figure S3.6 shows the distributions of the fitted LSTM own-treatment propensities at the four decision times $\tau_{1:4} = (1, 6, 11, 16)$, separately for treated and untreated units, at both $n = 50000$ and $n = 5000$. The two distributions overlap at every decision time and both sample sizes, so no treated or untreated unit receives a fitted propensity near zero or one. This overlap should not be read as a direct check of positivity. The fitted propensities are more concentrated than the true assignment probabilities: under the data-generating process, the true own-treatment propensities of the first replication range from about 0.03 to 0.77 at $\tau_1$ (1st and 99th percentiles), with a mean of 0.285, whereas the fitted propensities at $n = 50000$ are concentrated between about 0.33 and 0.58. Across the 100 replications, the standard deviation of the unit-level fitted propensity, averaged over the four decision times, has median 0.071 at $n = 5000$ and 0.073 at $n = 50000$, compared with 0.171 for the true propensity. The fitted assignment model thus shrinks propensities toward their mean, which reduces the variability of the inverse-propensity factor relative to the oracle weights. The comparison with the oracle-propensity estimator in the main text indicates the consequence of this shrinkage for the target estimates.

**Local support for the frequent treatment path.** To summarize effective local support for the frequent-path targets, we use the effective sample size (ESS) of the proposed estimator's final localized weights:

$$\mathrm{ESS}(x, d; h) = \frac{\left\{\sum_{i=1}^{n} \widehat{\omega}_{i,T;4}(h)\right\}^2}{\sum_{i=1}^{n} \left\{\widehat{\omega}_{i,T;4}(h)\right\}^2},$$

where the weights are defined in Section 3.4 of the main text. These weights combine own-treatment-path matching, joint spillover-path localization across the four decision times, and inverse-propensity adjustment. The joint localization can leave relatively few observations with appreciable weight even when the overall sample is large.

At $n = 5000$ and $h = 0.03$, the mean ESS across all $R = 100$ replications for the frequent own-treatment path is 1.24, 1.44, and 1.20 at $d^L$, $d^M$, and $d^H$, respectively, compared with 1.39

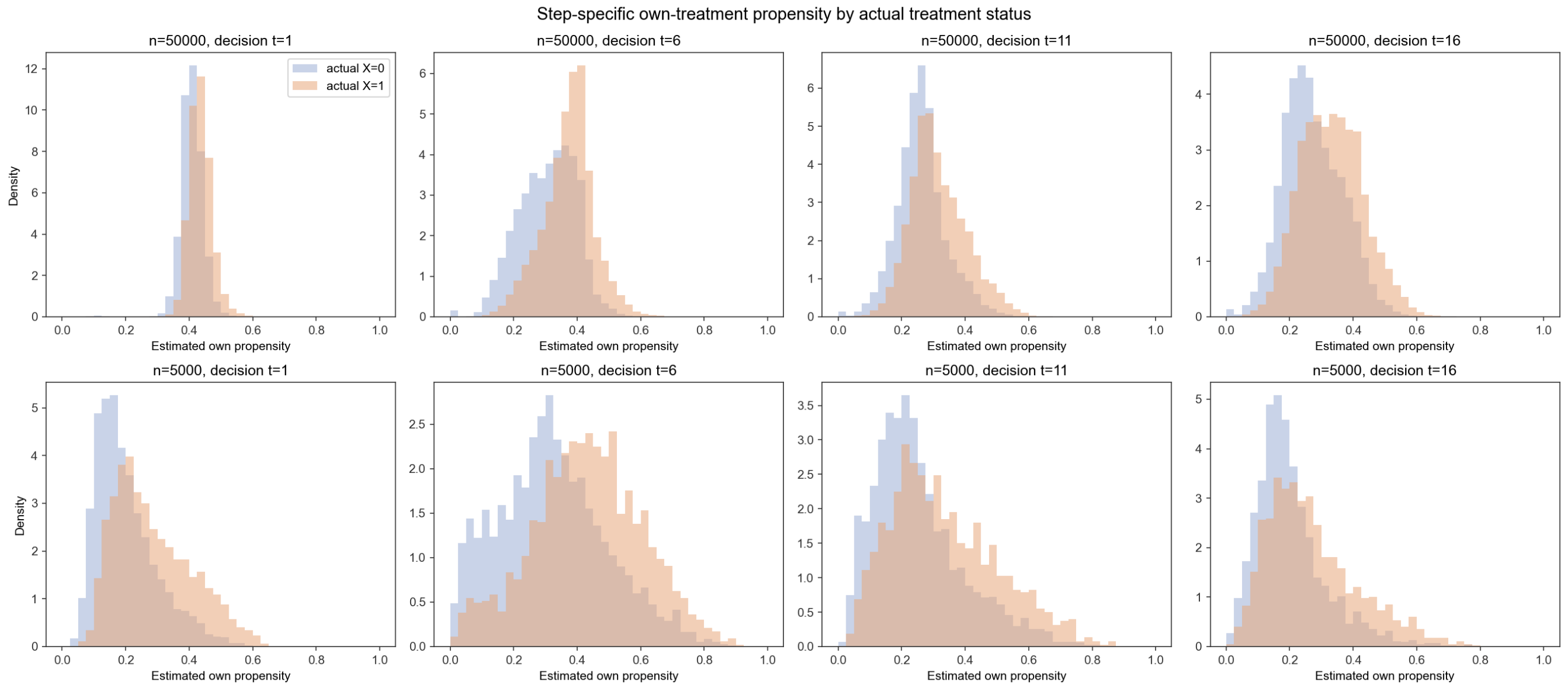


Figure S3.6: Fitted own-treatment propensity distributions by realized treatment status. Columns correspond to the four decision times $\tau_{1:4} = (1, 6, 11, 16)$; the top and bottom rows correspond to $n = 50000$ and $n = 5000$, respectively. Propensities are from the model fitted to the first Monte Carlo replication at each sample size.

to 3.46 for the other four paths. At $n = 50000$, the corresponding frequent-path values are 2.33, 2.60, and 1.80, compared with 3.15 to 15.02 for the other four paths. Thus, at each reference spillover level and at both sample sizes, the frequent path has the lowest mean ESS among the five representative own-treatment paths. This comparison motivates treating the three frequent-path effects as support stress tests and excluding them from the aggregate summaries of interval width and coverage at both sample sizes. All 13 causal contrasts, including these three effects, remain reported in the main text.

We also conducted a stress test with discrete spillover exposure, requiring exact matching of the full treatment and spillover paths without borrowing information from nearby exposure paths through kernel smoothing. At both $n = 5000$ and $n = 50000$, empirical support for the joint intervention paths was severely limited, resulting in highly unstable estimates.

### S3.1.6 Bandwidth Sensitivity of Estimation Accuracy and Interval Coverage

**Comparison settings.** We compare the four bandwidths $h = 0.02, 0.03, 0.04$, and $0.05$ at $n = 5000$, using $R = 100$ Monte Carlo replications for each bandwidth. The data-generating process, own-treatment paths, and reference spillover-exposure paths are held fixed, so that the comparison isolates the choice of localization bandwidth. All analyses use the complete four-decision window, $M = M_T = 4$, and the estimation and network-block bootstrap procedures described in Section S3.1.3. Tables S3.4–S3.6 report the complete results for $h = 0.02, 0.04$, and $0.05$. The corresponding results at the selected bandwidth $h = 0.03$ are presented in the main text.

**Estimation accuracy.** The proposed estimator has smaller absolute bias than OLS for all 28 targets, comprising 15 mean potential outcomes and 13 causal contrasts, at every bandwidth considered. Its MSE advantage is more sensitive to bandwidth: the proposed estimator has smaller MSE than OLS for 14 of the 28 targets at both $h = 0.02$ and $h = 0.03$, 16 targets at $h = 0.04$, and 17 targets at $h = 0.05$. Thus, the bias advantage over OLS persists throughout the comparison, while the number of targets with an MSE advantage increases at the larger bandwidths. Moreover, averaging MSE across the 28 targets gives 7.966, 6.874, 5.940, and 5.227 at $h = 0.02, 0.03, 0.04$, and 0.05, respectively, showing that the proposed estimator's average MSE decreases monotonically with bandwidth over the range considered.

**Interval coverage and local support.** Following the reporting convention in Section S3.1.5, we summarize interval performance across the ten causal contrasts that do not involve the frequent own-treatment path. Their average empirical coverage is 84.8%, 85.0%, 84.0%, and 83.5% at $h = 0.02, 0.03, 0.04$, and 0.05, respectively, with corresponding average interval widths of 5.421, 5.190, 4.669, and 4.220. Thus, average coverage is similar across the four bandwidths, while larger bandwidths yield narrower intervals on average.

The three frequent-path contrasts are examined separately as stress tests, given the limited local support documented in Section S3.1.5. For example, coverage for SE (frequent) is only 40%

at $h = 0.02$, compared with 63%, 62%, and 53% at $h = 0.03, 0.04$, and 0.05, respectively. Although increasing the bandwidth from 0.02 improves coverage for this contrast, further increases do not produce a monotonic improvement or bring coverage close to the nominal 95% level. These stress tests illustrate the inferential difficulty associated with limited local support, which bandwidth adjustment alone does not resolve over the range examined.

**Bandwidth choice.** The comparison favors $h = 0.03$ as the smallest bandwidth with adequate local support. At $h = 0.02$, the proposed estimator is unavailable in up to 48% of replications for some targets (Table S3.4), whereas at $h = 0.03$ it is valid in at least 99% of replications for every target. Among the bandwidths with full support, $h = 0.03$ gives the smallest absolute relative bias: averaged over the 28 targets, it is 0.039 at $h = 0.03$, compared with 0.042 at both $h = 0.04$ and $h = 0.05$, and is close to the 0.039 obtained at $h = 0.02$ over its valid replications. Larger bandwidths lower MSE through reduced variance and improve MSE relative to OLS for more targets, but at the cost of larger bias and without improving average coverage. We therefore use $h = 0.03$ in the main analyses.

#### S3.1.7 Sensitivity to Unmeasured Confounding: Effect Bounds and Tipping Points

**Analysis settings and robustness summaries.** We apply the joint-assignment sensitivity model in Section S1 to the numerical experiments. We computed sensitivity bounds for $Y_{i,T}$, with $T = 20$, in 100 independent networks at $n = 5000$ and, under an otherwise identical design, at $n = 50000$, using $M = 4$, bandwidth $h = 0.03$, and the 15 potential-outcome means and 13 causal effects specified in Section S3.1.2. The propensity model was refitted in each replication and held fixed across

$$\mathcal{G}_\Gamma = \{1, 1.01, 1.025, 1.05, 1.1, 1.2, 1.3, 1.5, 2, 3, 5\}.$$

The data-generating mechanism was unchanged; no additional latent confounder was introduced. With four decisions, the weight multipliers range over $[\Gamma^{-4}, \Gamma^4]$.

Table S3.4: Estimation of mean potential outcomes and causal effects at $n = 5000$ and $h = 0.02$, using $R = 100$ Monte Carlo replications.

| | | Proposed | | | | Oracle propensity | | OLS | |
|---|---|---|---|---|---|---|---|---|---|
| Target | GT | Bias (ARB) | MSE | Width | Cov. | Bias (ARB) | MSE | Bias (ARB) | MSE |
| *Panel A: Mean potential outcomes* | | | | | | | | | |
| never (low) | 90.397 | **-0.012 (0.000)** | 0.495 | 1.643 | 82% | -0.023 (0.000) | 0.537 | -0.585 (0.006) | **0.363** |
| never (mid) | 89.597 | -0.107 (0.001) | 0.597 | 1.691 | 85% | **-0.098 (0.001)** | **0.587** | -1.308 (0.015) | 1.721 |
| never (high) | 88.165 | +0.026 (0.000) | **0.720** | 1.713 | 90% | **+0.012 (0.000)** | 0.755 | -3.151 (0.036) | 9.943 |
| early (low) | 88.467 | **-0.150 (0.002)** | 1.028 | 1.868 | 77% | -0.177 (0.002) | 1.057 | -0.570 (0.006) | **0.342** |
| early (mid) | 87.227 | **-0.129 (0.001)** | 1.031 | 2.052 | 91% | -0.151 (0.002) | 1.020 | -0.853 (0.010) | **0.738** |
| early (high) | 85.400 | +0.298 (0.003) | **0.981** | 1.920 | 82% | **+0.292 (0.003)** | 0.993 | -2.301 (0.027) | 5.313 |
| late (low) | 74.407 | **+0.215 (0.003)** | 9.242 | 5.401 | 67% | +0.277 (0.004) | 9.203 | +0.694 (0.009) | **0.508** |
| late (mid) | 72.863 | **+0.013 (0.000)** | 7.884 | 5.793 | 76% | +0.025 (0.000) | 7.736 | +0.714 (0.010) | **0.527** |
| late (high) | 69.005 | **+0.371 (0.005)** | 9.284 | 5.620 | 74% | +0.468 (0.007) | 9.161 | +1.297 (0.019) | **1.702** |
| intermittent (low)† | 71.906 | **-1.062 (0.015)** | 10.516 | 3.483 | 38% | -1.106 (0.015) | 11.072 | +1.631 (0.023) | **2.689** |
| intermittent (mid) | 69.123 | -0.863 (0.012) | **7.295** | 5.019 | 74% | **-0.803 (0.012)** | 7.402 | +2.890 (0.042) | 8.371 |
| intermittent (high) | 65.051 | **-0.146 (0.002)** | **7.745** | 5.426 | 79% | -0.148 (0.002) | 8.208 | +3.687 (0.057) | 13.618 |
| frequent (low)† | 60.011 | **-3.135 (0.052)** | **14.850** | 2.136 | 14% | -3.252 (0.054) | 15.502 | -7.419 (0.124) | 55.079 |
| frequent (mid) | 56.336 | -2.104 (0.037) | **14.434** | 4.989 | 46% | **-1.988 (0.035)** | 16.551 | -5.267 (0.094) | 27.779 |
| frequent (high)† | 49.045 | +0.315 (0.006) | 7.406 | 3.906 | 54% | **+0.286 (0.006)** | 8.184 | -1.251 (0.026) | **1.608** |
| *Panel B: Causal contrasts* | | | | | | | | | |
| SE (never) | -2.233 | +0.038 (0.017) | **1.205** | 2.717 | 89% | **+0.036 (0.016)** | 1.218 | -2.566 (1.150) | 6.622 |
| SE (early) | -3.068 | **+0.474 (0.154)** | **2.045** | 3.269 | 92% | +0.506 (0.165) | 2.222 | -1.731 (0.564) | 3.033 |
| SE (late) | -5.402 | **+0.237 (0.044)** | 19.077 | 9.211 | 77% | +0.238 (0.044) | 18.397 | +0.603 (0.112) | **0.399** |
| SE (intermittent)† | -6.855 | **+1.005 (0.147)** | 20.199 | 7.980 | 80% | +1.155 (0.168) | 16.818 | +2.056 (0.300) | **4.262** |
| SE (frequent)† | -10.966 | **+4.046 (0.369)** | **27.483** | 5.774 | 40% | +4.256 (0.388) | 30.005 | +6.168 (0.562) | 38.075 |
| DE (early) | -2.370 | **-0.022 (0.009)** | 1.632 | 3.057 | 91% | -0.053 (0.022) | 1.624 | +0.455 (0.192) | **0.219** |
| TE (early) | -4.998 | **+0.322 (0.064)** | **1.367** | 2.927 | 91% | +0.328 (0.066) | 1.414 | -1.716 (0.343) | 3.002 |
| DE (late) | -16.734 | **+0.120 (0.007)** | 8.959 | 6.534 | 83% | +0.124 (0.007) | 8.665 | +2.022 (0.121) | **4.102** |
| TE (late) | -21.392 | **+0.395 (0.018)** | 9.715 | 6.494 | 80% | +0.513 (0.024) | 9.777 | +1.881 (0.088) | **3.591** |
| DE (intermittent) | -20.474 | -0.756 (0.037) | **7.543** | 5.800 | 82% | **-0.704 (0.034)** | 7.765 | +4.199 (0.205) | 17.654 |
| TE (intermittent) | -25.346 | -0.153 (0.006) | **8.001** | 6.217 | 83% | **-0.145 (0.006)** | 8.605 | +4.272 (0.169) | 18.314 |
| DE (frequent) | -33.261 | -2.011 (0.060) | **14.103** | 5.943 | 61% | **-1.903 (0.057)** | 16.430 | -3.959 (0.119) | 15.736 |
| TE (frequent)† | -41.353 | +0.357 (0.009) | 8.207 | 5.002 | 64% | **+0.338 (0.008)** | 9.164 | -0.667 (0.016) | **0.552** |

*Note:* GT denotes the Monte Carlo ground-truth value. Bias is the mean estimation error, MSE is the mean squared estimation error, and ARB is absolute bias divided by the absolute GT value. Bold identifies the method with the smallest absolute bias (and ARB) or MSE for each target. Width and coverage refer to the proposed estimator's 95% percentile intervals obtained from $B_{\text{boot}} = 100$ network-block bootstrap resamples. The oracle-propensity and OLS benchmarks are reported for point estimation only. SE, DE, and TE denote spillover, direct, and total effects, respectively. A † marks targets for which insufficient local support prevents calculation of the proposed estimate in more than 10% of replications. Point-estimation summaries use replications with valid estimates; interval width and coverage use replications with valid intervals.

Table S3.5: Estimation of mean potential outcomes and causal effects at $n = 5000$ and $h = 0.04$, using $R = 100$ Monte Carlo replications.

| | | Proposed | | | | Oracle propensity | | OLS | |
|---|---|---|---|---|---|---|---|---|---|
| Target | GT | Bias (ARB) | MSE | Width | Cov. | Bias (ARB) | MSE | Bias (ARB) | MSE |
| *Panel A: Mean potential outcomes* | | | | | | | | | |
| never (low) | 90.397 | -0.138 (0.002) | **0.258** | 1.262 | 79% | **-0.108 (0.001)** | 0.390 | -0.585 (0.006) | 0.363 |
| never (mid) | 89.597 | -0.138 (0.002) | 0.254 | 1.224 | 86% | **-0.122 (0.001)** | **0.226** | -1.308 (0.015) | 1.721 |
| never (high) | 88.165 | +0.037 (0.000) | **0.324** | 1.383 | 93% | **+0.017 (0.000)** | 0.370 | -3.151 (0.036) | 9.943 |
| early (low) | 88.467 | -0.277 (0.003) | 0.807 | 1.845 | 80% | **-0.239 (0.003)** | 0.706 | -0.570 (0.006) | **0.342** |
| early (mid) | 87.227 | **-0.154 (0.002)** | 0.499 | 1.675 | 88% | -0.168 (0.002) | **0.456** | -0.853 (0.010) | 0.738 |
| early (high) | 85.400 | +0.256 (0.003) | 0.761 | 1.734 | 90% | **+0.249 (0.003)** | **0.724** | -2.301 (0.027) | 5.313 |
| late (low) | 74.407 | **+0.062 (0.001)** | 6.271 | 5.182 | 75% | +0.209 (0.003) | 7.027 | +0.694 (0.009) | **0.508** |
| late (mid) | 72.863 | **+0.021 (0.000)** | 5.061 | 4.840 | 79% | +0.031 (0.000) | 5.167 | +0.714 (0.010) | **0.527** |
| late (high) | 69.005 | +0.342 (0.005) | 6.281 | 5.234 | 79% | **+0.238 (0.003)** | 6.321 | +1.297 (0.019) | **1.702** |
| intermittent (low) | 71.906 | -1.422 (0.020) | 7.169 | 4.867 | 58% | **-1.324 (0.018)** | 8.812 | +1.631 (0.023) | **2.689** |
| intermittent (mid) | 69.123 | -0.886 (0.013) | **4.368** | 4.298 | 73% | **-0.611 (0.009)** | 4.763 | +2.890 (0.042) | 8.371 |
| intermittent (high) | 65.051 | **-0.017 (0.000)** | **5.338** | 5.005 | 81% | +0.025 (0.000) | 6.396 | +3.687 (0.057) | 13.618 |
| frequent (low) | 60.011 | **-3.135 (0.052)** | **15.364** | 5.908 | 41% | -3.139 (0.052) | 16.499 | -7.419 (0.124) | 55.079 |
| frequent (mid) | 56.336 | -1.984 (0.035) | **13.405** | 5.771 | 54% | **-1.758 (0.031)** | 14.296 | -5.267 (0.094) | 27.779 |
| frequent (high) | 49.045 | **+0.145 (0.003)** | 6.278 | 5.182 | 82% | +0.499 (0.010) | 8.547 | -1.251 (0.026) | **1.608** |
| *Panel B: Causal contrasts* | | | | | | | | | |
| SE (never) | -2.233 | +0.175 (0.078) | **0.520** | 2.018 | 86% | **+0.125 (0.056)** | 0.690 | -2.566 (1.150) | 6.622 |
| SE (early) | -3.068 | +0.533 (0.174) | 1.711 | 2.884 | 90% | **+0.487 (0.159)** | **1.534** | -1.731 (0.564) | 3.033 |
| SE (late) | -5.402 | +0.280 (0.052) | 13.437 | 8.269 | 83% | **+0.028 (0.005)** | 13.256 | +0.603 (0.112) | **0.399** |
| SE (intermittent) | -6.855 | +1.404 (0.205) | 13.084 | 8.060 | 83% | **+1.348 (0.197)** | 15.344 | +2.056 (0.300) | **4.262** |
| SE (frequent) | -10.966 | **+3.280 (0.299)** | **21.251** | 9.346 | 62% | +3.638 (0.332) | 26.346 | +6.168 (0.562) | 38.075 |
| DE (early) | -2.370 | **-0.016 (0.007)** | 0.680 | 2.230 | 89% | -0.046 (0.019) | 0.600 | +0.455 (0.192) | **0.219** |
| TE (early) | -4.998 | +0.394 (0.079) | 1.011 | 2.365 | 86% | **+0.356 (0.071)** | **0.970** | -1.716 (0.343) | 3.002 |
| DE (late) | -16.734 | +0.159 (0.010) | 5.603 | 5.173 | 84% | **+0.153 (0.009)** | 5.624 | +2.022 (0.121) | **4.102** |
| TE (late) | -21.392 | +0.480 (0.022) | 6.690 | 5.653 | 79% | **+0.345 (0.016)** | 6.848 | +1.881 (0.088) | **3.591** |
| DE (intermittent) | -20.474 | -0.748 (0.037) | **4.452** | 4.658 | 80% | **-0.488 (0.024)** | 4.958 | +4.199 (0.205) | 17.654 |
| TE (intermittent) | -25.346 | **+0.121 (0.005)** | **5.897** | 5.379 | 80% | +0.132 (0.005) | 7.116 | +4.272 (0.169) | 18.314 |
| DE (frequent) | -33.261 | -1.846 (0.056) | **12.872** | 6.229 | 60% | **-1.635 (0.049)** | 14.272 | -3.959 (0.119) | 15.736 |
| TE (frequent) | -41.353 | **+0.283 (0.007)** | 6.684 | 5.662 | 87% | +0.606 (0.015) | 9.009 | -0.667 (0.016) | **0.552** |

*Note:* Metric definitions, boldface conventions, and interval construction follow Table S3.4. All 100 replications contribute to the reported summaries.

Table S3.6: Estimation of mean potential outcomes and causal effects at $n = 5000$ and $h = 0.05$, using $R = 100$ Monte Carlo replications.

| | | Proposed | | | | Oracle propensity | | OLS | |
|---|---|---|---|---|---|---|---|---|---|
| Target | GT | Bias (ARB) | MSE | Width | Cov. | Bias (ARB) | MSE | Bias (ARB) | MSE |
| *Panel A: Mean potential outcomes* | | | | | | | | | |
| never (low) | 90.397 | -0.167 (0.002) | **0.226** | 1.122 | 81% | **-0.124 (0.001)** | 0.333 | -0.585 (0.006) | 0.363 |
| never (mid) | 89.597 | -0.124 (0.001) | 0.148 | 1.001 | 88% | **-0.103 (0.001)** | **0.126** | -1.308 (0.015) | 1.721 |
| never (high) | 88.165 | +0.024 (0.000) | **0.218** | 1.226 | 92% | **+0.008 (0.000)** | 0.267 | -3.151 (0.036) | 9.943 |
| early (low) | 88.467 | -0.314 (0.004) | 0.670 | 1.697 | 81% | **-0.265 (0.003)** | 0.537 | -0.570 (0.006) | **0.342** |
| early (mid) | 87.227 | **-0.120 (0.001)** | 0.327 | 1.424 | 89% | -0.126 (0.001) | **0.290** | -0.853 (0.010) | 0.738 |
| early (high) | 85.400 | +0.259 (0.003) | 0.623 | 1.599 | 84% | **+0.244 (0.003)** | **0.560** | -2.301 (0.027) | 5.313 |
| late (low) | 74.407 | **+0.139 (0.002)** | 5.136 | 4.874 | 79% | +0.187 (0.003) | 5.889 | +0.694 (0.009) | **0.508** |
| late (mid) | 72.863 | **-0.039 (0.001)** | 3.540 | 4.337 | 83% | -0.080 (0.001) | 3.633 | +0.714 (0.010) | **0.527** |
| late (high) | 69.005 | +0.419 (0.006) | 5.422 | 4.926 | 80% | **+0.205 (0.003)** | 5.169 | +1.297 (0.019) | **1.702** |
| intermittent (low) | 71.906 | -1.413 (0.020) | 6.475 | 4.596 | 58% | **-1.271 (0.018)** | 8.026 | +1.631 (0.023) | **2.689** |
| intermittent (mid) | 69.123 | -0.862 (0.012) | **3.400** | 3.890 | 67% | **-0.477 (0.007)** | 3.841 | +2.890 (0.042) | 8.371 |
| intermittent (high) | 65.051 | -0.089 (0.001) | **4.350** | 4.693 | 83% | **+0.053 (0.001)** | 5.948 | +3.687 (0.057) | 13.618 |
| frequent (low) | 60.011 | **-3.100 (0.052)** | **15.210** | 5.726 | 42% | -3.155 (0.053) | 16.459 | -7.419 (0.124) | 55.079 |
| frequent (mid) | 56.336 | -1.888 (0.034) | **12.534** | 5.567 | 55% | **-1.566 (0.028)** | 13.302 | -5.267 (0.094) | 27.779 |
| frequent (high) | 49.045 | **+0.136 (0.003)** | 5.564 | 5.004 | 80% | +0.608 (0.012) | 8.683 | -1.251 (0.026) | **1.608** |
| *Panel B: Causal contrasts* | | | | | | | | | |
| SE (never) | -2.233 | +0.191 (0.086) | **0.429** | 1.751 | 83% | **+0.133 (0.059)** | 0.582 | -2.566 (1.150) | 6.622 |
| SE (early) | -3.068 | +0.573 (0.187) | 1.455 | 2.581 | 83% | **+0.509 (0.166)** | **1.205** | -1.731 (0.564) | 3.033 |
| SE (late) | -5.402 | +0.279 (0.052) | 11.206 | 7.607 | 82% | **+0.018 (0.003)** | 11.037 | +0.603 (0.112) | **0.399** |
| SE (intermittent) | -6.855 | **+1.324 (0.193)** | 11.358 | 7.430 | 82% | +1.324 (0.193) | 13.730 | +2.056 (0.300) | **4.262** |
| SE (frequent) | -10.966 | **+3.235 (0.295)** | **20.627** | 8.861 | 53% | +3.762 (0.343) | 27.410 | +6.168 (0.562) | 38.075 |
| DE (early) | -2.370 | **+0.004 (0.002)** | 0.423 | 1.842 | 87% | -0.023 (0.010) | 0.363 | +0.455 (0.192) | **0.219** |
| TE (early) | -4.998 | +0.426 (0.085) | 0.870 | 2.110 | 84% | **+0.368 (0.074)** | **0.778** | -1.716 (0.343) | 3.002 |
| DE (late) | -16.734 | +0.085 (0.005) | **3.769** | 4.526 | 87% | **+0.023 (0.001)** | 3.782 | +2.022 (0.121) | 4.102 |
| TE (late) | -21.392 | +0.585 (0.027) | 5.964 | 5.249 | 84% | **+0.329 (0.015)** | 5.648 | +1.881 (0.088) | **3.591** |
| DE (intermittent) | -20.474 | -0.739 (0.036) | **3.380** | 4.132 | 77% | **-0.373 (0.018)** | 4.016 | +4.199 (0.205) | 17.654 |
| TE (intermittent) | -25.346 | **+0.077 (0.003)** | **4.788** | 4.974 | 86% | +0.177 (0.007) | 6.584 | +4.272 (0.169) | 18.314 |
| DE (frequent) | -33.261 | -1.764 (0.053) | **12.192** | 5.844 | 57% | **-1.462 (0.044)** | 13.449 | -3.959 (0.119) | 15.736 |
| TE (frequent) | -41.353 | **+0.302 (0.007)** | 6.062 | 5.354 | 84% | +0.732 (0.018) | 9.205 | -0.667 (0.016) | **0.552** |

*Note:* Metric definitions, boldface conventions, and interval construction follow Table S3.4. All 100 replications contribute to the reported summaries.

We summarize directional sensitivity using the tipping point $\Gamma^*$ defined in Section S1.3, the smallest value on the prespecified grid at which the causal-effect sensitivity interval contains zero. If no tested value gives an interval containing zero, we report $\Gamma^* > 5$.

**Sensitivity of effect magnitudes and directions.** Increasing $\Gamma$ widens the bounds on causal effects and mean potential outcomes at both sample sizes (Figures S3.8 and S3.9). For $\Gamma > 1$, the figures show wider average sensitivity intervals at $n = 50000$ than at $n = 5000$. Effect magnitudes remain sensitive even when the bounds stay negative: stability of direction does not imply stability of magnitude.

The early direct effect is sensitive in direction: its interval includes zero in 52% of replications at $\Gamma = 2$ at $n = 5000$ and in 97% at $n = 50000$, and in all replications at $\Gamma = 5$ at both sizes. For the never, early, late, and intermittent paths, the median tipping point for the spillover effect falls from $\Gamma^* = 3$ to 2, from 5 to 2, from 3 to 1.5, and from 5 to 2, respectively, as the sample size increases (Figure S3.7; Table S3.7). The late, intermittent and frequent direct and total effects remain negative in every valid replication through $\Gamma = 5$ at both sizes. The early total effect also becomes more sensitive: it includes zero in 5% of replications at $\Gamma = 5$ at $n = 5000$ and in 65% at $n = 50000$.

**Local support and interpretation.** The frequent-path results are interpreted as support stress tests, consistent with Section S3.1.5. In these sensitivity runs, the median ESS for the frequent path, pooled over the 100 replications and three spillover levels, is 1.0 at $n = 5000$ and 2.1 at $n = 50000$. The corresponding median largest normalized weights are 0.98 and 0.64. These diagnostics concern the fitted weights on which the sensitivity bounds are conditional; a narrow sensitivity interval alone does not establish adequate local support.

Thus, allowing unmeasured confounding weakens conclusions about the early direct effect and the spillover effects more readily than those about the late, intermittent and frequent direct and total effects, and does so more at $n = 50000$. This comparison is conditional on the specified reweighting model; it does not establish unbiasedness under confounding. Zero inclusion means

Table S3.7: Sensitivity results. DE, SE and TE denote direct, spillover and total effects. Columns at fixed $\Gamma$ give percentages of intervals containing zero among $R_{\text{valid}}$ valid replications. Values of $\Gamma^* > 5$ are right-censored. At $n = 5000$, two frequent/low mean estimates were excluded because their unnormalized weight sums were below $10^{-8}$.

| | | $n = 5000$ | | | | $n = 50000$ | | | |
|---|---|---|---|---|---|---|---|---|---|
| Effect | Path | $R_{\text{valid}}$ | Median $\Gamma^*$ | $\Gamma = 2$ | $\Gamma = 5$ | $R_{\text{valid}}$ | Median $\Gamma^*$ | $\Gamma = 2$ | $\Gamma = 5$ |
| DE | early | 100 | 2 | 52 | 100 | 100 | 2 | 97 | 100 |
| | late | 100 | $> 5$ | 0 | 0 | 100 | $> 5$ | 0 | 0 |
| | intermittent | 100 | $> 5$ | 0 | 0 | 100 | $> 5$ | 0 | 0 |
| | frequent | 100 | $> 5$ | 0 | 0 | 100 | $> 5$ | 0 | 0 |
| SE | never | 100 | 3 | 48 | 97 | 100 | 2 | 94 | 100 |
| | early | 100 | 5 | 22 | 61 | 100 | 2 | 61 | 100 |
| | late | 100 | 3 | 46 | 89 | 100 | 1.5 | 89 | 100 |
| | intermittent | 100 | 5 | 31 | 67 | 100 | 2 | 65 | 100 |
| | frequent | 98 | $> 5$ | 7 | 26 | 100 | 5 | 24 | 76 |
| TE | early | 100 | $> 5$ | 0 | 5 | 100 | 5 | 1 | 65 |
| | late | 100 | $> 5$ | 0 | 0 | 100 | $> 5$ | 0 | 0 |
| | intermittent | 100 | $> 5$ | 0 | 0 | 100 | $> 5$ | 0 | 0 |
| | frequent | 100 | $> 5$ | 0 | 0 | 100 | $> 5$ | 0 | 0 |

that the conservative sensitivity bounds no longer rule out a null effect, not that a sampling confidence interval includes zero.

#### S3.1.8 Stress Tests under Burden-Dependent Spillover Responses and Alternative Treatment Interactions

**Alternative data-generating mechanisms.** We conduct additional numerical experiments to examine estimation under burden-dependent spillover benefits and alternative own-by-spillover treatment interactions. The network, baseline variables, covariate dynamics, and observed outcome updates follow the mechanisms in Section S3.1.1. In the latent-state equation, we replace the terms for accumulated spillover exposure and its interaction with accumulated own-treatment exposure

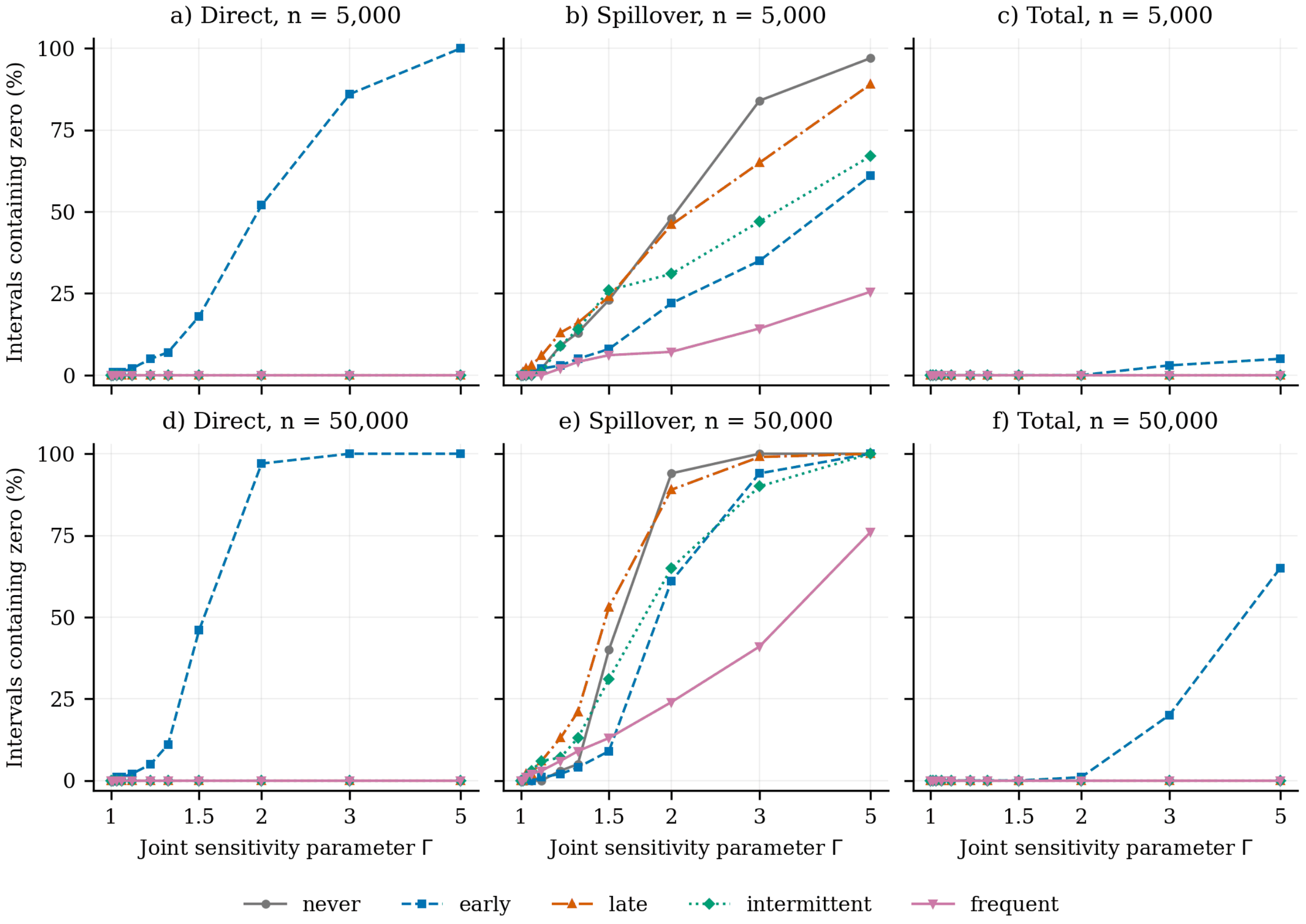


Figure S3.7: Percentage of the proposed estimator's sensitivity intervals containing zero, by joint sensitivity parameter Γ and own-treatment path. Rows give the sample sizes and columns the direct, spillover and total effects. Denominators are the valid-replication counts in Table S3.7.

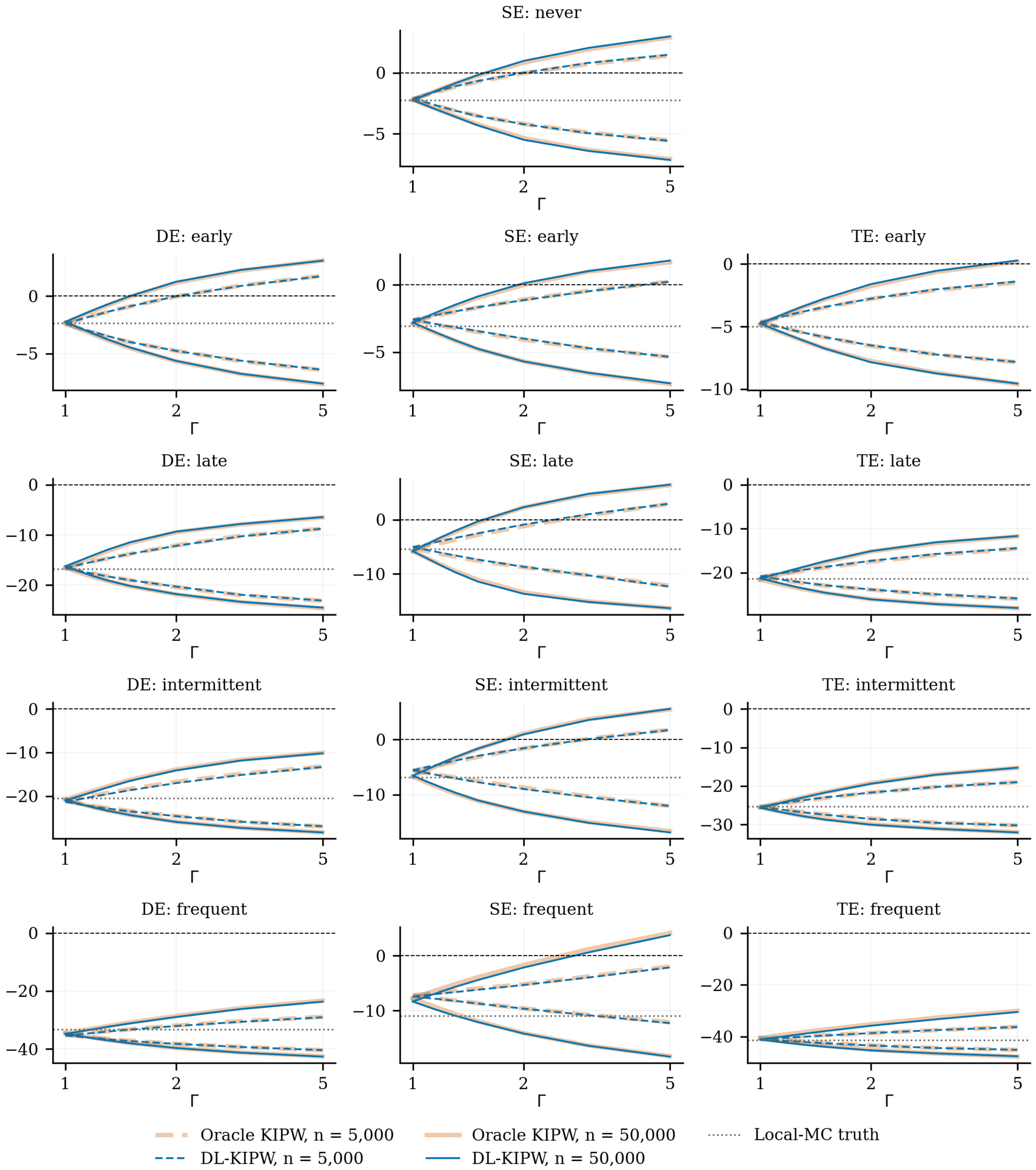


Figure S3.8: Mean sensitivity endpoints for the 13 causal effects, averaged over valid replications. Dashed lines are $n = 5000$ and solid lines $n = 50000$. Wide translucent orange lines use oracle weights (legend: Oracle KIPW) and thin blue lines use the proposed estimator's fitted weights (legend: DL-KIPW), drawn on top because the two nearly coincide. Dotted lines mark the approximate truth from local structural Monte Carlo (legend: Local-MC truth), which is common to both sample sizes. The black dashed horizontal line marks zero.

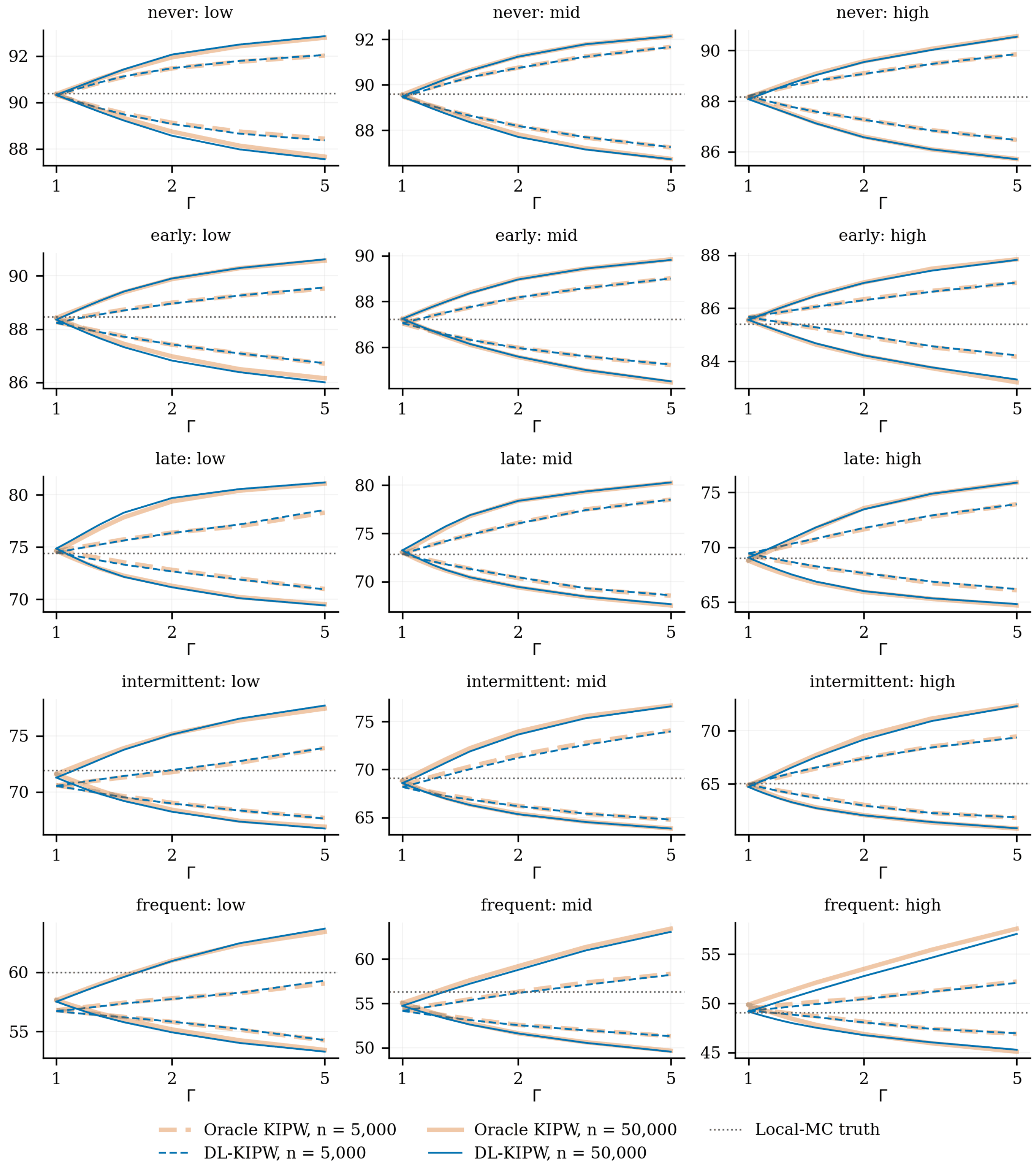


Figure S3.9: Mean sensitivity endpoints for the 15 potential-outcome means, averaged over valid replications. Dashed lines are $n = 5000$ and solid lines $n = 50000$. Wide translucent orange lines use oracle weights (legend: Oracle KIPW) and thin blue lines use the proposed estimator's fitted weights (legend: DL-KIPW), drawn on top because the two nearly coincide. Dotted lines mark the approximate truth from local structural Monte Carlo, which is common to both sample sizes.

by

$$\text{burden-modified spillover-treatment term} : -\beta_{XS}\{1+\eta_B(2q_{1,i,t}-1)\}\left(1-e^{-a_X\widetilde{D}_{i,t}}\right)$$
$$\text{own-treatment–spillover interaction term} : -\beta_{XD}\left(1-e^{-a_X\widetilde{X}_{i,t}}\right)\left(1-e^{-a_X\widetilde{D}_{i,t}}\right).$$

Here, $q_{1,i,t}$, $\widetilde{X}_{i,t}$, and $\widetilde{D}_{i,t}$ retain their definitions in Section S3.1.1. The parameter $\eta_B$ allows the spillover-treatment benefit to increase with the focal unit's lagged covariate burden. When $\eta_B = 0.75$, the corresponding multiplier lies between 0.25 and 1.75, preserving the direction of the spillover-treatment benefit.

We consider three configurations: burden-dependent spillover benefits without the own-by-spillover interaction (DGP A); the own-by-spillover interaction without burden-dependent effect modification (DGP B); and both mechanisms together (DGP C). Their parameter values are reported in Table S3.8. All three configurations use the treatment-assignment equation in Section S3.1.1 with $\theta_{W2} = 0$, retaining the linear worsening term and the other assignment components.

**Comparison settings and results.** Each configuration uses $n = 5000$, $R = 20$ replications, and an independently generated validation sample of 500 units per replication. The five representative own-treatment paths and the definitions of the causal contrasts remain unchanged. These experiments use a separate set of fixed reference spillover paths, shared across the three configurations and all replications. With entries rounded to three decimal places, these are

$$d^L = (0.200, 0.250, 0.111, 0.200),$$
$$d^M = (0.333, 0.375, 0.286, 0.300),$$
$$d^H = (0.429, 0.375, 0.333, 0.375).$$

We compare the proposed estimator, its oracle-propensity counterpart, and OLS as described in Section S3.1.3. The two weighting estimators use Gaussian localization with $h = 0.03$ and no truncation of the final localized weights. For each configuration and target regime, the Monte Carlo reference mean is obtained from 10,000 two-hop local structural draws under the corresponding

DGP.

Table S3.8 summarizes performance over the ten primary causal contrasts, excluding the three frequent-path contrasts. The proposed estimator has smaller average absolute bias than OLS under all three alternative DGPs, whereas OLS has smaller average MSE because of its low variance, as in the primary design.

Table S3.8: Point-estimation performance under alternative DGPs at $n = 5000$, $R = 20$, and $h = 0.03$.

| | | | Proposed | | Oracle-propensity | | OLS | |
|---|---|---|---|---|---|---|---|---|
| DGP | $\eta_B$ | $\beta_{XD}$ | Abs. bias | MSE | Abs. bias | MSE | Abs. bias | MSE |
| A | 0.75 | 0 | 0.370 | 7.317 | 0.317 | 6.882 | 1.314 | 2.470 |
| B | 0 | 4 | 0.567 | 5.927 | 0.580 | 5.961 | 1.425 | 3.368 |
| C | 0.75 | 4 | 0.357 | 5.359 | 0.234 | 5.041 | 1.464 | 3.542 |

*Note:* Bias and MSE are first calculated across replications for each contrast using its DGP-specific Monte Carlo reference value. The table reports the average absolute bias and average MSE over the ten primary contrasts, excluding the three frequent-path contrasts. All three DGPs use $\theta_{W2} = 0$.

## S3.2 COVID-19 Data Construction, Exposure Patterns, and Support Diagnostics

### S3.2.1 Data Sources, Policy Coding, and County-Network Construction

**Data sources and study panel.** We analyze a county-by-time panel covering 3108 counties across 48 U.S. states (excluding Alaska and Hawaii) and the District of Columbia (49 state-level units). Policies are implemented at the state level, whereas health outcomes and related covariates are observed at the county level.

COVID-19 case and death data are collected from the CDC, and policy data are collected from the Boston University COVID-19 State Policy Database. These records are linked to county covariates compiled from County Health Rankings & Roadmaps, county election records, and NOAA temperature data. County records are linked using five-digit FIPS codes, and state-level

policy measures are assigned to counties within the corresponding state. The reported analyses cover April to October 2020, with weekly outcome observations and monthly policy decisions.

**Policy categories.** We distinguish seven named state-level policy domains: Healthcare System, Masking, Business and Economic Restrictions, Reopening Phase, Education and Childcare, Gatherings or Venues, and Emergency or Governance. An additional Other or Uncategorized category contains other policy types not assigned to these domains. Starting from the raw policy records, we map each policy entry to the defined category label (see Table S3.9).

Table S3.9: Mapping from policy types to policy categories.

| **Policy category** | **Policy types** |
|---|---|
| Masking | • mask requirement<br>• mandate face mask use by all individuals in public facing businesses<br>• mandate face mask use by all individuals in public spaces |
| Business and Economic Restrictions | • non-essential businesses<br>• manufacturing<br>• construction<br>• wholesale trade<br>• agriculture<br>• personal care<br>• gyms<br>• bars<br>• food and drink<br>• entertainment |

Continued on next page.

| Policy category | Policy types (examples / raw labels) |
|---|---|
| Healthcare System | • medical<br>• nursing homes<br>• nursing home visitations<br>• alternative care facilities<br>• resumed elective medical procedures<br>• suspended elective medical dental procedures<br>• allow audio only telehealth<br>• allow expand medicaid telehealth coverage<br>• modify medicaid requirements with 1135 waivers (date of CMS approval)<br>• reopened ACA enrollment using a special enrollment period |
| Education and Childcare | • childcare (k-12)<br>• day care<br>• education<br>• higher education<br>• colleges & universities<br>• day camps / overnight camps<br>• residential overnight camps<br>• training jobs |
| Reopening Phase | • phase 1–5<br>• new phase<br>• phase 2, 3<br>• phase 3 step 2 |

Continued on next page.

| Policy category | Policy types (examples / raw labels) |
|---|---|
| Gatherings or Venues | • public gathering<br>• gatherings<br>• outdoor and recreation<br>• houses of worship<br>• museums<br>• libraries<br>• graduation<br>• graduation ceremony guidelines |
| Emergency or Governance | • state of emergency<br>• state of emergency / funds<br>• executive order<br>• updated guidelines<br>• election<br>• public health advisory system<br>• health risk status |
| Other or Uncategorized | Other policy types not assigned to the categories above. |

*Notes:* Policy types are grouped into categories based on their primary intent to mitigate COVID-19 transmission (non-pharmaceutical interventions; NPIs). Categories such as *Masking*, *Business and Economic Restrictions*, *Gatherings or Venues*, *Education and Childcare*, and selected *Healthcare System* measures are considered transmission-mitigating policies. The main analyses reported here focus on Business and Economic Restrictions and Education and Childcare. In contrast, *Emergency or Governance* and *Reopening Phase* largely reflect administrative actions or reopening milestones rather than direct mitigation and are therefore excluded from the primary exposure definition. *Other or Uncategorized* contains remaining labels not mapped above.

**State-month treatment intensity (0–1).** For each state, month, and policy category, we define a monthly treatment intensity as the average daily number of distinct active policy types over the

dates represented in that month's policy-activity records, divided by the total number of distinct policy types recorded across states in that category. We use this normalized average daily intensity as the state-month treatment measure.

**Binary treatment.** We construct a binary version of the treatment to study contrasts between relatively *high-intensity* and *low-intensity* policy months. Specifically, for each state and month, we compare the intensity to a pre-specified threshold of 0.2. If the intensity is at least 0.2, we label the state-month as treated ($X = 1$); otherwise we label it as untreated ($X = 0$). The same threshold is used for Business and Economic Restrictions and Education and Childcare. All counties within the same state share the same treatment status for a given month, and this status remains unchanged across that month's weekly observations.

**County network and spillover exposure.** County adjacency is defined using a geographic neighbor graph, and neighbors can be either within the same state or across state borders.

We construct this graph using the U.S. Census Bureau's county adjacency file, retaining connections among counties in the analysis panel. The network is held fixed throughout follow-up. For county $i$, spillover exposure at monthly decision time $\tau_m$ is the proportion of first-order neighbors assigned high-intensity treatment:

$$D_{i,\tau_m} = N_i^{-1} \sum_{j \in \mathcal{N}^1(i)} X_{j,\tau_m},$$

where $N_i = |\mathcal{N}^1(i)|$. Each neighbor receives equal weight, so $D_{i,\tau_m}$ takes values in $\{0, 1/N_i, \dots, 1\}$, consistent with the binomial component of the spillover-exposure model. Spillover exposure is set to zero for counties without neighbors.

**County-level outcomes.** The analysis outcome is the weekly case rate per 10,000 residents. For the descriptive distribution plots only, we also construct two monthly county-level rates:

(1) `new_cases_month_rate_10k`: new cases per month per 10,000 population;

(2) `new_deaths_month_rate_10k`: new deaths per month per 10,000 population.

The main text reports two outcome summaries for each endpoint month: the last-week case rate and the average weekly case rate within that month, both expressed per 10,000 residents. Weeks are assigned to months according to their week-ending dates.

**County-level covariates.** We adjust for baseline county characteristics capturing political environment, demographics, socioeconomic status, healthcare access/capacity, housing/environmental factors, baseline health risks, and movement patterns (see Table S3.10).

Moreover, the weekly death rate enters the adjustment set as a time-varying covariate. We also include monthly mean temperature as a time-varying covariate, with each month's temperature value repeated across the corresponding weekly observations. Adjustment further incorporates histories of case rates, own treatment, and spillover exposure, together with neighborhood summaries, as in the main analysis.

### S3.2.2 Missingness Patterns and Imputation

Missingness was concentrated in a small number of variables, primarily policy-related treatment measures (Figure S3.10). The largest missing shares were observed in Gatherings or Venues (40%), Reopening Phase (35%), Education and Childcare (33%), Masking (30%), Business and Economic Restrictions (23%), and Other or Uncategorized (20%). Mental Health Providers also showed moderate missingness (8%), whereas most other variables had little to no missing data.

The month-specific missingness pattern (Figure S3.11) was highest in March–April 2020 for Masking, Reopening Phase, Gatherings or Venues, and Education and Childcare, and reached its lowest level in May–June 2020 for most categories. Emergency or Governance and Healthcare System showed essentially no missingness across months. Only Masking remained low after June. For Business and Economic Restrictions, Gatherings or Venues, Reopening Phase, and Education and Childcare, missingness rose again from July onward. By December, missingness for Gatherings or Venues was above 50%, and missingness for Business and Economic Restrictions exceeded its March level. This indicates that missingness in policy treatment variables was temporally

Table S3.10: Baseline county-level covariates used for adjustment.

| **Domain** | **Covariates (county-level)** |
|---|---|
| Political / Governance | • Political affiliation / partisanship indicator |
| Demographics | • Total population<br>• Share of population aged 65 and older<br>• Rural population share |
| Socioeconomic Status | • Median household income<br>• Unemployment rate<br>• Children in poverty share<br>• Income inequality index |
| Healthcare Access / Capacity | • Uninsured share<br>• Other primary care provider supply<br>• Mental health provider supply<br>• Physical inactivity share |
| Housing / Environment | • Severe housing problems share<br>• Household overcrowding share<br>• Fine particulate matter ($PM_{2.5}$) air pollution |
| Baseline Health Risks | • Adult obesity prevalence<br>• Diabetes prevalence |
| Movement | • Driving alone to work share<br>• Long commute while driving alone share |

*Notes.* The term "other primary care providers" refers to primary care providers other than physicians, such as nurse practitioners and physician assistants.

structured, and likely reflected a combination of policy rollout timing, inconsistent early reporting, and evolving coding definitions.

For imputation, missing numeric covariate values, excluding the partisanship indicator, were replaced by the corresponding state-level mean, while missing treatment indicators were set to 0 by replacing missing policy intensities with zero before binarization. Missing partisanship indicators were filled using supplementary county-level presidential election records. Treatment imputation was particularly extensive in March–April 2020, when treatment coding was still evolving, especially for Masking, Reopening Phase, Gatherings or Venues, and Education and Childcare. It was also non-negligible in later months: for Business and Economic Restrictions, which uses September

and October endpoints, the share of missing policy intensities rose from July onward, so part of the untreated status in those months reflects zero imputation.

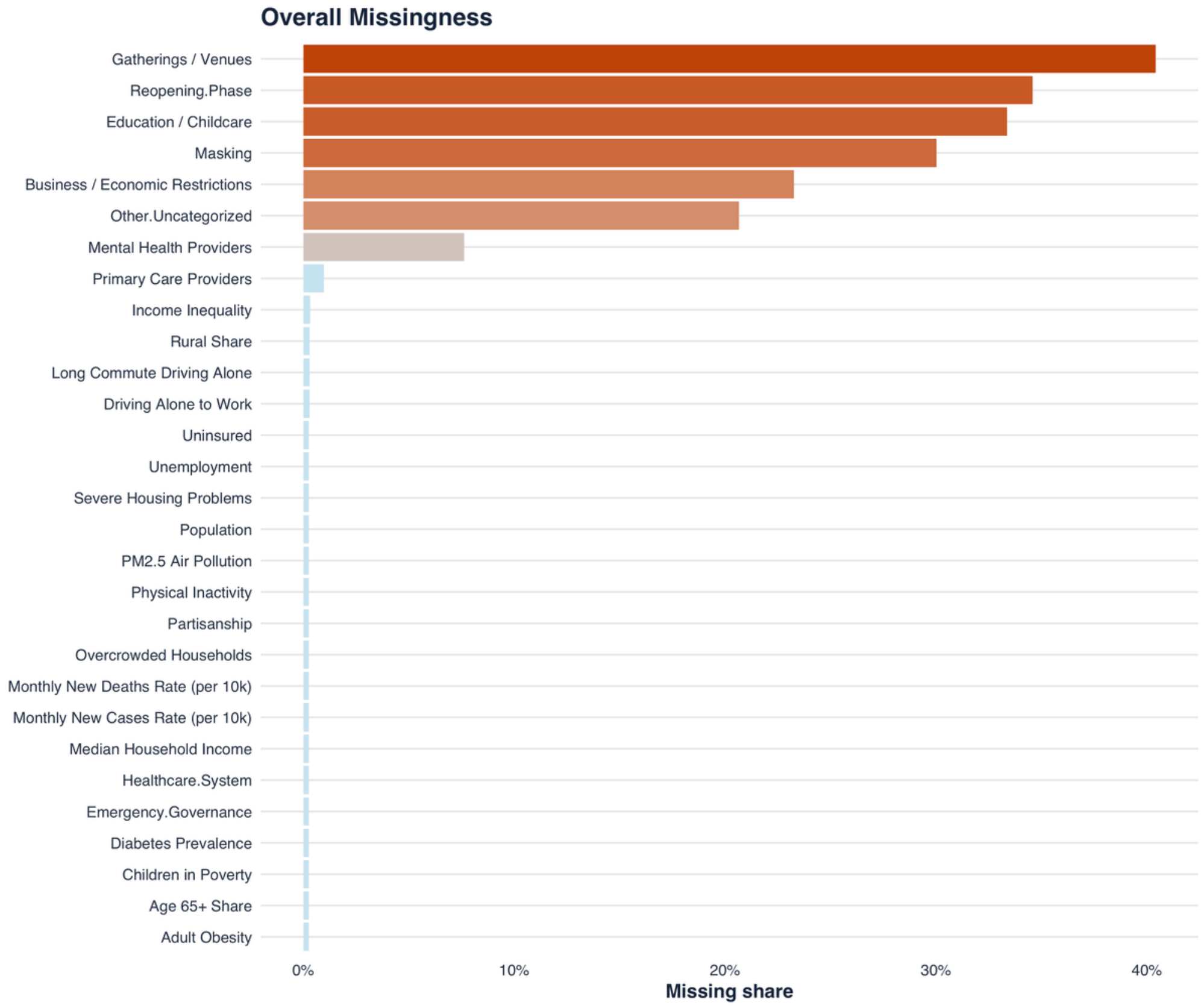


Figure S3.10: Missingness across county-level variables.

### S3.2.3 Temporal Patterns of Own and Neighboring Policy Exposure

Before binarization (Figure S3.12), the treatment variables exhibited clearly multimodal distributions, with concentrations around distinct intensity levels. These patterns motivate representing own treatment using discrete policy-intensity categories. The neighboring policy-intensity summaries shown in Figure S3.12 had distributions similar to those of own policy intensity, suggesting that neighboring policy environments were broadly aligned. These pre-binarization summaries are descriptive; the effect analysis uses binary own treatment and the equal-weight proportion of high-intensity neighbors defined above. The treatment distributions also varied across months,

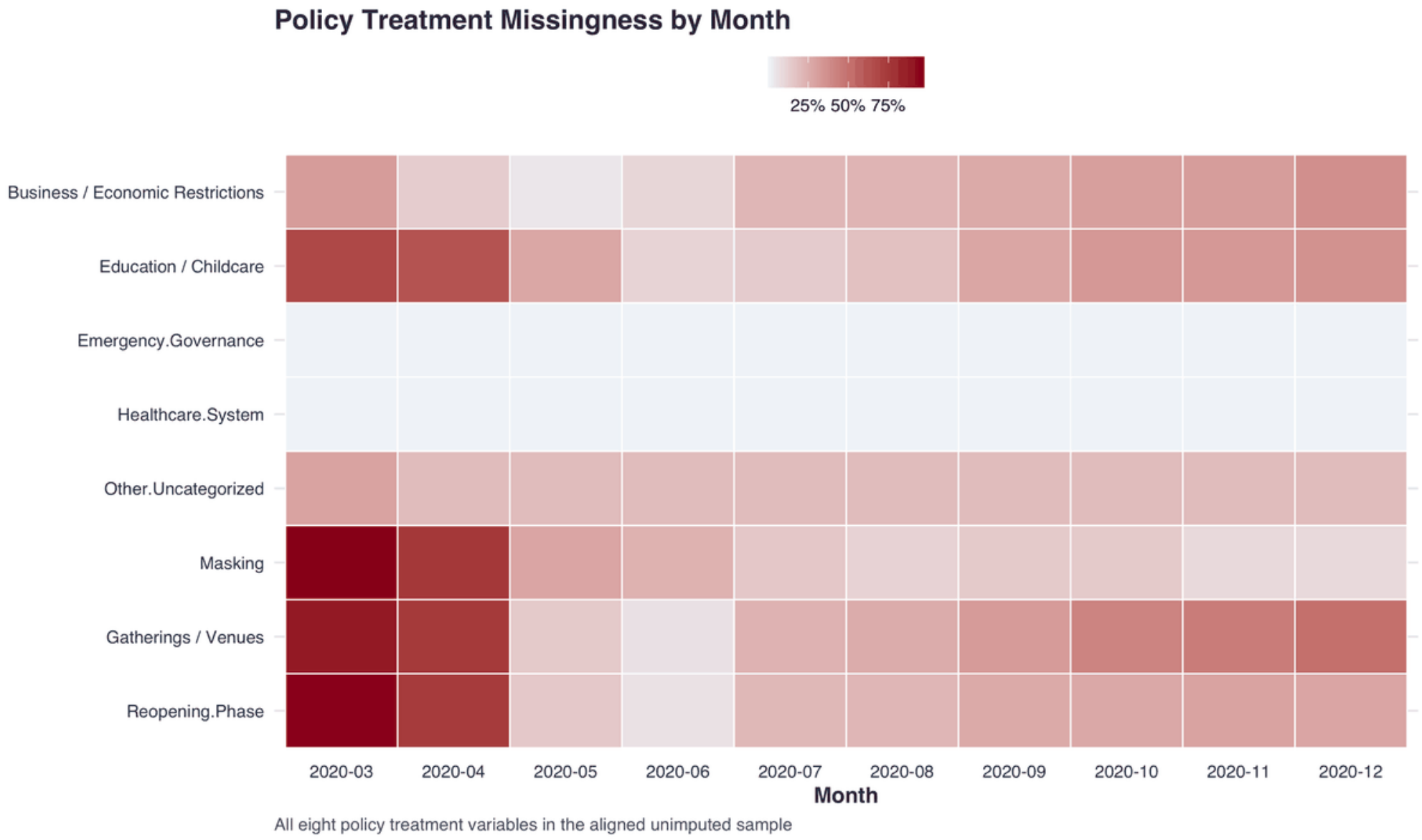


Figure S3.11: Policy treatment missingness by month. The figure displays March–December 2020, extending beyond the April–October 2020 period of the main analyses.

indicating that policy intensity changed over time during the study period. In addition, the distributional shapes differed across policy domains, suggesting heterogeneity in implementation patterns across different types of policies.

Binarization converted the treatment intensity variables into binary indicators using prespecified thresholds (Figure S3.13). Using the threshold 0.2 for Business and Economic Restrictions and Education and Childcare, the proportions of treated and untreated counties varied over time and differed across policy domains. Both treatment arms were present in every month, but the split was unbalanced in some months: treated counties were a small minority for Education and Childcare in March–April 2020, and untreated counties were a small minority for Business and Economic Restrictions in May 2020.

### S3.2.4 Outcome and Covariate Distributions across Months and Population Strata

We examine the distributions of county-level outcomes and covariates to characterize temporal variation and differences associated with county population size. Figures S3.14–S3.17 compare four

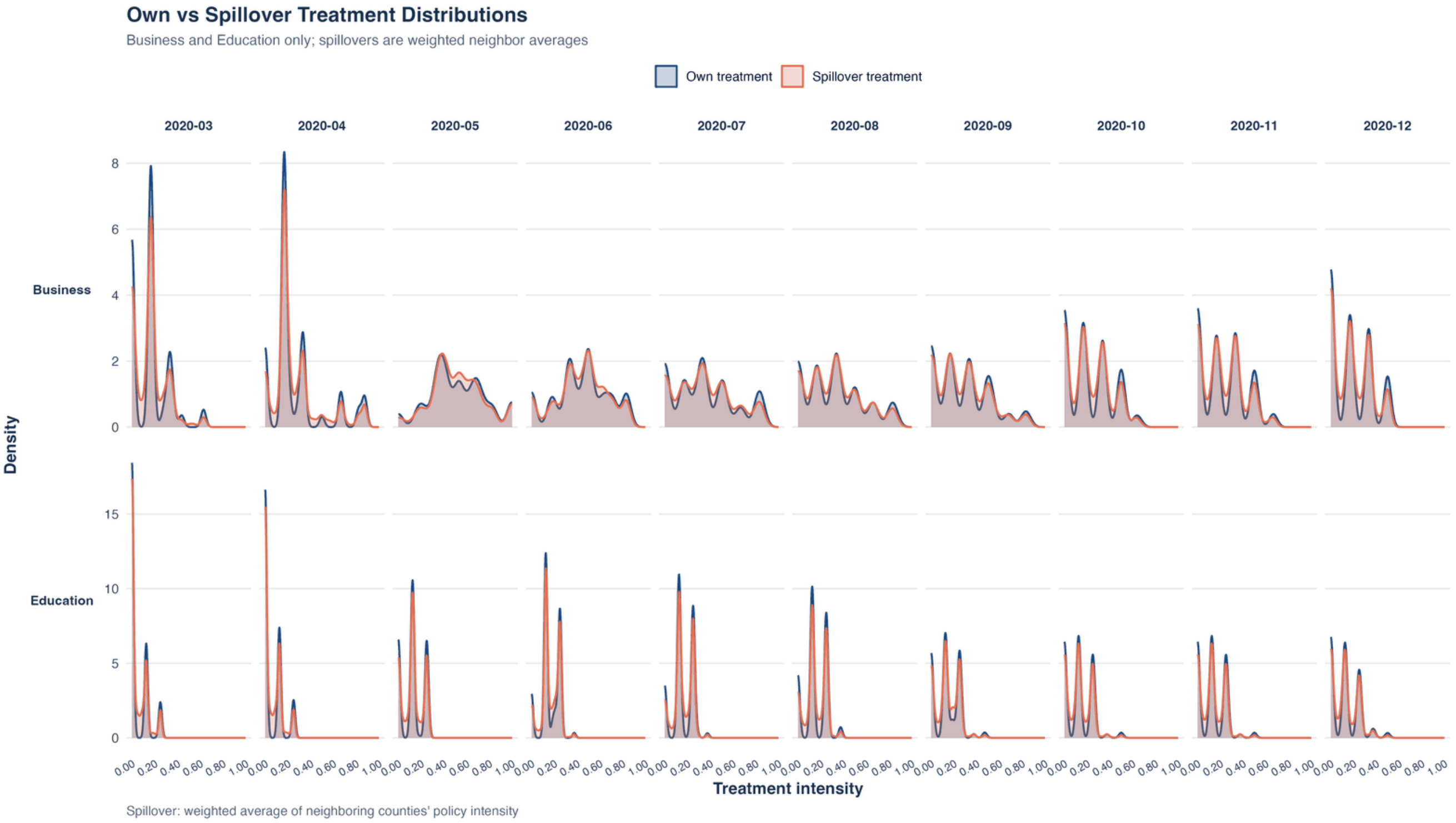


Figure S3.12: Own and neighboring policy-intensity distributions by month for Business and Economic Restrictions (top) and Education and Childcare (bottom), March–December 2020. Both own and neighboring intensities are shown before binarization for descriptive purposes; the effect analysis uses binary own treatment and the proportion of high-intensity neighbors.

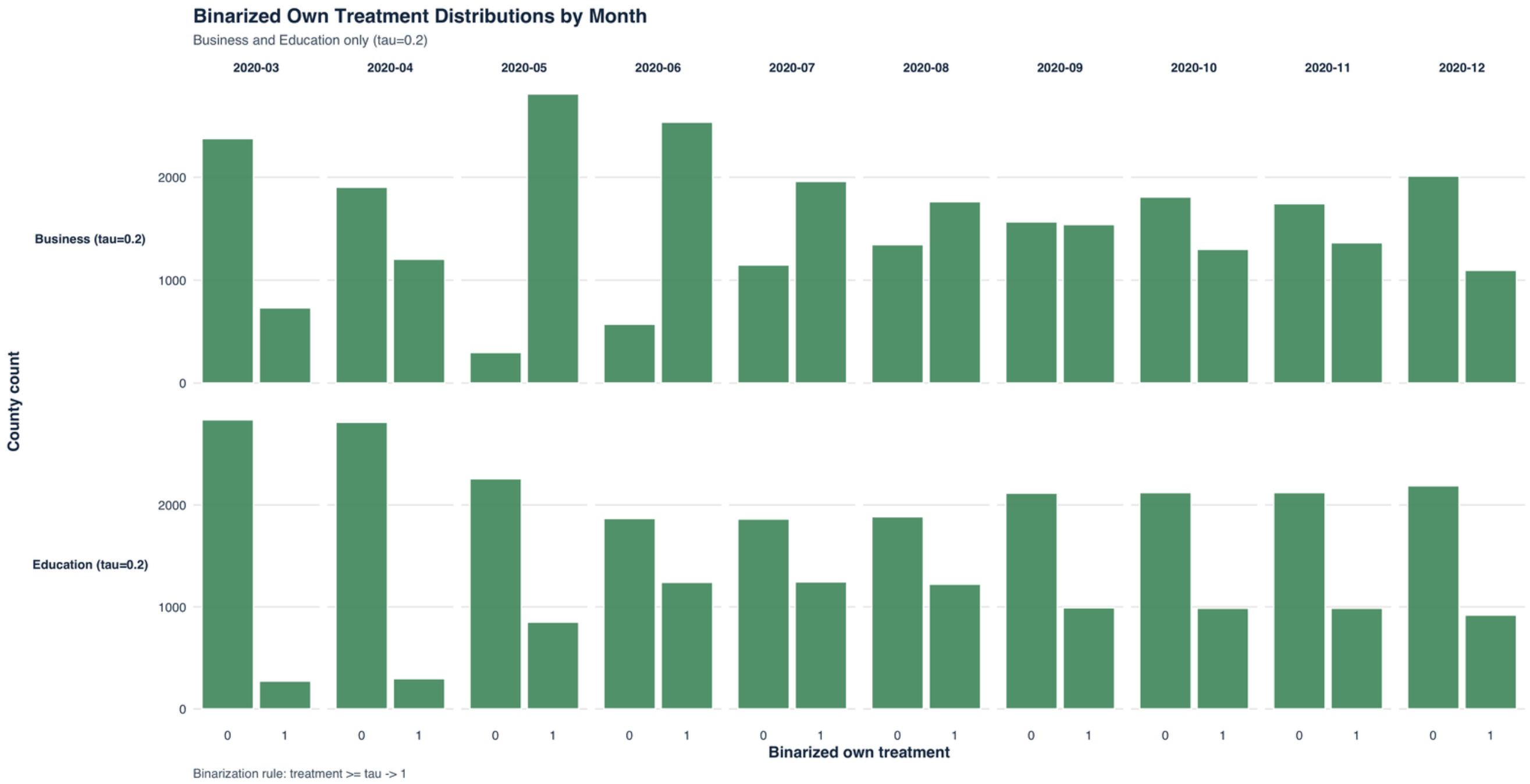


Figure S3.13: Monthly distributions of binary own treatment for Business and Economic Restrictions (top) and Education and Childcare (bottom), using the threshold 0.2 for both policy domains, March–December 2020.

population strata, labelled Q1 (small), Q2 (medium-small), Q3 (medium-large), and Q4 (large). Stratum membership is defined at the county level and held fixed across months.

**Outcome distributions.** Using the pseudo-log x-axis, the outcome distributions are more clearly visible near low values (Figure S3.14). Monthly New Cases Rate (per 10k) was more dispersed than Monthly New Deaths Rate (per 10k) and showed more pronounced month-to-month variation. Across the four population strata, the overall temporal pattern was broadly similar. The main stratification difference appeared in the case-rate distributions and changed over time. In April–June 2020, larger counties tended to be shifted slightly to the right and to have heavier upper tails, whereas smaller counties were more concentrated at low case-rate values. From September to November 2020, this ordering reversed: the largest-county stratum peaked to the left of the other strata, and the smallest-county stratum had the heaviest upper tail. In contrast, differences across population strata were smaller for death rates.

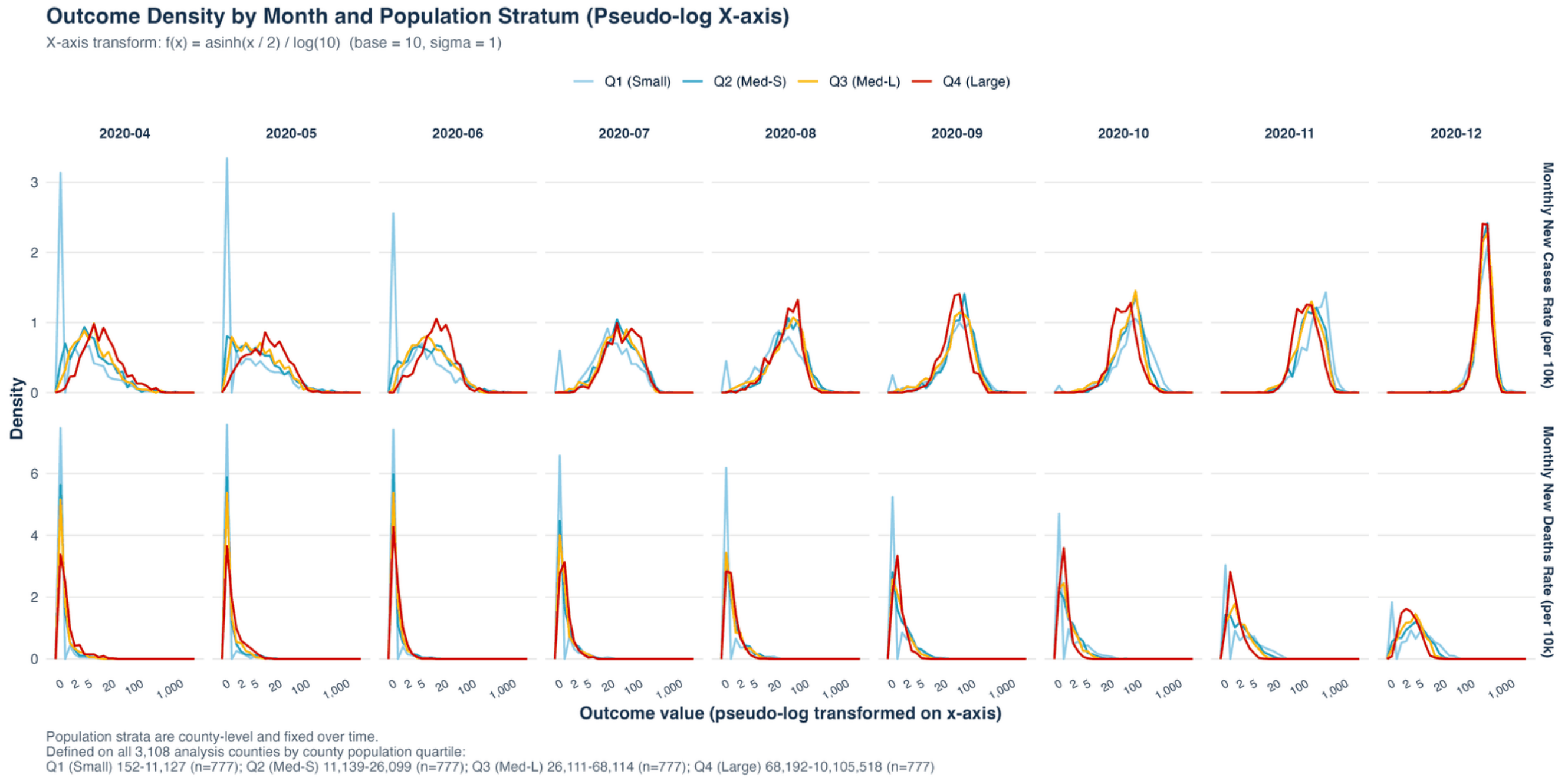


Figure S3.14: Outcome density distributions by month and population stratum. The figure displays April–December 2020.

**Covariate distributions.** Overall, several covariates showed clear stratification patterns, indicating that county size was associated with systematic demographic, socioeconomic, environmental, and political differences.

For the continuous county-level covariates (Figure S3.15), larger counties tended to have lower rural share, higher median household income, and lower shares of children in poverty and uninsured residents, whereas smaller counties were generally more rural. Socioeconomic disadvantage did not decrease monotonically with county size across the three smaller strata: the medium-small stratum (Q2) had the lowest median household income ($46,365 versus $47,713 in Q1) and the highest median shares of children in poverty (23.6% versus 20.2%) and unemployment, whereas the smallest stratum (Q1) had the highest uninsured share. Several health-related covariates also varied across strata. Larger counties tended to have slightly lower age 65+ share, while some risk-factor distributions, such as diabetes prevalence, had a lighter upper tail in the largest counties than in the middle strata. Small counties also had markedly lower PM2.5 concentrations. By contrast,

some variables, such as income inequality, adult obesity, and long commute driving alone, showed substantial overlap across strata, indicating weaker stratification.

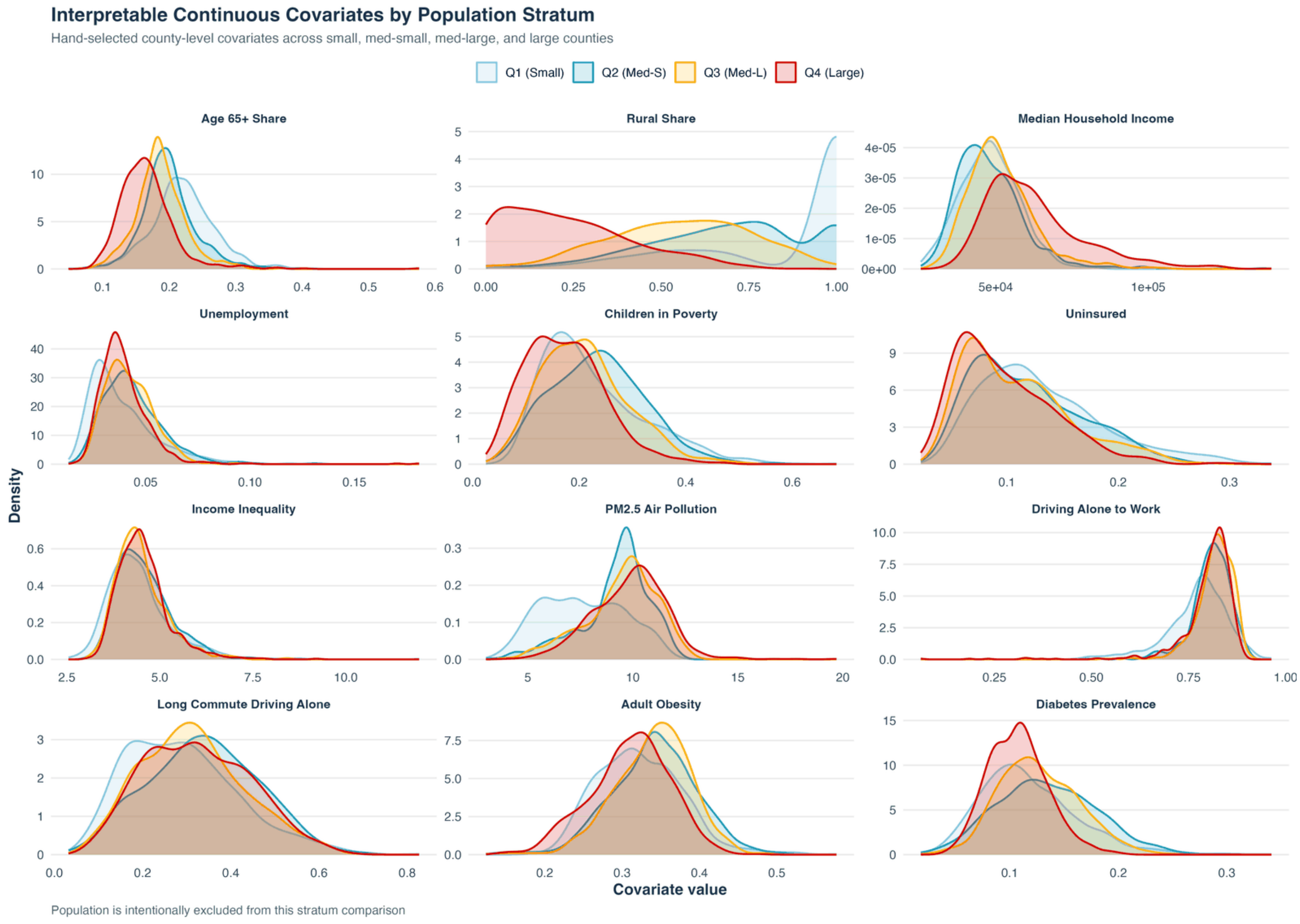


Figure S3.15: Selected continuous covariates by population stratum. The panels show a subset of the county-level covariates listed in the covariate table; population is excluded because it defines the strata.

For the time-varying temperature covariate (Figure S3.16), all four strata followed a very similar seasonal pattern over time, with distributions shifting upward in summer and downward in late fall and winter. Differences across strata were present but modest relative to the common month-to-month seasonal change.

For the discrete partisanship covariate (Figure S3.17), the population strata differed more clearly. Smaller and medium-sized counties were overwhelmingly coded as Republican, whereas

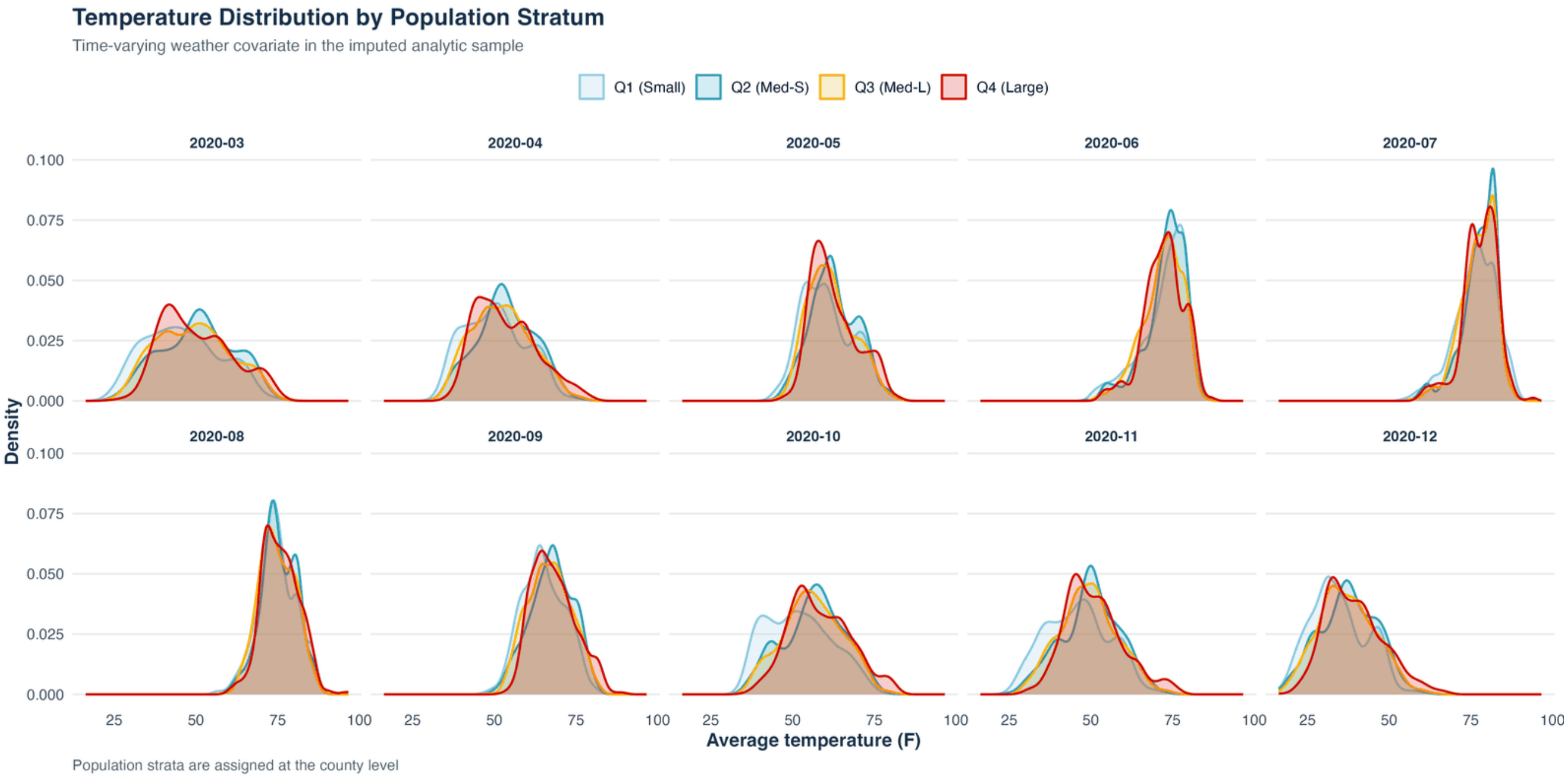


Figure S3.16: Temperature distribution by population stratum. The figure displays March–December 2020.

the largest counties had a substantially higher share coded as Democrat. This suggests that political orientation varied systematically with county size.

### S3.2.5 Propensity Overlap and Network-Block Bootstrap Inference

**Neural-model training.** We fit separate assignment models for Business and Economic Restrictions and Education and Childcare, using the architecture and joint training objective described in Sections 3.1–3.3 of the main text. The same training settings are used for both policy domains, as summarized in Table S3.11. Training and validation are split by county, while neighborhood summaries are computed on the full observed county network. As in Section S3.1.3, validation loss guides early stopping and selection of the fitted model retained for estimation. For each policy domain, training is repeated using 100 random seeds (42–141), with the same observed data and model specification. Within each run, validation loss determines the fitted model retained for estimation.

Table S3.11: Hyperparameter settings for the neural model and training.

| Category | Parameter(s) | Value |
|---|---|---|
| **Optimization** | Optimizer | AdamW |
| | Initial learning rate | $10^{-3}$ |
| | $L_2$ regularization (weight decay) | $10^{-4}$ |
| | Batch size; maximum epochs | 128; 120 |
| | Dropout rate | 0.1 |
| | Gradient clipping (max-norm) | 1.0 |
| | Mixed precision training | Disabled |
| **Scheduling** | Early stopping patience | 20 |
| | Learning-rate scheduler patience | 10 |
| **Architecture** | LSTM hidden dimension; number of layers | 128; 2 |
| | Dimensions of $(Z^X, Z^D, Z^C)$ | $(16, 16, 32)$ |
| | Spillover-exposure model | Zero–one-inflated binomial (ZOIB) |
| **Loss Weights** | Own-treatment loss weight $w_X$ | 1.0 |
| | Spillover-exposure loss weight $w_D$ | 1.0 |
| | Auxiliary reconstruction loss weight $w_I$ | 1.0 |

*Notes:* Training combines the own-treatment and spillover-exposure likelihood losses with the auxiliary Huber reconstruction loss and regularization, as described in Section 3.3 of the main text. The zero–one-inflated binomial model assigns separate probability masses to spillover-exposure levels 0 and 1 and uses a truncated binomial component for intermediate exposure levels.

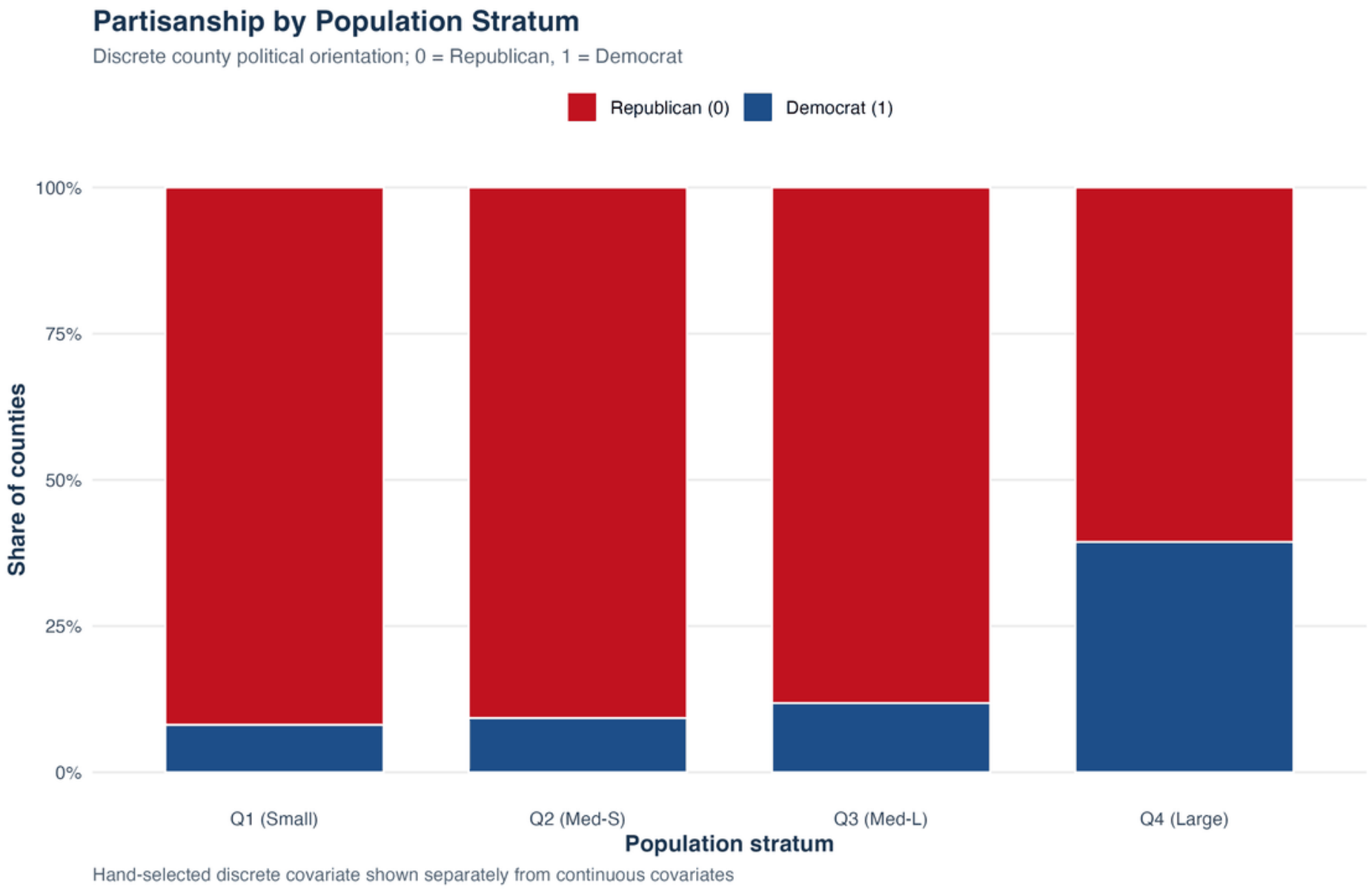


Figure S3.17: Partisanship by population stratum.

**Own-treatment propensity overlap.** Figure S3.18 shows the distributions of fitted own-treatment propensities among validation counties, separately for treated and untreated counties at each monthly decision time. The panels cover April–October 2020 for Business and Economic Restrictions and April–August 2020 for Education and Childcare. Decision steps in the panel titles index the weekly process time at which each monthly decision is made. Overlap varies across decision times and policy domains. For Business and Economic Restrictions, the two distributions overlap only at the April decision; from May through October, fitted propensities concentrate near zero for untreated counties and near one for treated counties. For Education and Childcare, untreated counties concentrate near zero already in April, the two distributions are partly mixed in May, and they are separated from June through August. This separation implies limited empirical support for the selected policy paths at most decision times, so the corresponding weights are driven by relatively few counties.

**Effect estimation and network-block bootstrap.** The effect analysis uses September and October 2020 endpoints for Business and Economic Restrictions and July and August 2020 endpoints

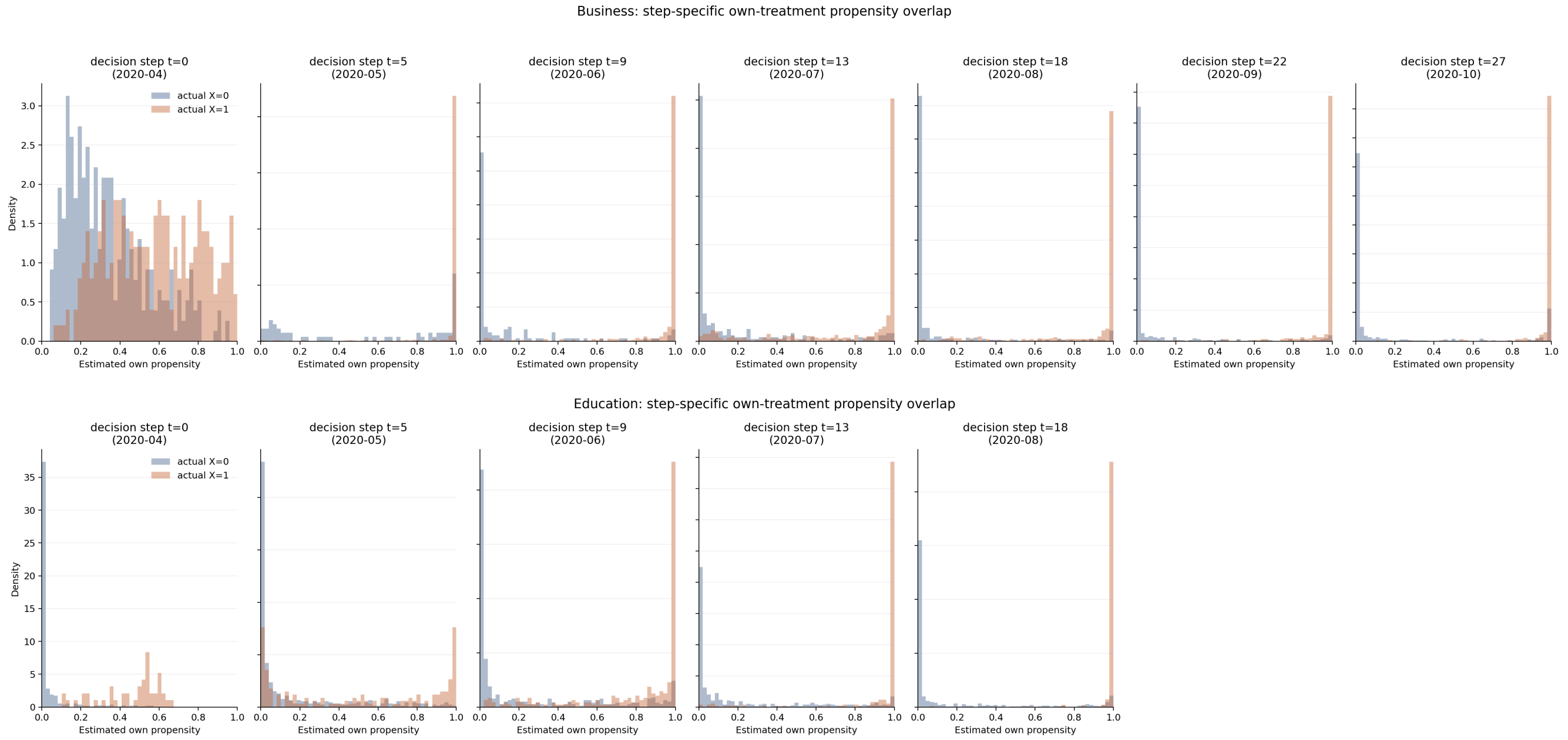


Figure S3.18: Step-specific own-treatment propensity overlap in real data analysis. Histograms show fitted propensities among validation counties, stratified by observed treatment status. The upper row corresponds to Business and Economic Restrictions, and the lower row to Education and Childcare.

for Education and Childcare. Target paths begin in April 2020 and satisfy $x_{\tau_m} = d_{\tau_m} \in \{0, 1\}$ at each monthly decision. Each selected path is compared with the joint low-intensity baseline, $x_{\tau_m} = d_{\tau_m} = 0$ throughout the corresponding history, for both the last-week and average weekly case rates. We estimate these effects on the full county network using the self-normalized estimator in Section 3.4 of the main text, with Gaussian localization and bandwidth $h = 0.0833$.

Within each of the 100 refits, we apply the fixed-model network-block bootstrap described in Section S3.1.3, holding the fitted assignment models, observed county network, constructed histories, and localized weights fixed during resampling.

For the application, we apply randomized 1-hop-max clustering to the full county network three times, obtaining three alternative partitions. Each partition assigns every county to one of a set of disjoint blocks. For each refit, $B_{\text{boot}} = 1000$ resamples use these partitions in turn, sampling blocks with replacement from the partition used for that resample. Joint recomputation of the

self-normalized means follows Section S3.1.3.

For each policy domain, endpoint month, and outcome summary, the target and baseline means are recomputed jointly within each resample, and their differences yield the bootstrap policy-effect estimates. The reported point estimate is the mean of the 100 refit-specific estimates. The reported 95% interval is formed from the 2.5th and 97.5th percentiles of the 100,000 pooled bootstrap effect estimates.

# References


Louis HY Chen and Qi-Man Shao. Normal approximation under local dependence. 2004.

Georgia Papadogeorgou, Kosuke Imai, Jason Lyall, and Fan Li. Causal inference with spatio-temporal data: Estimating the effects of airstrikes on insurgent violence in iraq. *Journal of the Royal Statistical Society: Series B (Statistical Methodology)*, 84(5):1969–1999, 2022.

Paul R Rosenbaum. Sensitivity analysis for certain permutation inferences in matched observational studies. *Biometrika*, 74(1):13–26, 1987.



\end{document}